\documentclass{article} %
\usepackage{iclr2027_conference,times}

\usepackage{amsmath,amsfonts,bm}

\def\eqref#1{equation~\ref{#1}}

\def\1{\bm{1}}

\DeclareMathAlphabet{\mathsfit}{\encodingdefault}{\sfdefault}{m}{sl}
\SetMathAlphabet{\mathsfit}{bold}{\encodingdefault}{\sfdefault}{bx}{n}

\usepackage{hyperref}
\usepackage{url}
\usepackage{float}
\usepackage{graphicx}
\usepackage{booktabs}
\usepackage{subcaption}
\usepackage{wrapfig}
\usepackage{multirow}
\usepackage{xltabular}
\usepackage{longtable}
\usepackage{caption}
\usepackage{needspace}

\title{MixCards: Evaluating Structured \\ Descriptions of Multi-Source Sound Scenes}

\author{Sripathi Sridhar$^{1*}$, Yujin Kim$^{2*}$, Giovana Morais$^{2*}$, Liang-Yuan Wu$^{2}$, \\ \bf Mehmet Atilay Kucukoglu$^{2}$, Prem Seetharaman$^{3}$, Lloyd May$^4$, \\ \bf Mark Cartwright$^{1\dagger}$, Magdalena Fuentes$^{2\dagger}$ \\[4pt]
\small $^{1}$ New Jersey Institute of Technology \enspace $^{2}$ New York University \enspace $^{3}$ Adobe Research \enspace $^{4}$ Monash University\\[2pt]
}

\iclrfinalcopy %
\begin{document}

\maketitle
\lhead{Preprint}
\renewcommand{\thefootnote}{\fnsymbol{footnote}}
\footnotetext[1]{Equal contribution. \textsuperscript{$\dagger$}Equal advising.}
\renewcommand{\thefootnote}{\arabic{footnote}}

\vspace{-8pt}
\begin{abstract}
Audio captioning benchmarks typically evaluate a single unstructured free-text description of an audio scene, providing limited insight into how a model fails to understand polyphonic audio. A plausible caption may omit a source, or conflate multiple sources into one, or assign an attribute of one source to another, and conventional captioning evaluation cannot locate these errors. We introduce MixCardBench, a benchmark for structured captioning of polyphonic audio, which evaluates
models in two stages: first, identifying and temporally localizing the sources in a scene, and second, generating a structured MixCard describing the semantic, temporal, and acoustic properties of each source. The benchmark contains 4884 ecologically plausible synthetic soundscapes with up to nine overlapping speech, music and sound effects sources built from high-quality recordings and source-level annotations, where each source has a reference MixCard. We evaluate seven state-of-the-art large audio language models and
find substantial limitations in current models, which struggle considerably with speech and sound effects timing and detection, with the strongest model reaching only around 60\% mean card-field score. Performance varies significantly across attributes, exposing capability differences that are hidden by aggregate caption scores. We assess the quality and ecological validity of MixCardBench through human validation, distribution matching and a listening study, and show that human corrections have little effect on model rankings, and our synthesized soundscapes are closer to in-the-wild YouTube videos than source guided but otherwise random synthesis. 
The benchmark dataset\footnote{\url{https://doi.org/10.5281/zenodo.23702059}} and code\footnote{\url{https://github.com/Sound-Captioning-for-Accessibility/mixcardbench}} are publicly available. Demo examples are at \url{https://mixcards-anon.github.io/mixcards-demo/}.
\end{abstract}

\setlength{\textfloatsep}{12pt plus 2pt minus 2pt}
\setlength{\abovecaptionskip}{5pt}
\section{Introduction}
\vspace{-5pt}
How we describe sound is dependent on why we are describing it and who/what is consuming these descriptions. Different consumers have different needs. Sound designers want sounds described by attributes like descriptive adjectives, visual context, and complementary sounds to aid search through SFX libraries \citep{sridhar2026audiocards}. Deaf and hard of hearing (DHH) people want sounds described in ways that match their preferences and needs \citep{may2025choices,huang2025captune,findlater2019deaf}.

Given the scale of online video uploaded every day \citep{wojcicki2020youtube15}, AI models increasingly caption sound for accessibility rather than people \citep{may2024unspoken}. Even in professional captioner workflows, automatic speech recognition (ASR) is now often used to draft captions that are subsequently edited by humans \citep{3playmedia2024}. However, captions for sound accessibility need to communicate both speech and non-speech information (NSI), i.e., all of the audio events other than the lexical content of dialogue, including sound effects (e.g., ``[Dog whines]''), music (e.g., ``[Uptempo bebop]''), and extra-speech information (e.g., ``[Whispering] I love you.''). Current models that caption NSI either have limited vocabularies \citep{chaudhuri_2017,expressivecap_google_2024} or are trained on general-purpose captions written by crowdworkers or generated by LLMs~\citep{kim2019audiocaps,drossos2020clotho,mei2024wavcaps}, with no audience or purpose in mind, whereas DHH viewers' needs for NSI vary with context and genre~\citep{may2025choices,huang2025captune}. Accessibility research has identified structured media representations as a path for building personalizable media descriptions that meet users' varied needs and preferences \citep{morris2018rich, may2024towards}, but existing benchmarks do not evaluate LALMs' abilities to infer such structured descriptions of audio. 

\raggedbottom

\begin{figure}[t]
    \centering
    \includegraphics[width=\linewidth]{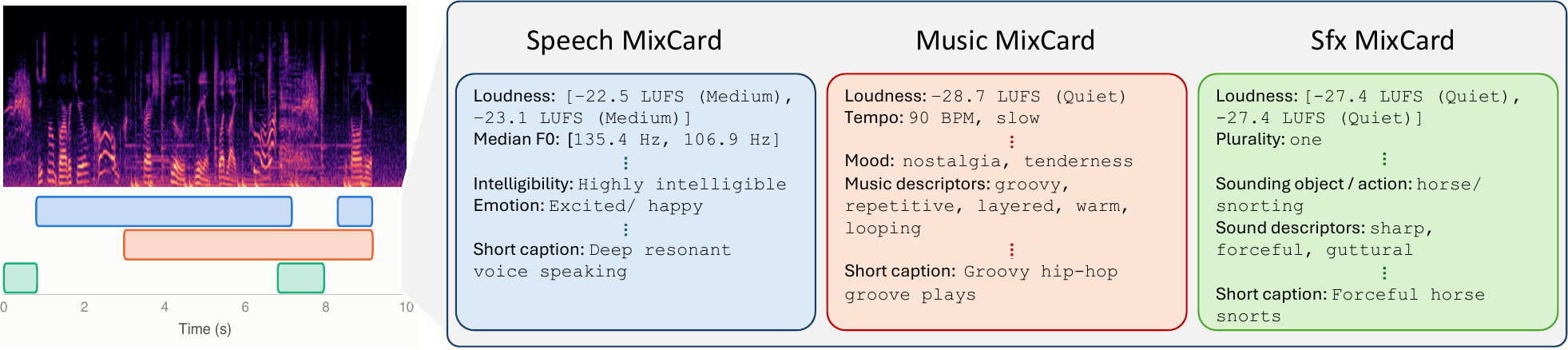}
    \caption{We propose MixCardBench, a benchmark for  multi-source structured captioning where a model detects sound sources and generates a structured description per source (i.e. MixCard). We depict an example from our benchmark and its corresponding speech, music, and sound effects ground truth MixCards. We include only a subset of MixCard fields here due to space constraints.}
    \label{fig:mixcards_fields_overview}
\end{figure}%

It is also not just people who consume descriptions of audio. Currently, the strongest audio reasoning systems first describe audio in text before reasoning on the textual output, either in single-pass caption-to-LLM cascades \citep{kumarTACTimestampedAudio2026, ma2026omni} or in audio tool-using agents \citep{wijngaard2025audiotoolagent}. 
In these systems, audio reasoning performance is bottlenecked by audio captioning performance \citep{lu2026auditory}. 
However, for polyphonic audio, caption fidelity often fails in ways that are challenging to detect and localize in unstructured text. A fluent caption may omit a source, merge two sources, or bind an attribute to the wrong source. Standard captioning metrics, whether n-gram based~\citep{liu2017improved}, embedding-based~\citep{zhou2022can,dixit2025mace}, or LLM-judged~\citep{wu2025clair}, reduce a caption's quality to a single score against the audio or reference captions, and that score does not reveal which source was missed or which property was wrong. Multiple-choice audio understanding and reasoning benchmarks have the same limitation: a wrong answer shows that a model failed somewhere, but it does not localize the failure point~\citep{sakshi2025mmau,kumar2026mmaupro,ma2026mmar}. Thus, there is currently a bottleneck in the performance of audio reasoning in large audio-language models that the field cannot adequately measure.

In this paper, we present MixCardBench, a data generation pipeline and benchmark for structured captioning of ecologically valid polyphonic soundscapes of speech, music, and sound effects. 
MixCardBench proposes a structured approach to captioning where each source is described by domain-specific (music, speech, and sound effects) field schemas that span semantic, descriptive, temporal, and acoustic information (see Figure~\ref{fig:mixcards_fields_overview}); and it includes an evaluation protocol that assesses each of the fields independently and can aggregate the results hierarchically up to the clip level. With MixCardBench we can pinpoint where models' audio understanding capabilities are failing both globally and on specific clips, shedding light on the perception failures that caption-LLM cascades and audio tool-using agents depend on and establishing measurable progress towards more accessible, personalizable audio captioning. 

To do this, we developed a scalable data-generation pipeline. It uses AudioSet Strongly-Labeled annotations, a large library of isolated source clips, and multiple stages of automated quality checks to produce ecologically valid soundscapes.  Because each soundscape is built from isolated stems, we can automatically extract detailed information about every source and present it as a structured caption.
With this approach, we aim to mitigate the effects of synthetic mixtures, which have been found to be prone to generalization errors such as overestimating rare classes like ``gunshots'' \citep{kumarTACTimestampedAudio2026}. In our analysis of our data, we found our mixtures more similar to AudioSet than random Scaper-based synthetic mixtures \citep{salamon2017scaper}, and we found evaluation on our generated annotations closely matched evaluation on a human-edited subset, reflecting the high quality of the annotations.
In analyzing our benchmark results, we find the benchmark challenging for all six LALMs we examined, with the best model achieving a mean card-field score of about 60\%. Performance varies across fields and pinpoints the strengths and weaknesses of large audio language models that traditional captioning and audio understanding 
metrics would hide.

\section{Related work}
\vspace{-5pt}
\noindent \textbf{Captioning benchmarks and metrics. } 
Automated audio captioning is a well-established task used to evaluate audio understanding, typically evaluated on single-sentence captions \citep{drossos2020clotho, kim2019audiocaps}. Models successfully use LLM-augmented captions to scale training data and achieve strong performance on single-sentence captioning benchmarks \citep{mei2024wavcaps, sunAutoACDLargescaleDataset2024, bai2025audiosetcaps}.
However, single-sentence captions make it challenging to evaluate models' audio understanding in a fine-grained manner. Reasoning benchmarks do evaluate models on specific audio understanding skills in closed-vocabulary and open-vocabulary settings \citep{sakshi2025mmau, kumar2026mmaupro, ma2026mmar, wuMMACMassiveMultidimensional2026}, including focus on multi-source scenes \citep{chen26aa_interspeech}, but do not expose failure points in the model reasoning, i.e., a wrong answer for ``Is there a dog barking in this file?" could mean a failure to identify the sounding object (the dog), the sounding action (barking), or both.

\noindent \textbf{Time-aligned and Structured captions. }%
Dense captioning, involving captioning individual objects in an image \citep{Johnson_2016_CVPR} and describing relationships between objects in a scene \citep{krishnaVisualGenomeConnecting2017}, is well-studied in the vision community \citep{Johnson_2017_CVPR, urbanek2024picture77}.
In the audio domain, models \citep{kumarTACTimestampedAudio2026} and benchmarks \citep{primusTACOSTemporallyalignedAudio2025, xie2025audiotime} have been developed to perform and evaluate time-aligned captioning of multi-source scenes. TAC \citep{kumarTACTimestampedAudio2026} does dense captioning of all sound events in a scene with timestamps, but does not generate structured metadata per source that would enable fine-grained evaluation along semantic axes. 
Structured captioning has been explored in sound design \citep{sridhar2026audiocards}, music \citep{bukey2026rethinking} and general sound \citep{taheri2025sarlm}, but only for single-source or global descriptions. Others \citep{jiangSciPhiLargeLanguage2025} develop multi-source captioning models, but their output fields relate only to acoustic and spatial attributes, not semantics.

\section{MixCardBench}
\label{sec:mixcards}

\begin{figure}
    \centering
    \includegraphics[width=\linewidth]{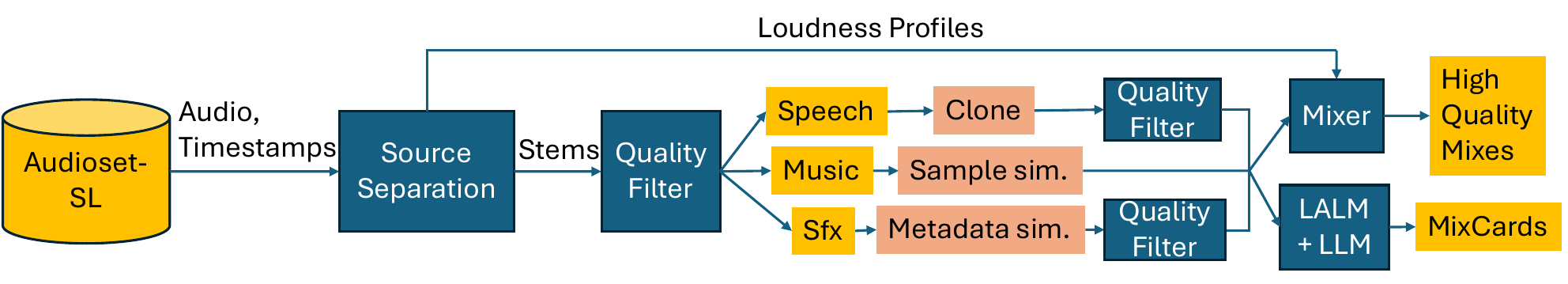}
    \caption{Our pipeline used to generate open, high-quality, ecologically-valid soundscapes containing speech, music, and sound effects sources. We use SAM Audio for source separation. We use Vevo-Timbre to clone separated speech into public domain voices. We use MULE embeddings to find similar music as the original AudioSet mixture. We use ontology mapping and sound event detection using PE-A-Frame and PE-AV to sample sources matching the target AudioSet event class.}
    \label{fig:overall-pipeline}
\end{figure}

\vspace{-5pt}
\subsection{Overview and Design principles}
\textbf{Ecological validity.} MixCardBench is intended to measure model capabilities that should transfer to naturally occurring audio. %
Implausible timing, co-occurrence, or levels in the soundscape synthesis could make the benchmark measure synthesis artifacts rather than real-world source reasoning. MixCardBench instead respects co-occurrence of events as they happen in the wild, features examples where context and timing of events make sense together (e.g. a phone ringing followed by a person answering it), and has mixtures that reflect relative loudness of events as they occur in real settings. We use AudioSet-SL \citep{hershey2021audiosetstrong} eval as our reference dataset to guide our soundscape synthesis, which features audio and temporally strong annotations (onset and offset of events) from YouTube videos, fairly approximating real world data. An evaluation of MixCardBench's ecological validity is presented in Section~\ref{sec:bench_validation}.%

\textbf{Open data.} Our benchmark will be open under a non-commercial research license, including the mixtures and their associated structured descriptions. AudioSet-SL itself cannot be redistributed independently and is subject to dataset drift as source videos become unavailable over time. We therefore reconstruct the soundscapes using high-quality open speech, music, and sound-effects datasets, while preserving the clip-level structure of AudioSet-SL through its event timing, co-occurrence patterns, and relative loudness profiles. %
We illustrate the reconstruction pipeline in Figure \ref{fig:overall-pipeline}, and describe its process in Section~\ref{ssec:benchmark_construction}. 

\textbf{Source-level structured descriptions in polyphonic settings.} 
MixCardBench provides a ground-truth structured caption, or \textit{MixCard}, for each source in the audio, enabling evaluation of both source detection and its semantic, temporal, and acoustic properties. Such supervision is difficult to obtain from natural soundscapes, where source-specific attributes are entangled in the mixture. We define separate schemas for music, speech, and sound effects (Figure~\ref{fig:mixcards_fields_overview}), including free-text descriptions, object and action labels, loudness, timing, frequency profiles, and category-specific attributes such as genre, mood, instruments, and tempo. These fields are automatically derived from the isolated sources using signal processing, specialized models, and LALMs, as described in Section~\ref{ssec:benchmark_construction} and Appendix~\ref{app:B_mixcardbench_audio_construction}. This source- and field-level granularity allows us to localize model failures beyond source detection. Complete schemas are provided in Appendix~\ref{app:C_mixcards_schema}.

\textbf{Multi-level diagnostic evaluation.} We evaluate each MixCard field using a metric appropriate to its data type and aggregate scores hierarchically at the card, sound-category, and benchmark levels. This supports both overall model comparison and fine-grained analysis of source- and field-specific failures. The complete evaluation protocol is described in Section~\ref{ssec:eval_protocol}.

\textbf{MixCardBench Overview.} Our benchmark has a total of 4884 ecologically plausible synthetic soundscapes, each of them 10 seconds long, for a total of 13.6 hours of audio. Polyphony ranges from 1 to 9 sources, where 52.5\% of soundscapes contain a single source,  47.5\% contain two or more sources; 27.7\% contain at least three sources and 13.6\% at least four. We have a total of 9618 MixCards, where 46.7\% of them are sound effects, 41.7\% are music, and 11.6\% are speech. More details and figures can be found in Appendix~\ref{app:E_mixcardbench_stats}.

\subsection{MixCardBench Audio Pipeline}
\label{ssec:benchmark_construction}

We generate synthetic soundscapes by reconstructing the event structure of AudioSet-SL recordings using high-quality source datasets and Scaper \citep{salamon2017scaper}, an open-source software for soundscape synthesis. 

\textbf{Source datasets.}
For each reference recording, we retrieve its event annotations and use SAM Audio~\citep{shi2026sam} to separate the audio into speech, music, and residual. These separated stems are used to guide the ecologically-plausible synthesis of soundscapes and replace each event with a corresponding source from one of the following datasets: Pro Sound Effects (PSE) \citep{pse2026}, a professional sound-effects library; LibriTTS-R \citep{koizumiLibriTTSRRestoredMultiSpeaker2023} and VCTK \citep{yamagishi2019cstr}, which provide speech recordings for voice resynthesis; and the Free Music Archive (FMA) \citep{defferrad2017fma}, which provides music recordings. For all datasets, we sample from the official test sets where possible, or create our own test split for MixCardBench. Details can be found in Appendix \ref{app:B_mixcardbench_audio_construction}.

\textbf{Label processing.} AudioSet-SL annotations require preprocessing before synthesis. We merge nearby instances of the same class to avoid over-segmenting single events, and remove redundant parent labels when they overlap child events by more than 80\%, preventing artificial inflation of source count. See Appendices \ref{app:A_annotation-cleanup} and \ref{app:judge-filtering} for details.

\textbf{Ontology mapping.} We map AudioSet classes to the UCS ontology using an automated alignment pipeline, then manually review and correct the mappings. We also replace mapped classes that are not represented in the source library with the nearest available class and drop the small number of classes that cannot be reliably mapped. Details can be found in Appendix~\ref{app:A_annotation-cleanup}.

\textbf{Source separation.} We use SAM Audio \citep{shi2026sam} to extract reference speech, music, and sound-effect stems from AudioSet-SL mixtures. Its timestamp-conditioned prompting allows us to leverage AudioSet-SL onset/offset annotations during separation. We map AudioSet classes to the three source categories and separate them iteratively, with further details provided in Appendix~\ref{app:source_separation}.

\textbf{Loudness mixing templates.} To preserve realistic relative source levels, we estimate the loudness of the separated speech, music, and sound-effect stems from each AudioSet-SL recording using LUFS, and transfer these relative levels to the corresponding replacement sources in the reconstructed soundscape. Because sound effects are separated as a single consolidated stem, we first mix their replacement sources and apply a shared gain adjustment to match the reference sound-effects loudness. We use a peak limiter on the final mixture to prevent clipping while preserving these relative source levels. More details in Appendix~\ref{app:loudness_mixing}.

\textbf{SFX Processing.} To identify suitable replacement events within PSE recordings, we use PE-A-Frame \citep{vyasPushingFrontierAudiovisual2025} to localize candidate occurrences of each target AudioSet-SL class within its mapped UCS categories. Because PE-A-Frame provides only coarse activity regions, we refine event boundaries using a background-relative RMS detector and construct candidate pools for each class. We then use PE-AV \citep{vyasPushingFrontierAudiovisual2025} to compute text-audio similarity and retain the top 10\% most semantically relevant candidates. For each AudioSet-SL event, we select candidates with similar duration and sample a matching segment, with additional padding and fades to preserve event onsets and avoid abrupt transitions. More details in Appendix~\ref{app:sfx}.

\textbf{Speech processing.} To preserve the linguistic and paralinguistic content of the original speech while replacing speaker identity, we resynthesize separated AudioSet-SL speech using Vevo-Timbre \citep{zhangVevoControllableZeroShot2024, amphion} with public voices from LibriTTS-R \citep{koizumiLibriTTSRRestoredMultiSpeaker2023} and VCTK \citep{yamagishi2019cstr}. The separated speech conditions the speaking style and emotion, while a reference utterance provides the target timbre. We apply a series of quality filters to remove problematic cases, including overlapping singing, multi-speaker speech, non-English segments, and examples with poor transcription or audio quality, using a transcription-diarization pipeline \citep{bain2022whisperx}, WER, and DNSMOS. These filters retain 32\% of the resynthesized speech. More details in Appendix~\ref{app:speech}.

\textbf{Music processing.} To preserve high-level musical characteristics such as mood and genre, we use MULE \citep{mccallum2022mule} to retrieve acoustically similar music from the FMA test set \citep{defferrad2017fma}. We first identify the FMA track with the highest cosine similarity to the reference music, and then select the most similar segment within that track while matching the duration of the original event. We mask silent regions before retrieval to reduce matches to silent segments. Once inserted into the soundscape, we recompute the event onset and offset from its RMS envelope and zero-crossing points. More details in Appendix~\ref{app:music}.

\subsection{MixCards Generation}
\label{ssec:mixcards_generation}

We propose three types of MixCards: speech, music, and sound effects. Each unique sound class in the sound effects category is assigned a separate MixCard, while speech and music sources are described in one MixCard per category. The fields for each card type are presented in Appendix~\ref{app:C_mixcards_schema}. We generate MixCards through a multi-stage pipeline. First, we populate each card with measured signal features (e.g., event timestamps, loudness, frequency profiles, and pitch) and available source metadata (e.g. genre). Next, we use an LALM (Qwen3-Omni-30B-A3B-Instruct) to generate semantic and perceptual attributes from the isolated source stems, prompting it separately for each field. Additional attributes, such as music tempo and speech intelligibility and emotion, are estimated using domain-specialized models (e.g., BeatThis! \citep{foscarin2024beatthis} for tempo). To reduce redundancy in the LALM-generated tags, we manually construct closed vocabularies for selected fields, either by consolidating generated tags or adopting established vocabularies (e.g., GEMS-9 for music mood). Finally, an LLM (Qwen3.6-35B-A3B-FP8) verifies and refines the generated tags, and generates short and descriptive captions. We then enforce vocabulary constraints and output one structured JSON per MixCard. The complete generation pipeline is described in Appendix~\ref{app:D_mixcard-generation}.

\subsection{Evaluation protocol}
\label{ssec:eval_protocol}

We evaluate model predictions in two stages: Stage 1) sound event detection and temporal localization, and Stage 2) evaluation of the structured descriptions for each detected source. For the first stage, following TAC~\citep{kumarTACTimestampedAudio2026}, we evaluate source detection and temporal localization using event-based F1 with a 1-second onset and offset tolerance, and segment-based F1 with 100-ms windows. For each predicted-reference event pair, we compute LLM-based semantic similarity and perform optimal bipartite matching with a 0.5 threshold.

For the second stage, we adapt the structured caption evaluation framework of~\cite{wu2026evaluation}, extending it to speech, music, and sound effects. We evaluate each MixCard field independently using a metric appropriate to its data type. Numerical fields (e.g., loudness, pitch, and tempo) are evaluated using mean absolute error normalized by the field's median absolute deviation, while ordinal fields (e.g., frequency profiles and loudness levels) are scored according to the distance between predicted and reference levels. For semantic fields, we use an LLM judge to compare predicted and reference tag sets (e.g., sounding objects, actions, and music descriptors), accounting for valid synonyms while penalizing missing and spurious attributes. Free-text captions are evaluated using the same judge across four criteria: correctness, relevance, completeness, and clarity~\citep{kumar2026mmaupro}. In \citet{wu2026evaluation}, the authors found that LLM-based evaluation of free-text descriptions is more robust to meaning-preserving paraphrases, avoiding penalizing valid linguistic variations as errors, a desirable property for our evaluation. All field scores are normalized to $[0,1]$ so they are comparable, and aggregated hierarchically to obtain card-level, sound-category, and overall benchmark scores, enabling both overall model comparison and fine-grained analysis of source-specific errors. We use Llama-3-8B-Instruct \citep{llama3modelcard} as the judge. Details about the evaluation protocol can be found in Appendix~\ref{app:G_eval_protocol}.

\begin{table}[ht]
\centering
\small
\caption{Mean card field score (all fields) on the 565  validated cards, scored against the original GT (Orig.) and the validated GT (Val.).}
\label{tab:validated_overall}
\begin{tabular}{lrrrrrrrrr}
\toprule
 & \multicolumn{3}{c}{Speech} & \multicolumn{3}{c}{Music} & \multicolumn{3}{c}{Sound effects} \\
\cmidrule(lr){2-4} \cmidrule(lr){5-7} \cmidrule(lr){8-10}
Model & Orig. & Val. & $\Delta$ & Orig. & Val. & $\Delta$ & Orig. & Val. & $\Delta$ \\
\midrule
Qwen3-Omni & 0.639 & 0.645 & $+0.006$ & 0.595 & 0.575 & $-0.020$ & 0.598 & 0.588 & $-0.011$ \\
Qwen2.5-Omni & 0.475 & 0.462 & $-0.013$ & 0.470 & 0.465 & $-0.005$ & 0.509 & 0.508 & $-0.001$ \\
AF 3 & 0.031 & 0.030 & $-0.001$ & 0.045 & 0.042 & $-0.003$ & 0.128 & 0.139 & $+0.010$ \\
Music Flamingo & 0.307 & 0.307 & $+0.001$ & 0.305 & 0.293 & $-0.013$ & 0.298 & 0.305 & $+0.007$ \\
AF Next & 0.339 & 0.346 & $+0.007$ & 0.307 & 0.307 & $-0.001$ & 0.471 & 0.475 & $+0.003$ \\
Kimi-Audio & 0.430 & 0.441 & $+0.011$ & 0.357 & 0.350 & $-0.007$ & 0.328 & 0.331 & $+0.003$ \\
\bottomrule
\end{tabular}
\label{tab:validated_performance}
\end{table}

\section{Benchmark Validation}
\label{sec:bench_validation}
\vspace{-5pt}
\begin{wraptable}{r}{0.32\textwidth}
\vspace{-11pt}
\centering
\small
\caption{Synthetic soundscape distributions vs. AudioSet-SL.}
\label{tab:eco-kad-fad}
\vspace{-4pt}
\begin{tabular}{lcc}
\textbf{Dataset} & \textbf{KAD} $\downarrow$ & \textbf{FAD} $\downarrow$ \\ \hline
Ours      & \textbf{3.26} & \textbf{1.28} \\
Variant 1 & 4.00 & 1.74 \\
Variant 2 & 4.48 & 1.86 \\
Variant 3 & 6.29 & 2.16
\end{tabular}
\vspace{-4pt}
\end{wraptable}

\textbf{Annotation and Soundscape Quality.} We selected 300 soundscapes for human validation, approximately preserving the distribution of speech, music, and sound effects combinations in the full generated dataset prior to the final benchmark filtering ($\sim$17K soundscapes). In total, a team of eight human audio experts evaluated 839 distinct MixCards (158 music, 161 speech, and 520 sound effects).
Reviewers assessed the correctness of the generated tags for each field, either accepting or editing them, and discarded cards deemed unsuitable. Overall, 65 cards (7.7\%) were discarded, with drop rates of 7.0\%, 8.7\%, and 7.7\% for music, speech, and sound effects, respectively. In summary, human reviewers dropped the cards in which 1) stems were inaudible or contained implausible audio content, 2) stems contained more than one sounding object when they were supposed to have only one, or 3) the audio content did not match the assigned card label (e.g., stems containing speech data from FMA which were mapped to music). These reflect mostly synthesis errors rather than card generation errors. We identified problematic sound effects classes from dropped cards and applied this filter on the overall eval set, bringing down our effective error drop rate in validation to 5.6\%. Among reviewed cards (i.e. the 774 that were not dropped), 41.2\% of music, 14.5\% of speech, and 25.4\% of sound effects annotations were edited. Perceived audio quality (e.g. muffled, room reverb) was the most frequently edited field across all three categories, followed by music descriptors (e.g. repetitive, minimalist) and genre for music, intelligibility for speech, and sound descriptors (e.g. sharp, bright) and temporal descriptors (e.g. periodic) for sound effects. To further assess the quality of the automatically generated annotations, we evaluate the models against both the original MixCards and their human-corrected versions using our proposed evaluation protocol. This analysis quantifies the magnitude of these errors in terms of the benchmark's evaluation metrics, the potential impact of annotation inaccuracies on model evaluation, and identifies which fields are most affected. Table~\ref{tab:validated_performance} presents the results of this comparison. Model rankings are unchanged for all three card types, and changes in mean card score are small, with an absolute difference of at most $0.020$ (Qwen3-Omni on music), comparable to evaluation noise.

\begin{wrapfigure}{r}{0.35\linewidth}
    \centering
    \includegraphics[width=\linewidth]{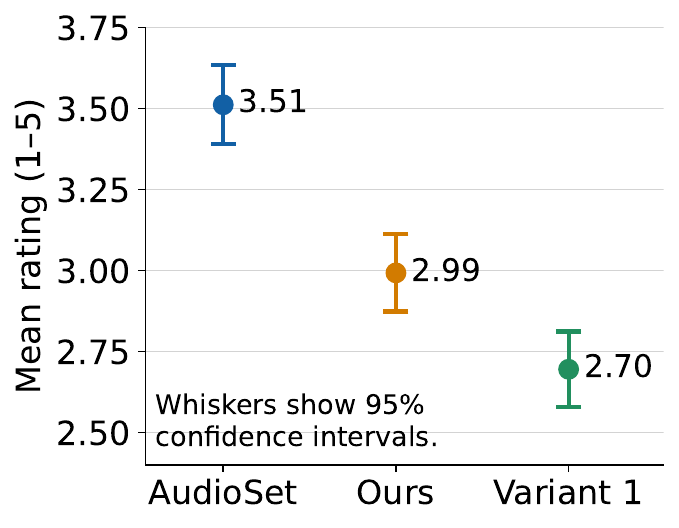}
    \caption{Mean ecological validity.}
    \label{fig:listening study}
\end{wrapfigure}

\textbf{Ecological Validity. }We conduct a quantitative study to verify the ecological validity of soundscapes generated by our synthesis pipeline. We compare against baselines containing mixes with increasing levels of randomness, while still using our source datasets. Our pipeline uses AudioSet-SL labels and timestamps, while Variants 1-3 do not. Variant 1 uses exactly the same clips used in our synthesized soundscapes, Variant 2 uses the same filtered candidate pools for music and SFX, and speech clips from where we cloned voices (i.e. prosody is not related to the soundscape), while Variant 3 uses our source datasets (PSE, FMA, and speech datasets) but with completely random sampling. We compare these soundscape dataset distributions against AudioSet-SL and present results in Table \ref{tab:eco-kad-fad}. We find that our dataset is closest to the AudioSet-SL distribution. Additionally, we conduct a qualitative listening study  for which we randomly selected 60 clips, 20 clips each from AudioSet-SL, 20 from MixCardBench, and 20 from Variant 1 which is the second closest synthesis to AudioSet-SL after ours. Twenty-one participants were asked to rate each clip on a 5-point plausibility scale (from \texttt{1-Not at all plausible} to \texttt{5-Extremely plausible}) in response to the question \textit{``How plausible is it that this audio clip could have come from a YouTube video?''} We report the mean of the ratings for each subset in Figure \ref{fig:listening study}. Our reconstructed soundscapes are rated as more plausible than the randomized variant, supporting the benefit of preserving real-world event structure. They remain less plausible than the original AudioSet-SL recordings. 

\section{Results and Discussion}
\vspace{-3pt}
\subsection{Experimental Setup}

\textbf{Inference. }To evaluate models on MixCardBench, we define a two-stage inference procedure for predicting the structured description of an audio scene. (1) \textbf{Event detection.} For each audio clip, the model identifies all audible sources, classifies them as speech, music, or sound effects, and predicts their corresponding onset/offset timestamps. Speech and music are treated as separate categories, while all other sound events are classified as sound effects. (2) \textbf{MixCard generation.} For each detected source, the model generates a complete MixCard in JSON format, predicting all fields defined by the corresponding category-specific schema in a single inference step. We adopt a two-stage inference procedure because preliminary experiments showed that separating event detection from MixCard generation improved source identification and adherence to the required output format.

\textbf{Models.} We evaluate seven state-of-the-art large audio language models (LALMs): Qwen3-Omni-30B-A3B-Instruct, Qwen2.5-Omni-7B, Audio Flamingo 3, Music Flamingo, Audio Flamingo Next, Kimi-Audio-7B-Instruct, and Gemini-3.1-Pro. All models are evaluated zero-shot using the same two-stage protocol.

\subsection{Main Results}
\begin{figure}[ht]
    \centering
    \includegraphics[width=\linewidth]{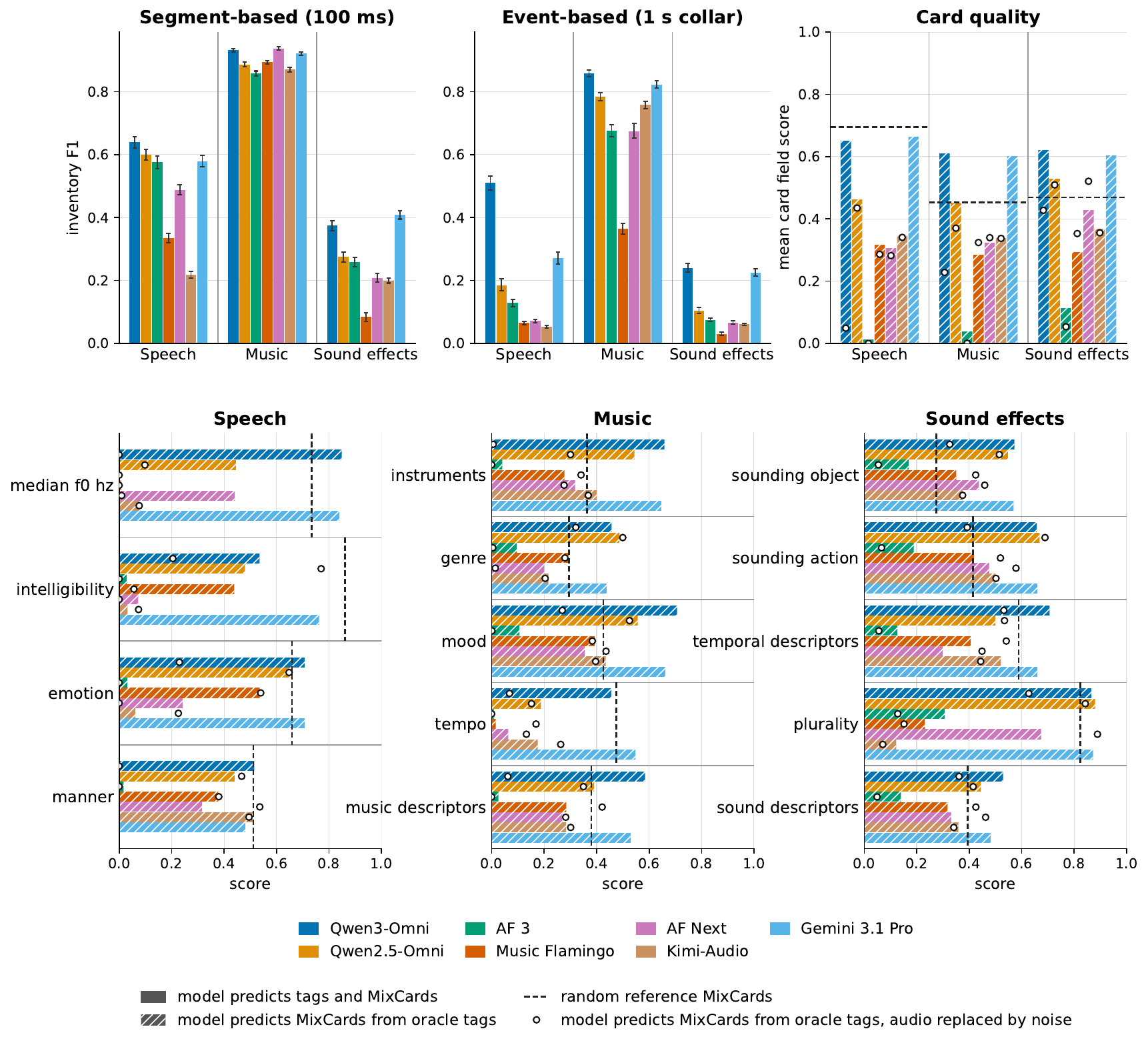}
    \caption{ Benchmark performance. Top: source detection and mean MixCard score by category. Bottom: category-specific MixCard field scores. Error bars show 95\% confidence intervals.}
    \label{fig:overview_unique}
\end{figure}

\vspace{-2pt}
\textbf{MixCardBench is challenging.} Figure~\ref{fig:overview_unique} summarizes models' performance across the two evaluation stages: Stage 1, sound event detection and temporal localization, and Stage 2, MixCards generation for speech, music and sound effects. Models detect and localize music relatively well, but perform substantially worse in speech and sound effects. Structured MixCard generation is also challenging. The strongest model achieves only around 60\% mean card-field score.

\textbf{Performance varies significantly across MixCard attributes, exposing differences typically obscured in a single aggregated caption.} Figures~\ref{fig:overview_unique} and \ref{fig:common_fields} showcase one of the most important findings of our benchmark: model performance differs widely across MixCard fields. Fields such as plurality in sound effects show strong performance compared to sound descriptors or sounding objects, which are substantially harder. Manner of speech is the hardest field for speech, and surprisingly, genre and tempo for music given their high-level perceptual salience. Performance also varies across fields shared by all card types, showing that models do not exhibit a uniform ability to describe the same properties across different source types.

\textbf{Source detection and localization is a major bottleneck.} Music is detected and localized more reliably than speech and sound effects. The larger event-based drop for speech and sound effects indicates that temporal localization remains a major bottleneck, particularly for sound effects.

\textbf{Qwen3-Omni and Gemini 3.1 Pro are the strongest models, with minor differences in individual scores}. Gemini 3.1 Pro has the higher segment-based F1 (0.701 vs. 0.648) and Qwen3-Omni the higher event-based F1 (0.535 vs. 0.434). With oracle tags, i.e. conditioned on ground truth stage 1 predictions, their mean MixCard scores are close (0.619 vs. 0.609). Qwen3-Omni helped generate some MixCard fields, which may bias the benchmark in its favour. However, Gemini 3.1 Pro, which was not used to build the benchmark, performs on par, suggesting that Qwen3-Omni's performance cannot be fully attributed to this bias. Both models nevertheless leave substantial room for improvement. On the other hand, Audio Flamingo 3 performs poorly on MixCard generation, and inspection of its outputs reveals that it frequently leaves card fields blank.

\textbf{Correct source tags alone do not lead to better MixCards}. With oracle tags, Stage 2 scores change by at most 0.035 compared with predicted tags, and still range from 0.070 (AF3) to 0.619 (Qwen3-Omni) (Figure \ref{fig:oracle_predicted_unique_fields}). Models thus also struggle to describe source attributes from the audio, beyond detecting the sources.

\textbf{The strongest models rely more on the audio.} We compare against random reference MixCards, drawn at random from dataset MixCards of the same type (dashed lines in Figure \ref{fig:overview_unique}), and against oracle-tag predictions with the audio replaced by Brown noise (circles). Random reference MixCards beat most models on numerical and quantized fields, such as median F0, intelligibility, tempo and plurality, and on the overall speech card score (0.694 vs. 0.662 for the best model). This reflects dataset priors from our construction and filtering pipeline. Qwen3-Omni and Gemini 3.1 Pro beat this baseline on the semantic and descriptive music and sound effects fields. With noise input, Qwen3-Omni scores lower on all 29 fields, while the other open models change by less than 0.1 in card score, suggesting they rely mainly on oracle tags and language priors. The audio matters least for sound effects, whose oracle tag already names the sound class.

\begin{figure}[ht]
    \centering
    \includegraphics[width=0.95\linewidth]{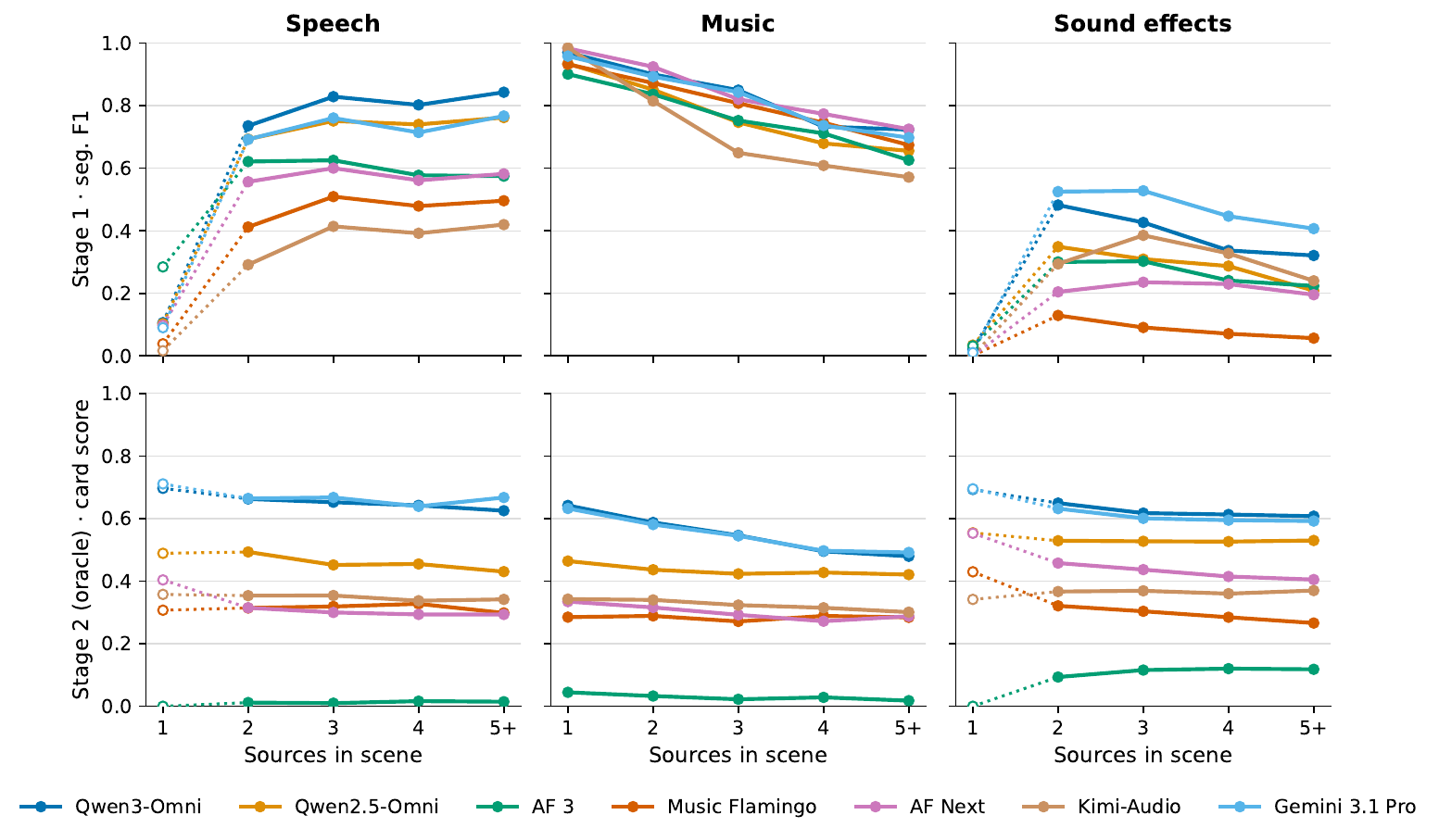}
    \caption{Effect of polyphony on Stage-1 segment F1 and Stage-2 (oracle) card scores by domain.}
    \label{fig:polyphony_modality}
\end{figure}

\subsection{How do models fail in card generation?} %

Across all card categories, models are strongest in numerical fields (e.g. loudness, frequency profile), weaker on semantic and temporal fields (e.g. sounding object), and particularly weak on free-text fields (descriptive or short captions). 
Errors in semantic fields for most models consist of missing sources or attributes, and less so hallucinating them. Models miss 51.4\% of reference attributes on average, compared with 35.6\% spurious attributes, with genre frequently omitted in music and sounding-object predictions particularly prone to spurious attributes for sound effects. Failures are only weakly shared across models, suggesting that models share some field-level weaknesses but often fail on different examples. Details can be found in Appendix~\ref{app:extended_results}.

\subsection{How does polyphony affect performance?}
\vspace{-2pt}

We look at the effect of polyphony in both stages (source detection and description). Source detection as per Segment-based F1 drops systematically across models as the number of sources increases in the recording, and if the number of overlapping sources increases, suggesting that the models struggle considerably in polyphonic settings. This drop in F1 is mostly driven by recall, which steadily drops (i.e. models miss more sources when there are more sources present), while precision remains more steady, suggesting the models do not hallucinate more sources with increased polyphony. Regarding the description of identified sources (Stage 2), performance remains comparatively stable as polyphony increases, indicating that scene complexity primarily affects source identification rather than description quality once a source has been detected.

\textbf{Breakdown of detection performance.} As polyphony increases, music F1 decreases consistently and sound effects F1 slightly declines, while speech F1 improves or stays stable. To better understand these trends, we plot the source SNR and domain prevalence as a function of polyphony in Figure \ref{fig:polyphony_modality_snr_prevalence}. We find that the combination of these trends could explain variation in detection performance across models as a function of polyphony and source category.

\subsection{Source detection failures}

Music is generally the easiest source to identify. Most models achieve high precision and recall, showing that music is detected when present and rarely hallucinated. Speech patterns differ per model, but most models achieve high recall, but lower precision (except AF3 which is the opposite), indicating that this category is mostly over-predicted. Same as shown in Figure~\ref{fig:overview_unique}, sound effects is the most challenging category, with low precision and recall for all models. This suggests models are both missing events, and hallucinating them. 
\begin{wrapfigure}[11]{r}{0.45\linewidth}
    \vspace{-4pt}
    \centering
    \includegraphics[width=\linewidth]{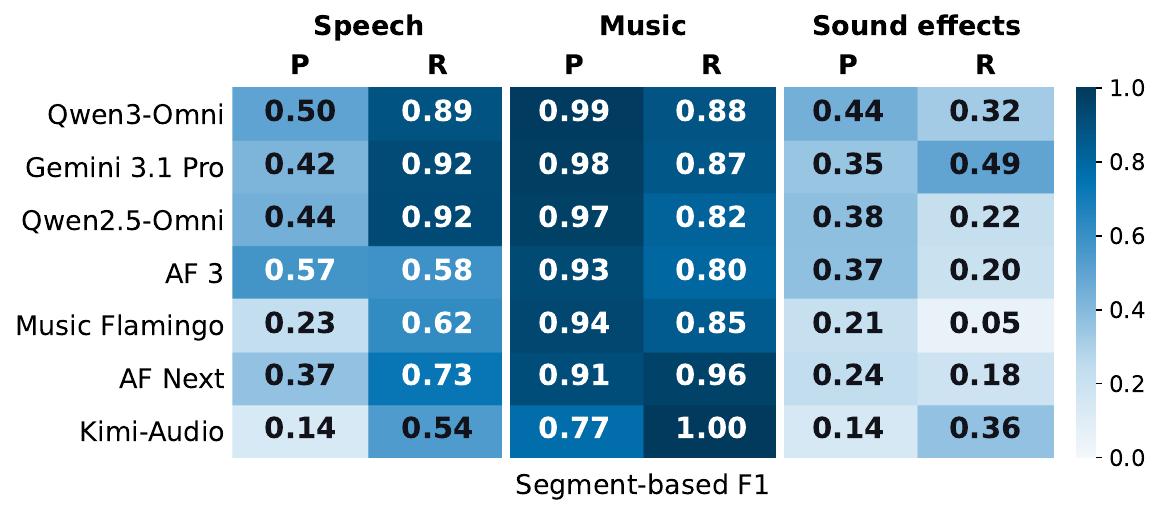}
    \caption{P/R heatmap on segment-based F1.}
    \label{fig:precision_recall_segment}
\end{wrapfigure}
\vspace{-0.1pt}
In terms of models, Kimi seems to prioritize recall over precision, recovering most speech, music and sound effects but at the cost of many false positives, while in contrast Qwen3-Omni seems more balanced in terms of precision and recall. Gemini 3.1 Pro also favors recall over precision, and is slightly worse than Qwen3-Omni at speech but better at sound effects. We also observed that the number of missing sources increased with polyphony across models, but the number of hallucinated sources remained similar. More details can be found in Appendix~\ref{app:extended_results}. In summary, source identification errors seem to be highly dependent on the category and the model, with sound effects being the clear bottleneck.

\section{Conclusions, Limitations and Future Work}
\vspace{-5pt}
We introduce MixCardBench, a benchmark for evaluating structured, source-level descriptions of polyphonic audio containing speech, music, and sound effects. MixCardBench measures whether models identify individual
sources and correctly describe their semantic, temporal, and acoustic properties. Our evaluation of six LALMs shows that this is challenging: models localize music relatively well, but struggle substantially more with speech and sound effects, and the strongest model achieves only around 60\% mean card-field score. Performance also varies considerably across MixCard fields, revealing model strengths and failure modes that would be obscured by a single caption-level score.
Finally, our human validation shows that correcting automatically generated MixCards has little
effect on model rankings or aggregate performance and the main quality bottleneck lies in synthesis, while our ecological-validity analysis shows that preserving real-world event structure produces soundscapes closer to AudioSet-SL than increasingly randomized synthesis variants.

MixCardBench also has several limitations that motivate future work. (i) Its coverage is constrained by the source datasets, with some music and sound-effect categories, including non-Western music, underrepresented; future versions should broaden source diversity.
(ii) Soundscape reconstruction can introduce synthesis artifacts, and attributes inferred from isolated sources may be perceived differently in the final mixture; improving synthesis and context-aware annotation is therefore an important direction. (iii) The notion of a source can be ambiguous, and the current MixCards do not model changes in source attributes over time or multiple source instances with different attributes; future extensions could represent source hierarchy
and temporally varying attributes. (iv) Since reference annotations are limited to 10s, our benchmark does not address source tracking and continuity across 10s boundaries in longer clips. (v) Despite filtering, some multi-speaker speech may remain, motivating stronger speech processing and explicit support for multiple speakers. (vi) Finally, Qwen-family models contribute to parts of the MixCard generation pipeline, which may favor related models in the evaluation; future versions should further reduce model-family dependence in annotation generation and expand evaluation to a broader set of LALMs.

\setlength{\textfloatsep}{20pt plus 2pt minus 4pt}
\setlength{\abovecaptionskip}{10pt}
\subsection*{AI Use Statement}
In this work, we used generative AI tools for feedback on research methodology, implementing methods, generating synthetic data, and cleaning generated data. We have not used generative AI tools for developing conceptual frameworks, proposing or refining hypotheses, or interpreting results. Formulating or proving mathematical claims, translation, and qualitative analysis are not applicable to this work. Additionally, we used generative AI tools for suggesting experimental parameters, creating and editing software code, creating and modifying scientific figures, creating artifacts (audio review tools and dashboards), drafting parts of the paper (tables), checking methods details in the paper against our code, summarizing and analysing existing literature, identifying relevant literature, and sourcing information. We have reviewed all AI-assisted work. We take responsibility for the final content of this work, including text, claims or artifacts produced with the aid of generative AI.

\subsection*{Ethics Statement}
MixCardBench is intended for research on audio understanding and structured audio captioning.
Our synthesis pipeline uses AudioSet-SL as a reference for the temporal structure, co-occurrence,
and relative loudness of real-world sound events. We do not redistribute the original AudioSet-SL
recordings. Instead, we reconstruct the soundscapes using replacement speech, music, and sound
effects from the source datasets described in Section~\ref{ssec:benchmark_construction}. For speech, we preserve linguistic and paralinguistic characteristics while replacing speaker
identity through timbre transfer using reference voices from the speech datasets described in the
paper. This reduces direct dependence on the identity of speakers in the original AudioSet
recordings, but the linguistic content of the source speech is retained. In particular, speech has not
been filtered for obscenities or vulgar language, and users of the benchmark may therefore
encounter such content. The source datasets also reflect biases in their coverage. For example, some musical traditions and sound-event categories are underrepresented, and MixCardBench should not be interpreted as representative of the full diversity of real-world audio. These biases should be considered when using the benchmark to draw conclusions about general audio-understanding capabilities. MixCardBench is designed for research evaluation and is released under a non-commercial
research license. Users are expected to respect the licenses and terms associated with the underlying
source datasets and to consider the potential limitations and harms of deploying models evaluated
on this benchmark in downstream applications.

\subsection*{Reproducibility Statement}
We have made several efforts to ensure that MixCardBench and our results are reproducible.
Section~\ref{ssec:benchmark_construction} and Figure~\ref{fig:overall-pipeline} give an overview of the synthesis pipeline,
and the appendices describe each step in full: Appendix~\ref{app:A_annotation-cleanup} covers the
AudioSet-SL annotation cleanup and the mapping of AudioSet classes to UCS categories, and
Appendix~\ref{app:B_mixcardbench_audio_construction} covers source selection, filtering and soundscape synthesis,
with file and track counts after every filtering step. The 70/30 train/test split of the PSE
library is stratified by UCS subcategory and seeded, and every random draw during synthesis
is seeded, so the pipeline produces the same soundscapes from the same inputs.
Appendix~\ref{app:C_mixcards_schema} defines every MixCard field and its vocabulary.
Appendix~\ref{app:D_mixcard-generation} lists the card generation steps and the model that produces
each field. Appendix~\ref{app:J_prompts} copies every prompt verbatim from our code, both for
ground-truth generation and for benchmark inference. The ground-truth models
(Qwen3-Omni-30B-A3B-Instruct and Qwen3.6-35B-A3B-FP8) and all six evaluated LALMs are
publicly available open-weight checkpoints. Section~\ref{ssec:eval_protocol} describes
the two-stage inference protocol, Appendix~\ref{app:G_eval_protocol} defines the scoring metric
for each field type, and Appendix~\ref{app:extended_results} reports the full results.
Section~\ref{sec:bench_validation} describes the human validation, the synthesis variants used for
the distribution comparison, and the listening study. We release the mixtures and their
reference MixCards, as well as the evaluation code, for reproducibility. 

\bibliographystyle{iclr2027_conference}
\bibliography{zotero,references}
\newpage
\appendix 
\section{AudioSet Annotation cleanup}
\label{app:A_annotation-cleanup}

We build our sound-event annotations from the AudioSet Strong labels (train and eval splits), which provide, for each 10-second segment, an annotation of $(\text{class}, t_{\text{start}}, t_{\text{end}})$ triplets aligned to the AudioSet ontology. These annotations exhibit three recurring issues that make them unsuitable for direct use as ground truth: (i) temporally overlapping annotations from an ontology class and its ancestor (e.g., ``Guitar'' and ``Music'' both annotated on the same time span), (ii) fragmented annotations, where a single perceptual event is split into several short, closely-spaced segments of the same class (e.g. ``Footsteps'' annotated to every single footstep), and (iii) noisy or unreliable segments that a downstream separation/judge model cannot substantiate. We address these issues through a five-stage cleanup pipeline followed by a music-label consolidation step and a final quality-filtering step based on judge scores.

\subsection{Ontology parsing and out-of-ontology filtering}
\label{sec:ontology-parsing}

We first parse the AudioSet ontology (a tree of sound classes) into a flat ancestry table by recursively traversing it, recording for every class its full list of ancestors (root to immediate parent) and its depth (class level) in the tree. This table is used throughout the pipeline to answer ``is class $A$ an ancestor of class $B$?'' questions.

Because the ontology reuses leaf names under multiple parents (e.g., \emph{growling} appears both under dog sounds and under cat sounds), we additionally flag these duplicate children separately, since collapsing them to a single class would discard the disambiguating parent context. 

We then merge the AudioSet Strong annotations with their display names and coarse categories (\emph{speech}, \emph{music}, \emph{sfx}). A small fraction of AudioSet class labels ($33$ classes, $\approx 1\%$ of annotated segments) do not correspond to any node in the ontology tree. We manually remap the subset of these labels that are simple naming variants of existing ontology nodes (e.g., \emph{``Vehicle horn, car horn, honking, toot''} $\rightarrow$ \emph{``Vehicle horn, car horn, honking''}), and drop the remaining out-of-ontology annotations entirely, since they cannot be placed in the class hierarchy used by later stages.

The full list of out-of-ontology annotations (before our remapping):
\texttt{{'Alert',
 'Audio logo',
 'Bathroom sounds',
 'Bicycle, tricycle',
 'Blender, food processor',
 'Canidae, wild dogs, wolves',
 'Carbon monoxide detector, CO detector',
 'Cart',
 'Chain',
 'Crockery breaking and smashing',
 'Dong, bong',
 'Ducks, geese, waterfowl',
 'Electric rotor drone, quadcopter',
 'Error signal',
 'Glass chink, clink',
 'Glass shatter',
 'Gurgling, bubbling',
 'Keypress tone',
 'Kitchen and dining room sounds',
 'Knife',
 'Lock',
 'Mechanical bell',
 'Pant (dog)',
 'Paper rustling',
 'Power saw, circular saw, table saw',
 'Refrigerator',
 'Ringing tone, ringback tone',
 'Shower',
 'Slurp, drinking straw',
 'Stairs',
 'Stomp, stamp',
 'Stream, river',
 'Studio recording',
 'Sweeping',
 'Tap dance',
 'Tire squeal, skidding',
 'Unknown sound',
 'Unmodified field recording',
 'Vehicle horn, car horn, honking, toot',
 'Video game sound',
 'Washing machine',
 'White noise, pink noise',
 'Windscreen wiper, windshield wiper'}}

\subsection{Removing redundant parent/child annotations}
\label{sec:redundant-removal}

AudioSet Strong frequently emits overlapping annotations for a class and one of its ontology ancestors over (approximately) the same time span, e.g., a \emph{Female speech} event co-occurring with a broader \emph{Speech} event. We treat the more specific (child) label as more informative and remove the redundant ancestor annotation whenever they cover the same interval.

\begin{figure}
    \centering
    \includegraphics[width=\linewidth]{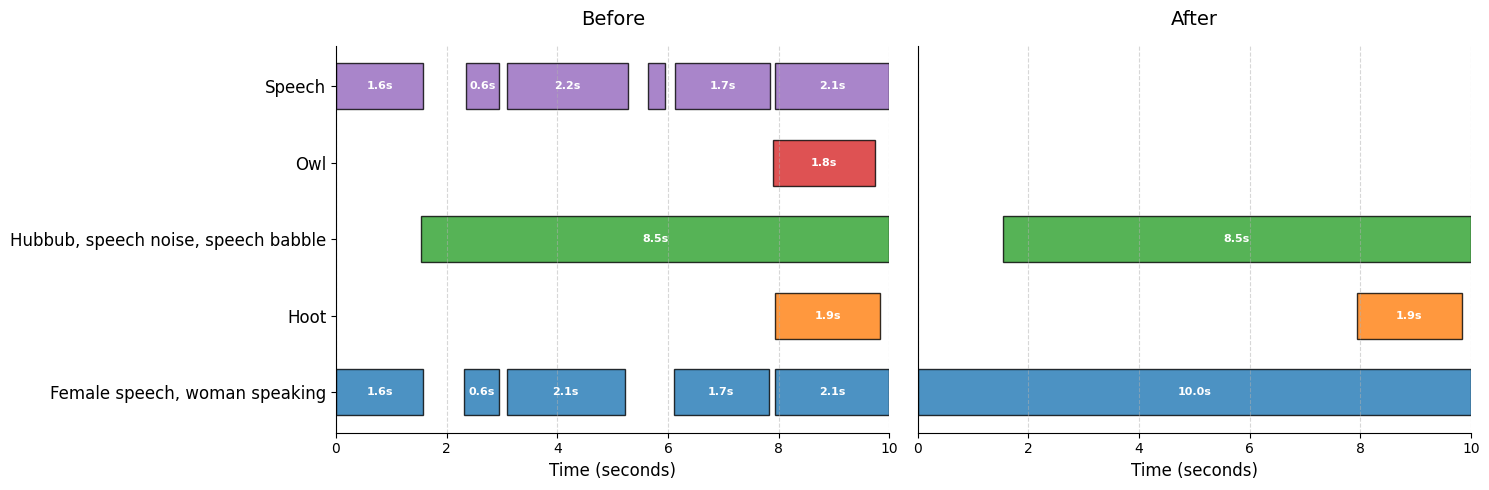}
    \caption{Before and After annotations of AudioSet sample ``KxzuJRh9GaQ\_30000''. \emph{Speech} is the parent class of \emph{Female speech, woman speaking}, therefore it is dropped as they have clearly identical annotations. The same happens with \emph{Owl} and \emph{Hoot}.}
    \label{fig:drop_redundand}
\end{figure}

For each segment, we build an interval tree over all $(t_{\text{start}}, t_{\text{end}})$ spans of its annotations~\citep{cormen1990introduction}. For every annotated event $c$ (the ``child'' candidate) we query all annotations $p$ whose interval overlaps $c$'s. An overlapping annotation $p$ is considered for removal only if:
\begin{itemize}
  \item $p \neq c$ in class label, and
  \item $p$'s class is a listed ancestor of $c$'s class in the ontology ancestry table (i.e., they lie on the same root-to-leaf branch), and
  \item the fraction of $c$'s duration covered by $p$ exceeds a fixed threshold,
\end{itemize}
\begin{equation}
  \text{overlap\_rate} = \frac{|[t^c_{\text{start}}, t^c_{\text{end}}] \cap [t^p_{\text{start}}, t^p_{\text{end}}]|}{t^c_{\text{end}} - t^c_{\text{start}}} > 0.8 .
\end{equation}
When this holds, the parent annotation $p$ is marked for removal (the child $c$ is always kept). Segments with a single annotation are left untouched. Only true ontology-lineage pairs are affected by this rule; overlapping annotations from unrelated classes are not touched here. We exemplify this workflow in Figure \ref{fig:drop_redundand}.

We provide the full list of classes assigned to different parents: \texttt{['Air horn, truck horn', 'Ambulance (siren)', 'Beatboxing', 'Bell',
'Bicycle bell', 'Bleat', 'Buzz', 'Car alarm',
'Change ringing (campanology)', 'Chant', 'Children shouting',
'Chime', 'Chirp, tweet', 'Choir', 'Church bell', 'Clapping',
'Clip-clop', 'Clunk', 'Cowbell', 'Crack', 'Crackle',
"Dental drill, dentist's drill", 'Ding-dong', 'Doorbell',
'Fire engine, fire truck (siren)', 'Growling', 'Hiss', 'Howl',
'Hubbub, speech noise, speech babble', 'Jet engine', 'Jingle bell',
'Kettle whistle', 'Knock', 'Mantra', 'Police car (siren)',
'Rattle', 'Snap', 'Squeak', 'Squish', 'Tap', 'Thump, thud',
'Thunk', 'Tick', 'Toot', 'Tuning fork',
'Vehicle horn, car horn, honking', 'Whoosh, swoosh, swish', 'Wind chime', 'Wind noise (microphone)']}

\subsection{Temporal smoothing of fragmented annotations}
\label{sec:temporal-smoothing}

\begin{figure}
    \centering
    \includegraphics[width=\linewidth]{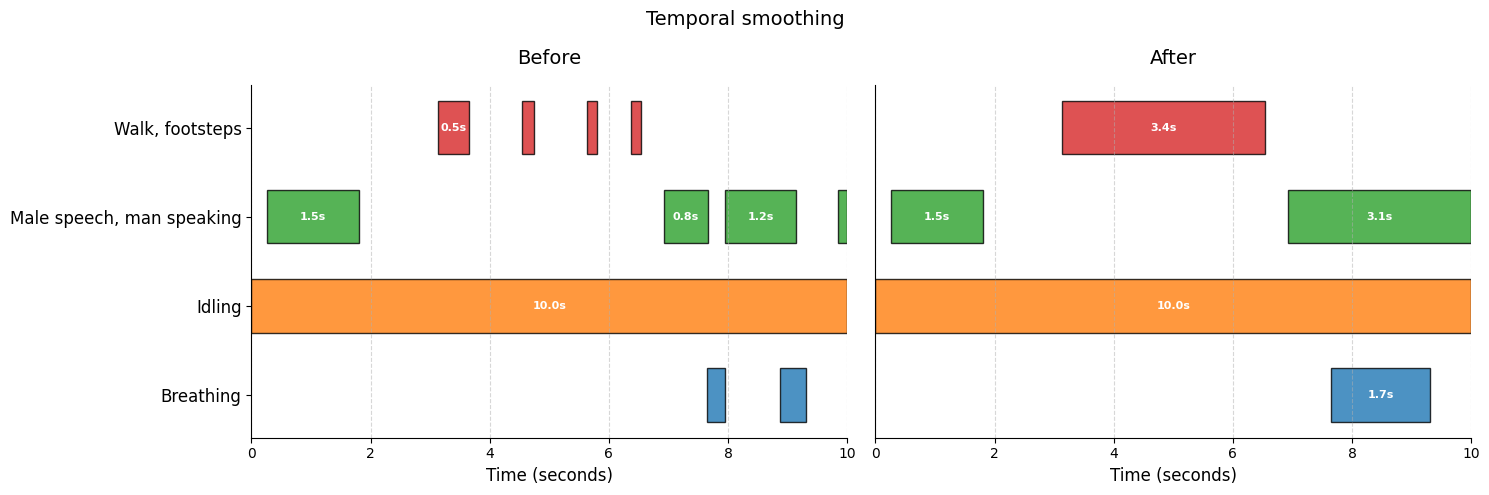}
    \caption{Before and After annotations of AudioSet sample ``xAyFvqkINEA\_30000''.}
    \label{fig:temporal_smoothing}
\end{figure}

A single acoustic event is often split by the automatic strong-labeling process into several short, same-class segments separated by small gaps. We merge these fragments per $(\text{segment\_id}, \text{class})$ group: after sorting a group's events by start time, we start a new block whenever the gap to the previous event's end time exceeds a threshold $\tau = 1.0$ s,
\begin{equation}
  \text{new\_block}_i = t^{\text{start}}_i > t^{\text{end}}_{i-1} + \tau ,
\end{equation}
and assign block indices via a cumulative sum over this boolean flag. Each resulting block is collapsed into a single annotation spanning $[\min t^{\text{start}}, \max t^{\text{end}}]$ over its members, with the remaining metadata (class, segment id, split, coarse category) taken from the first event in the block. Only annotations sharing the same class within a segment are merged; different classes are left as separate annotations.

\subsection{Foreground/background labeling}
\label{sec:fg-bg}

We assign each (smoothed) annotation a \emph{foreground} or \emph{background} role using a simple duration heuristic: since each AudioSet clip is $10$ s long, an event whose duration exceeds $9$ s is treated as effectively spanning the whole clip and is labeled \emph{background}; all other events are labeled \emph{foreground}. The \emph{Music} class is excluded from this rule (music annotations are always kept as-is here and handled separately by the music-consolidation step below):
\begin{equation}
  \text{label}(x) =
  \begin{cases}
    \text{background} & \text{if } (t^{\text{end}}_x - t^{\text{start}}_x) > 9\text{s} \text{ and class}(x) \neq \text{``Music''} \\
    \text{foreground} & \text{otherwise.}
  \end{cases}
\end{equation}

\subsection{Music label consolidation}
\label{sec:music-consolidation}

Because the ontology contains many fine-grained descendants of \emph{Music} and \emph{Singing} (instruments, genres, vocal styles, etc.), which frequently co-occur and overlap within a clip, we consolidate them into a single \emph{Music} annotation per contiguous span. We first collect the full set of ontology descendants of \emph{Music} and of \emph{Singing} (at any depth, via the ancestry table), together with the two root labels themselves, to form a \emph{music-label} set. Within each segment, annotations whose class falls in this set are sorted by start time and merged whenever they touch or overlap, i.e.\ using a zero-gap tolerance ($t^{\text{start}}_i \le t^{\text{end}}_{i-1}$, as opposed to the $1.0$s tolerance used for same-class smoothing in Section~\ref{sec:temporal-smoothing}), and relabeled as a single \emph{Music} event; non-music annotations pass through unchanged.

\subsection{Summary}
\label{sec:cleanup-summary}

The full AudioSet cleaning pipeline can be summarized as:
\begin{enumerate}
  \item Parse the AudioSet ontology into an ancestry table; merge with AudioSet Strong labels.
  \item Drop the $\approx 1\%$ of annotations whose class falls outside the ontology (after remapping simple naming variants).
  \item Remove ancestor annotations that are $>80\%$ covered by a same-branch descendant annotation (Section~\ref{sec:redundant-removal}).
  \item Merge same-class fragments separated by $\le 1.0$s gaps into single events (Section~\ref{sec:temporal-smoothing}).
  \item Label events longer than $9$s (excluding \emph{Music}) as background, and the rest as foreground (Section~\ref{sec:fg-bg}).
  \item Consolidate all descendants of \emph{Music}/\emph{Singing} into unified, overlap-merged \emph{Music} events (Section~\ref{sec:music-consolidation}).
\end{enumerate}

Our final clean version of AudioSet Strong has 16,975 unique segment\_ids, totaling 82,084 events.

\subsection{Ontology mapping}

We generate an initial mapping from AudioSet ontology classes to UCS using OntoAligner \citep{giglou2025ontoaligner}, using a mixture of all-MiniLM-L6-v2 sentence embeddings \citep{reimers-2019-sentence-bert} and Mistral-7B-v0.3 \citep{jiang2023mistral7b}. We then refine the mapping using Claude Opus by identifying the top-$k$ candidate UCS classes for each AudioSet ontology class through semantic matching and keyword lookup against the candidate classes' keywords and synonyms. We then manually review and update incorrect mappings, for $<10\%$ of the classes. Furthermore, we replace mapped classes that are unrepresented in our data with the nearest available class, e.g. ``Designed, Rhythmic" is replaced by ``Designed, Synthetic". This process covers $>90\%$ of the classes in AudioSet-SL. For the remaining classes, which we found that are not part of the AudioSet ontology, we either manually fix them and map them to UCS (e.g.``Blender, food processor'' becomes ``Blender''), or we drop them (e.g. ``unmodified field recording"). In total, we drop 33 classes, which represents less than 1\% of total annotations. 

\section{MixCardBench Construction Details}
\label{app:B_mixcardbench_audio_construction}

We describe now the quality filter and strategies to select high-quality sources to build MixCardBench. 

\subsection{Source Selection and Synthesis}
\subsubsection{AudioSet-SL clip selection}
\label{app:judge-filtering}

We discard clips whose annotations cannot be substantiated by an
audio-language judge model. Each coarse stem (\emph{speech}, \emph{music},
\emph{sfx}) obtained with SAM Audio is scored against a textual description
of its expected content with SAM Audio Judge. This gives four scores per
stem: \emph{overall}, \emph{recall}, \emph{precision}, and
\emph{faithfulness}. We retain a segment only if
\begin{equation}
  \min_{\text{stem}} \left(\text{judge\_recall}\right) \ge 4.0 ,
\end{equation}
a threshold chosen from a retention-curve sweep over $[0, 5]$ that trades
off dataset size against annotation reliability. Filtering is applied at
the segment level: if a segment fails the threshold, all of its
annotations are dropped. Table~\ref{tab:audioset-filters} lists the full
cascade, including the speech gates of Appendix~\ref{app:speech}.

\begin{table}[H]
\centering
\caption{Cascaded quality filters applied to AudioSet-SL eval.}
\label{tab:audioset-filters}
\begin{tabular}{lr}
\toprule
Step & Files \\
\midrule
AudioSet-SL eval & 16,975 \\
SAM Audio Judge, min.\ recall $\geq 4.0$ & 9110 \\
Speech gates (Table~\ref{tab:speech-filters}) & 5478 \\
Drop \emph{Narration, monologue}, \emph{Conversation}, \emph{Whispering}, \emph{Speech synthesizer} & 5180 \\
Drop segments with breathing + speech & 4933 \\
\midrule
Used for generating MixCardBench & 4933 \\
\bottomrule
\end{tabular}
\end{table}

\subsubsection{Source Separation}
\label{app:source_separation}
We extract target sources from AudioSet-SL mixtures to extract relevant information for our benchmark datasets. We use SAM Audio \citep{shi2026sam}, a state-of-the-art target source extraction model. In addition to text prompts, SAM Audio also accepts timestamp prompts, which allows us to leverage AudioSet-SL annotations as a prior for stronger source separation. For each audio, we use the AudioSet ontology to map class names to one of speech, music or sound effects categories. We map all sounds in the speech subtree of the AudioSet ontology as well as ``chatter" and ``hubbub" to speech, all music and singing classes to music, and everything else to sound effects. Since SAM Audio requires one text prompt at a time, we use an iterative approach, first separating speech sources, followed by music, then treating the remaining signal as sound effects. 

\subsubsection{Speech resynthesis}
\label{app:speech}

In order to retain the paralinguistics and semantics of the original speech, we extract the speech source(s) from the selected AudioSet-SL mixture, and resynthesize it using a voice in the public domain. We use Vevo-Timbre \citep{zhangVevoControllableZeroShot2024, amphion} for  timbre-transfer with voices from LibriTTS-R \citep{koizumiLibriTTSRRestoredMultiSpeaker2023} and VCTK \citep{yamagishi2019cstr}. We use LibriTTS-R test-clean, dev-clean and all of VCTK, i.e. 186 speakers in total. We use the separated speech to condition  the emotion and speaking style of the synthesized speech, and provide a reference speech file from a random open voice of the same gender as the speaker prompt. We only use utterances with $3s \leq$ duration $\leq 10s$ to provide sufficient conditioning. We mask out non-speech segments in the source signal using a binary mask with 5ms fade in and fade out. We apply a series of filters to reject low quality synthetic sources, such as dropping sources with overlapped singing, AudioSet sources containing babble or hubbub which are challenging to resynthesize due to multiple overlapping voices, and dropping non-english segments. Additionally, we use a transcription-diarization pipeline \citep{bain2022whisperx} on both the original separated speech and the resynthesized speech. We compute DNSMOS on the original and resynthesized speech and WER between the two transcripts, and drop examples with a language mismatch or multiple speakers, WER $> 0.3$, and $\Delta$ DNSMOS$_{\text{OVL}} < -0.3$. We apply these filters in a cascade, retaining 32\% of the resynthesized speech.
Table~\ref{tab:speech-voices} gives
the speaker pool, and Table~\ref{tab:speech-filters} gives the number of
segments surviving each quality gate.

\begin{table}[H]
\centering
\caption{Speaker source files.}
\label{tab:speech-voices}
\begin{tabular}{lr}
\toprule
Step & Speakers \\
\midrule
LibriTTS-R (2456) + VCTK (110) & 2566 \\
Test pool: LibriTTS-R test-clean (39) + dev-clean (40) + VCTK (110) & 189 \\
\midrule
Used for generating MixCardBench & 186 \\
\bottomrule
\end{tabular}
\end{table}

\begin{table}[H]
\centering
\caption{Cascaded quality filters applied to AudioSet-SL segments containing
speech. Each row gives the number of segments remaining after the filter.}
\label{tab:speech-filters}
\begin{tabular}{lr}
\toprule
Step & Segments \\
\midrule
AudioSet-SL segments with speech & 6962 \\
No babble/hubbub labels & 6546 \\
No singing overlapping speech & 6312 \\
WER computation succeeds & 5994 \\
Cloned-voice language matches original & 5435 \\
$\geq 3$ reference words & 4772 \\
WER $\leq 0.3$ & 2520 \\
$\Delta$DNSMOS$_{\text{OVL}} \geq -0.3$ & 2390 \\
Single speaker & 2389 \\
English & 2222 \\
\bottomrule
\end{tabular}
\end{table}

\subsubsection{Music retrieval}
\label{app:music}
Our goal is to preserve high-level musical characteristics, such as mood and genre, when replacing the original music in the AudioSet-SL soundscapes. To this end, we use MULE \citep{mccallum2022mule} to retrieve acoustically similar music from the Free Music Archive (FMA) test set \citep{defferrad2017fma}. In preliminary experiments comparing MULE to other state-of-the-art music embedding models such as MusicFM \citep{won2024foundation}, we found that MULE retrieved tracks that better preserved the mood and genre of the reference music. We therefore compute a single MULE embedding per clip by average pooling, to identify the FMA track with the highest cosine similarity to the original music. Then, we identify the highest moment of similarity within the clip by comparing embeddings at the frame level (with embeddings computed every 2 seconds) and estimate the highest similarity within the clip  to select the most similar segment within the retrieved track, matching the duration of the reference event. To avoid matching music events to silent segments, we identify and mask silent regions in the AudioSet-SL using an RMS threshold of $-70$ dB before computing cosine similarity. However, some silent segments may still be selected, particularly for events shorter than 3 seconds, which is the embedding window size. Once a replacement event is added to the soundscape, we recompute its onset and offset using its RMS envelope and the corresponding zero-crossing points. Replacement music is retrieved from the FMA test split with MULE, and Table~\ref{tab:fma} reports the size of the retrieval pool and the number of distinct tracks that were used.

\begin{table}[H]
\centering
\caption{Music source files.}
\label{tab:fma}
\begin{tabular}{lr}
\toprule
Step & Tracks \\
\midrule
FMA large, test split & 11,263 \\
\midrule
Distinct FMA tracks used & 2024 \\
\bottomrule
\end{tabular}
\end{table}

\subsubsection{Sound-effect selection}
\label{app:sfx}
\begin{figure}[H]
    \centering
    \includegraphics[width=\linewidth]{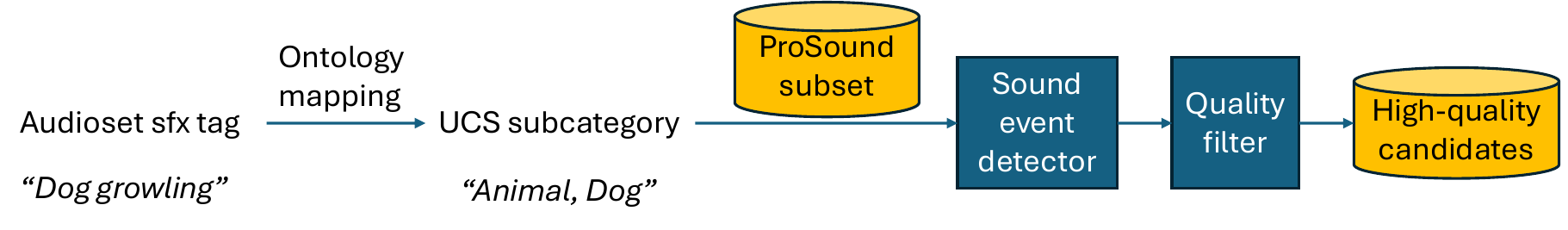}
    \caption{Sound-effect sampling: ontology mapping and PE-A-Frame
    detection build candidate pools, RMS onset detection isolates
    individual events, and PE-AV text--audio similarity keeps the most
    semantically relevant candidates.}
    \label{fig:sfx-sampling}
\end{figure}

For sound-effect sampling we use Pro Sound Effects (hereafter PSE), a high-quality professional sound effects library, to build ecologically valid soundscapes. PSE is organized according to the Universal Category System (UCS) \citep{ucs2026}, a public project that offers a framework for consistent categorization of sound effects. We partition PSE into train and test splits in a 70/30 ratio within every UCS subcategory to prevent source file leakage into training data. To reconstruct the AudioSet-SL soundscapes, we need to identify suitable replacement sound effect events within the PSE recordings. However, in our first iterations of the synthesis, based on the ontology mapping alone and random sampling, we found that these recordings may contain multiple occurrences of the target event, as well as other sounds, making it necessary to identify individual event boundaries and ensure that the selected segments match the target event semantically. We use PE-A-Frame \citep{vyasPushingFrontierAudiovisual2025}, an open-vocabulary sound event detection model, to locate target AudioSet-SL events inside the source files for its mapped UCS categories and subcategories. While PE-A-Frame provides coarse boundaries for target event activity, it does not isolate individual event onsets. We use a background-relative RMS detector with 3dB/0dB onset/offset thresholds to pinpoint individual onsets and offsets with 0.04s resolution within the PE-A-Frame spans, and build candidate event pools for each AudioSet-SL class. However, we found that this pipeline does not narrow down to the most semantically valid candidates, particularly for short events. We use PE-AV \citep{vyasPushingFrontierAudiovisual2025} to compute text-audio similarity between each (AudioSet-SL query, candidate event) pair and pick the top 10 percentile events per query. We find that this verification stage leads to significantly more semantically relevant samples in the generated soundscapes. For an AudioSet-SL event, we find candidate events with the nearest duration and sample uniformly randomly with 0.1s resolution. Multiple AudioSet-SL events of the same class are matched to multiple candidate events from the source file which the first event matched to. We apply a 20ms onset padding and a 100ms offset padding to ensure onsets and decay tails are included, as well as a 10ms fade-in and fade-out to avoid abrupt changes in the soundscape. Figure~\ref{fig:sfx-sampling} illustrates the sound-effect sampling
pipeline. Table~\ref{tab:prosound} reports the number of PSE files remaining after
each stage.

\begin{table}[H]
\centering
\caption{Cascaded quality filters applied to sound-effect source files.}
\label{tab:prosound}
\begin{tabular}{lr}
\toprule
Step & Files \\
\midrule
Our test split of PSE & 324,493 \\
Ambience / archived removed & 278,364 \\
Files that AudioSet-SL queries map to & 146,153 \\
PE-A-Frame event detection & 143,847 \\
RMS crops inside detected spans & 132,184 \\
Crop duration $\leq 10$\,s & 118,552 \\
PE-AV similarity, top 10\% per query & 29,711 \\
\midrule
Used in MixCardBench & 2483 \\
\bottomrule
\end{tabular}
\end{table}

\subsection{Soundscape Synthesis}

\subsubsection{Loudness mixing templates}
\label{app:loudness_mixing}
We construct ecologically valid mixing templates for events using the separated speech, music, and sound effects stems described above. For each AudioSet-SL recording, we compute the LUFS value of each stem using \texttt{pyloudnorm} \citep{steinmetz2021pyloudnorm}, and use these loudness values to determine the relative loudness of sources in the synthetic soundscape. Since SAM Audio produces a single consolidated sound effects stem containing all sound-effects events, we first combine the replacement sound-effects sources from PSE together, and measure the loudness of the resulting mixture. We then compute the difference between this value and the loudness of the reference sound effects stem, applying the same gain adjustment to each individual sound-effects source. To prevent clipping, we apply a peak limiter with a threshold of $-0.1$ dBFS to the final mixture. Scaper typically renormalizes the entire signal, but we found that renormalization altered the loudness of sources that had already been matched to their reference levels, often making speech excessively quiet. We apply the limiter's gain envelope to each individual source to ensure that their sum remains equal to the final mixture, preserving mixture consistency while preventing clipping and maintaining loudness levels closer to those of the reference recording.

\section{MixCards schemas}
\label{app:C_mixcards_schema}
\begingroup
\footnotesize
\setlength{\tabcolsep}{6pt}
\renewcommand{\arraystretch}{1.15}
\begin{longtable}{@{}p{0.21\linewidth} p{0.35\linewidth} p{0.40\linewidth}@{}}
\caption{MixCard field schema. Common fields are shared across all three card types, while category-specific fields exist only in MixCards for those categories. Examples are taken from dataset cards.}
\label{tab:mixcard-fields} \\
\toprule
\textbf{Field} & \textbf{Description} & \textbf{Example} \\
\midrule
\endfirsthead

\multicolumn{3}{@{}l}{\small\tablename~\thetable{} (continued)} \\
\toprule
\textbf{Field} & \textbf{Description} & \textbf{Example} \\
\midrule
\endhead

\midrule
\multicolumn{3}{r@{}}{\emph{Continued on next page}} \\
\endfoot

\bottomrule
\endlastfoot

\multicolumn{3}{@{}l}{\textbf{Common: signal features}} \\*
\midrule
\addlinespace[1pt]
(start, stop) tuples & Onset--offset for every instance of the event. & \texttt{[00:00.000, 00:04.075]; [00:09.636, 00:09.996]} \\*
Loudness & Integrated LUFS per instance with a six-level label (very loud / loud / medium / quiet / very quiet / extremely quiet). & \texttt{[-27.8 LUFS, Quiet]; [-26.8 LUFS, Medium]} \\*
Frequency profile & Energy across 10 bands (125\,Hz to 10\,kHz), each on a seven-level scale (very-low / low / medium-low / medium / medium-high / high / very-high). & \texttt{[medium-high; medium-high; medium-high; medium-low; low; low; low; low; low; low]; [medium-high; medium-high; medium-high; medium-high; medium-low; low; low; low; medium-low; medium-low]} \\
\addlinespace[1pt]
\midrule
\multicolumn{3}{@{}l}{\textbf{Common: free-text fields}} \\*
\midrule
\addlinespace[1pt]
Short caption & Brief caption of the event. & ``Monotonous mid voice speaking'' (speech); ``Driving house music blasts'' (music); ``Rapid, sharp dog barks'' (SFX) \\*
Descriptive caption & Full descriptive caption. & ``A mid-pitched voice speaks in a monotonous delivery, stating `This is only a test' and `Testing smoke' in two brief segments separated by silence.''; ``Multiple dogs bark rapidly and sharply in a dry, low-fidelity recording.'' \\
\addlinespace[1pt]
\midrule
\multicolumn{3}{@{}l}{\textbf{Common: tag field}} \\*
\midrule
\addlinespace[1pt]
Perceived audio quality & Tags on overall perceived quality (production, technical, channel/source). & ``low-fidelity, bass-heavy''; ``clear, reverb-heavy''; ``clear'' \\
\midrule
\multicolumn{3}{@{}l}{\textbf{Speech-specific}} \\*
\addlinespace[1pt]
Median F0 (Hz) & Median fundamental frequency per instance. & \texttt{[192.6]; [217.5]} \\*
Intelligibility & Intelligibility level (highly / mostly intelligible, degraded intelligibility), optionally with a channel tag (clean channel / poor channel quality). & ``highly intelligible, clean channel'' \\*
Emotion & One tag from five classes (neutral, calm, excited/happy, angry/stressed, sad/subdued). & neutral, excited/happy, angry/stressed, sad/subdued \\*
Manner of speech & Timbre, tone, pace, delivery characteristics. & sleepy, monotone, whispered \\*
Transcript & ASR transcript, one segment per utterance. & ``This is only a test.''; ``Testing smoke.'' \\
\midrule
\multicolumn{3}{@{}l}{\textbf{Music-specific}} \\*
\addlinespace[1pt]
Music descriptors & Descriptive adjectives (not mood or genre). & layered, sustained, harmonic, evolving \\*
Instrument(s) & Instruments present. & electric guitar, piano, vocals, whistling \\*
Genre tag(s) & Genre labels (FMA taxonomy). & Electronic, House \\*
Mood & Mood tags (nine GEMS categories). & amazement, solemnity, tenderness \\*
Tempo & BPM estimate with a five-level tempo label (very slow / slow / medium / fast / very fast), or no tempo. & \texttt{[165, very fast]; [80, slow]; [121, fast]} \\
\midrule
\multicolumn{3}{@{}l}{\textbf{Sound-effects-specific}} \\*
\addlinespace[1pt]
Sounding object (noun) & Source object. & dog, dogs, wooden door, crowd, alarm clock \\*
Sounding action (verb) & Action producing the sound. & barking, beeping, breaking \\*
Temporal descriptors & Time-course characteristics. & fade out, intermittent, sustained \\*
Number of sources / plurality & Count of concurrent sources (not repetitions over time). & none / one / multiple \\*
Sound descriptors / adjectives & Descriptive adjectives (not mood or genre). & harsh, hurried, guttural \\
\end{longtable}
\endgroup

\section{MixCards generation methods}
\label{app:D_mixcard-generation} 
We describe the overall MixCard generation pipeline below. See Table \ref{tab:mixcard-generation-models} for additional details about algorithms and models used to estimate each field.

\begin{enumerate}
    \item \textbf{Input.} Each of the 4884 generated ten-second soundscapes comes with per-event stems. We sum the stems of the same source into one stem for speech, one for music, and one for each sound effects class in the soundscape, leaving out the background noise. Each of these stems gets its own MixCard. We compute every field from the stems instead of the mixture, so each field is estimated at the highest possible signal-to-noise ratio. Signal features are measured on the stems at their native sample rate, while all models receive them as mono 16 kHz audio.
    \item \textbf{Recompute annotations.} We recompute the annotations from the stems and drop classes without a matched source file.
    \item \textbf{Copy ground-truth metadata.} We copy metadata such as the FMA genre.
    \item \textbf{Signal features.} We compute energy-based timestamps, loudness, frequency profile, and median F0. 
    \item \textbf{Task-specific models.} We estimate intelligibility, emotion, tempo, and genre when it is unavailable in FMA.
    \item \textbf{Numeric to ordinal mapping.} Loudness, frequency profile, intelligibility, emotion, and tempo are mapped from numeric values to pre-defined ordinal labels.
    \item \textbf{Transcription.} For speech, we use WhisperX to detect the language, diarize, and transcribe with timestamps.
    \item \textbf{Audio language model.} We estimate all remaining fields, except captions, with Qwen3-Omni-30B-A3B-Instruct. To minimize hallucinations, each field is inferred independently, except sounding object and sounding action, which share one. Each prompt carries the field definition and the available text context for the stem: the measured fields, the FMA genre for music, and the PSE category, keywords, and description for sound effects. Some fields are open vocabulary (e.g. manner of speech, sound descriptors), others use a manually defined vocabulary (e.g. mood, music descriptors, perceived audio quality), and sounding object and action are chosen only from terms in the PSE metadata that the model can hear (see Appendix~\ref{app:J_prompts}).
    \item \textbf{Caption and clean-up.} We process one MixCard at a time with Qwen3.6-35B-A3B-FP8, giving it the card fields plus the transcript for speech and the PSE metadata for sound effects. It first cleans up the estimated fields, checking for field definition mismatches, redundancy, and contradictions; it can only move a tag to a better field or drop it, never add or reword one. It then writes the short and descriptive captions from the cleaned fields. A second pass with the same model removes proper nouns from sound effects cards.
    \item \textbf{Vocabulary gate.} We drop tags that fall outside the defined vocabulary of each closed-vocabulary field. If no tags remain, the field keeps its tags from the audio language model.
\end{enumerate}

\begingroup
\small
\begin{xltabular}{\linewidth}{@{}llX@{}}
\caption{MixCard field generation. Fields appear in the same order as the field schema above. The model column names where a field originates; every field then passes through the LLM consolidation pass. Qwen-Omni: Qwen3-Omni-30B-A3B-Instruct (listens to the isolated stem). LLM: Qwen3.6-35B-A3B-FP8 (reads the card fields and source metadata as text).}
\label{tab:mixcard-generation-models} \\
\toprule
\textbf{Field} & \textbf{Field type} & \textbf{Model} \\
\midrule
\endfirsthead

\multicolumn{3}{@{}l}{\small\tablename~\thetable{} (continued)} \\
\toprule
\textbf{Field} & \textbf{Field type} & \textbf{Model} \\
\midrule
\endhead

\midrule
\multicolumn{3}{r@{}}{\emph{Continued on next page}} \\
\endfoot

\bottomrule
\endlastfoot

\multicolumn{3}{@{}l}{\textbf{Common: signal features}} \\*
(start, stop) tuples          & Numeric            & RMS envelope \\*
Loudness                      & Numeric            & pyloudnorm (integrated LUFS) \citep{steinmetz2021pyloudnorm}. See detail below. \\*
Frequency profile             & Numeric            & Band energy (10 bands). See detail below. \\
\addlinespace
\multicolumn{3}{@{}l}{\textbf{Common: free-text fields}} \\*
Short caption                 & Free text          & LLM \\*
Descriptive caption           & Free text          & LLM \\
\addlinespace
\multicolumn{3}{@{}l}{\textbf{Common: tag field}} \\*
Perceived audio quality       & Closed vocab tags  & Qwen-Omni \\
\midrule
\multicolumn{3}{@{}l}{\textbf{Speech-specific}} \\*
Median F0 (Hz)                & Numeric            & librosa pYIN \citep{mauch2014pyin} \\*
Intelligibility               & Closed vocab tags  & TorchAudio SQUIM \citep{kumar2023squim} \\*
Emotion                       & Closed vocab tags  & audEERING wav2vec2 \citep{wagner2023dawn} \\*
Manner of speech              & Open vocab tags    & Qwen-Omni \\*
Transcript                    & Free text          & WhisperX \citep{bain2022whisperx} \\
\midrule
\multicolumn{3}{@{}l}{\textbf{Music-specific}} \\*
Music descriptors             & Closed vocab tags  & Qwen-Omni \\*
Instrument(s)                 & Closed vocab tags  & Qwen-Omni \\*
Genre tag(s)                  & Closed vocab tags  & FMA metadata (ensemble model fallback) \\*
Mood                          & Closed vocab tags  & Qwen-Omni (GEMS-9 \citep{jacobsen2025gems9}) \\*
Tempo                         & Numeric            & BEAT THIS! \citep{foscarin2024beatthis} \\
\midrule
\multicolumn{3}{@{}l}{\textbf{Sound-effects-specific}} \\*
Sounding object (noun)        & Open vocab tags    & Qwen-Omni \\*
Sounding action (verb)        & Open vocab tags    & Qwen-Omni \\*
Temporal descriptors          & Closed vocab tags  & Qwen-Omni \\*
Number of sources / plurality & Closed vocab tags  & Qwen-Omni \\*
Sound descriptors / adjectives & Open vocab tags   & Qwen-Omni \\
\end{xltabular}
\endgroup

\begin{figure}[H]
    \centering
    \includegraphics[width=\linewidth]{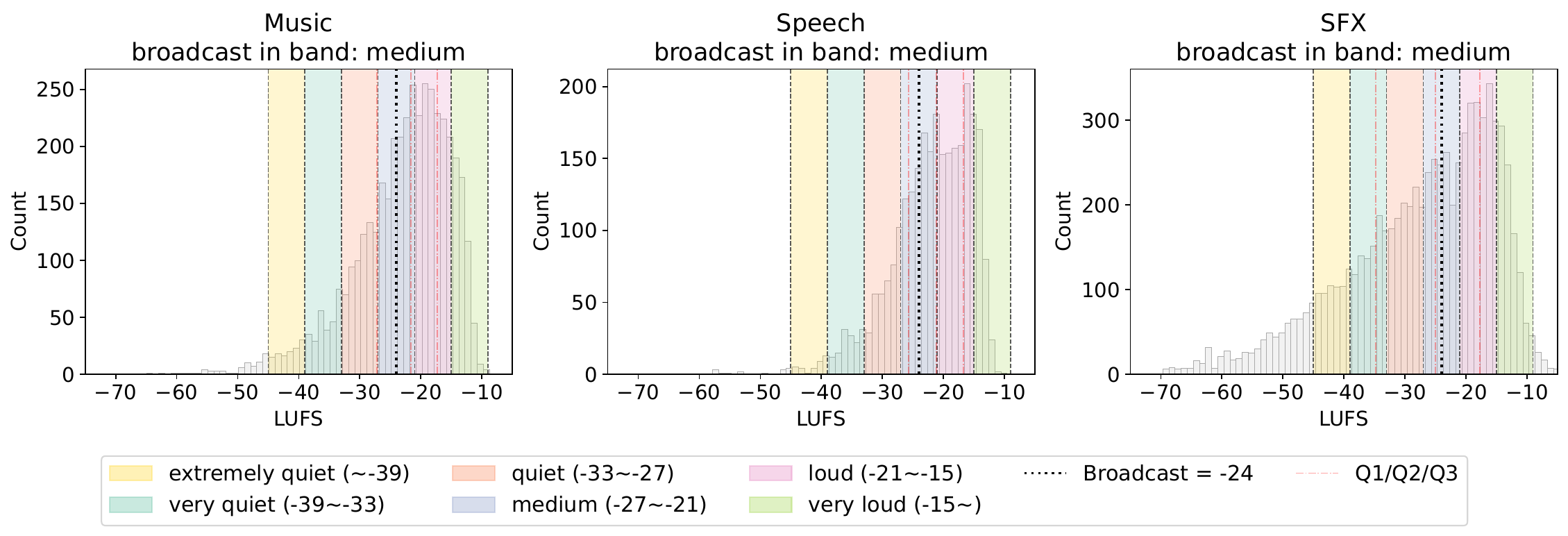}
    \caption{LUFS distribution and discretized bands per domain.}
    \label{fig:lufs-distribution}
\end{figure}

\textbf{Loudness (LUFS).} We compute the LUFS distribution of all events in our data and use
$-24$\,LUFS, a standard broadcast reference level, as the \emph{medium}
level. We discretize loudness into six labels with a 6\,dB bandwidth:
\emph{extremely quiet}, \emph{very quiet}, \emph{quiet}, \emph{medium},
\emph{loud}, and \emph{very loud}. Figure~\ref{fig:lufs-distribution}
shows the resulting distribution per domain.

\begin{figure}[H]
    \centering
    \includegraphics[width=\linewidth]{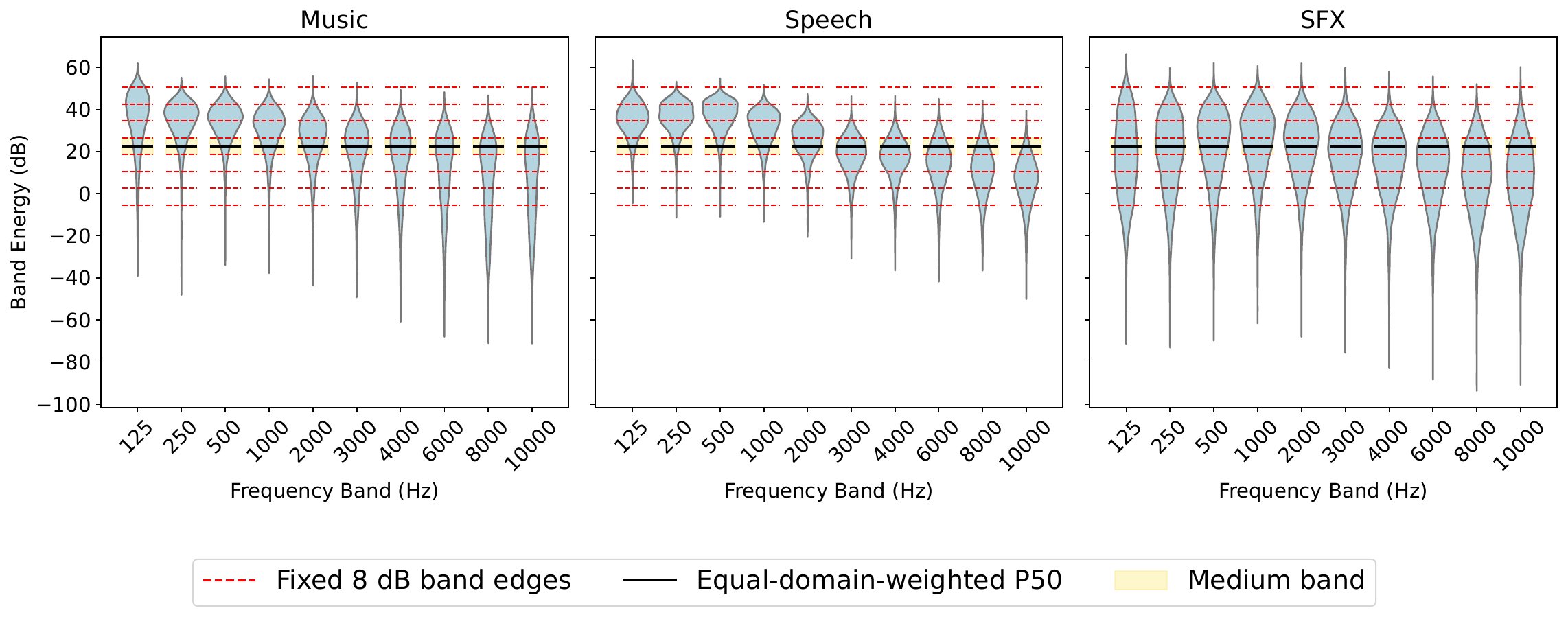}
    \caption{Audiometry frequency profile distribution and discretized
    levels per domain.}
    \label{fig:frequency-band}
\end{figure}

\textbf{Audiometry frequency profile.} We measure band energy at ten audiometric frequencies (125, 250, 500,
1000, 2000, 3000, 4000, 6000, 8000, and 10000\,Hz) for music, speech, and
sound effects. Because the domains differ in sample count, we take the
equal-weighted median across domains as the \emph{medium} reference. We
then discretize into seven levels with a fixed 8\,dB bandwidth shared by
all domains: \emph{very low}, \emph{low}, \emph{medium-low},
\emph{medium}, \emph{medium-high}, \emph{high}, and \emph{very high}.
Figure~\ref{fig:frequency-band} shows the resulting distribution.

\section{MixCardBench Statistics}
\label{app:E_mixcardbench_stats}
\begin{table}[H]
\centering
\caption{MixCardBench card count per audio category. In total, we have 9618 cards.}
\label{tab:mixcardsbench}
\begin{tabular}{lrr}
\toprule
Event type & Scenes & Cards \\
\midrule
Speech & 1116 & 1116\\
Music & 4009 & 4009\\ 
Sound effects & 2269 & 4493\\
\midrule
Total & 4884 (13.6h) & 9618\\
\bottomrule
\end{tabular}
\end{table}

\begin{figure}[ht]
    \centering
    \begin{subfigure}[b]{0.48\linewidth}
        \centering
        \includegraphics[width=\linewidth]{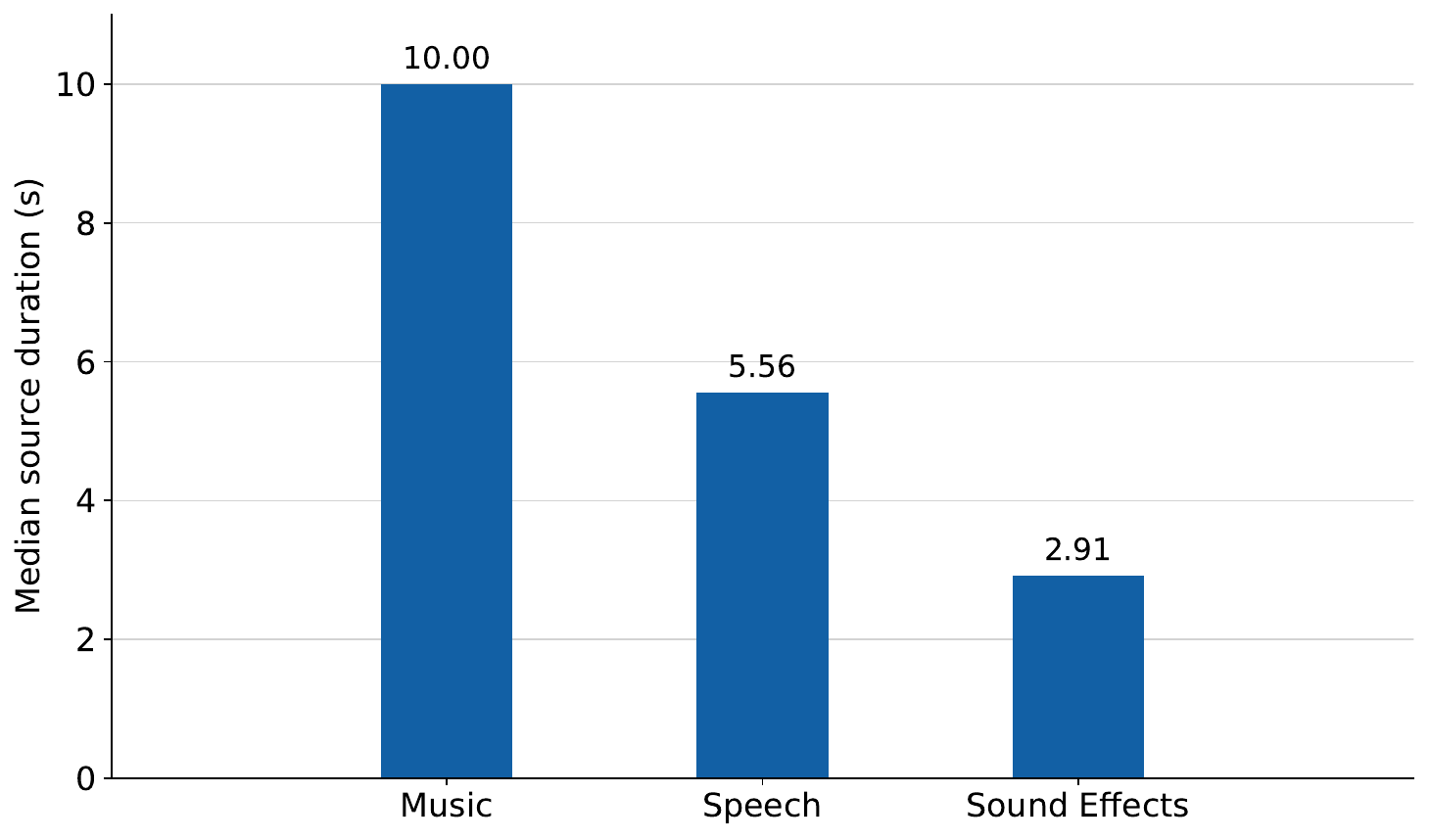}
        \caption{Median duration per source category.}
        \label{fig:duration_per_categories}
    \end{subfigure}
    \hfill
    \begin{subfigure}[b]{0.48\linewidth}
        \centering
        \includegraphics[width=\linewidth]{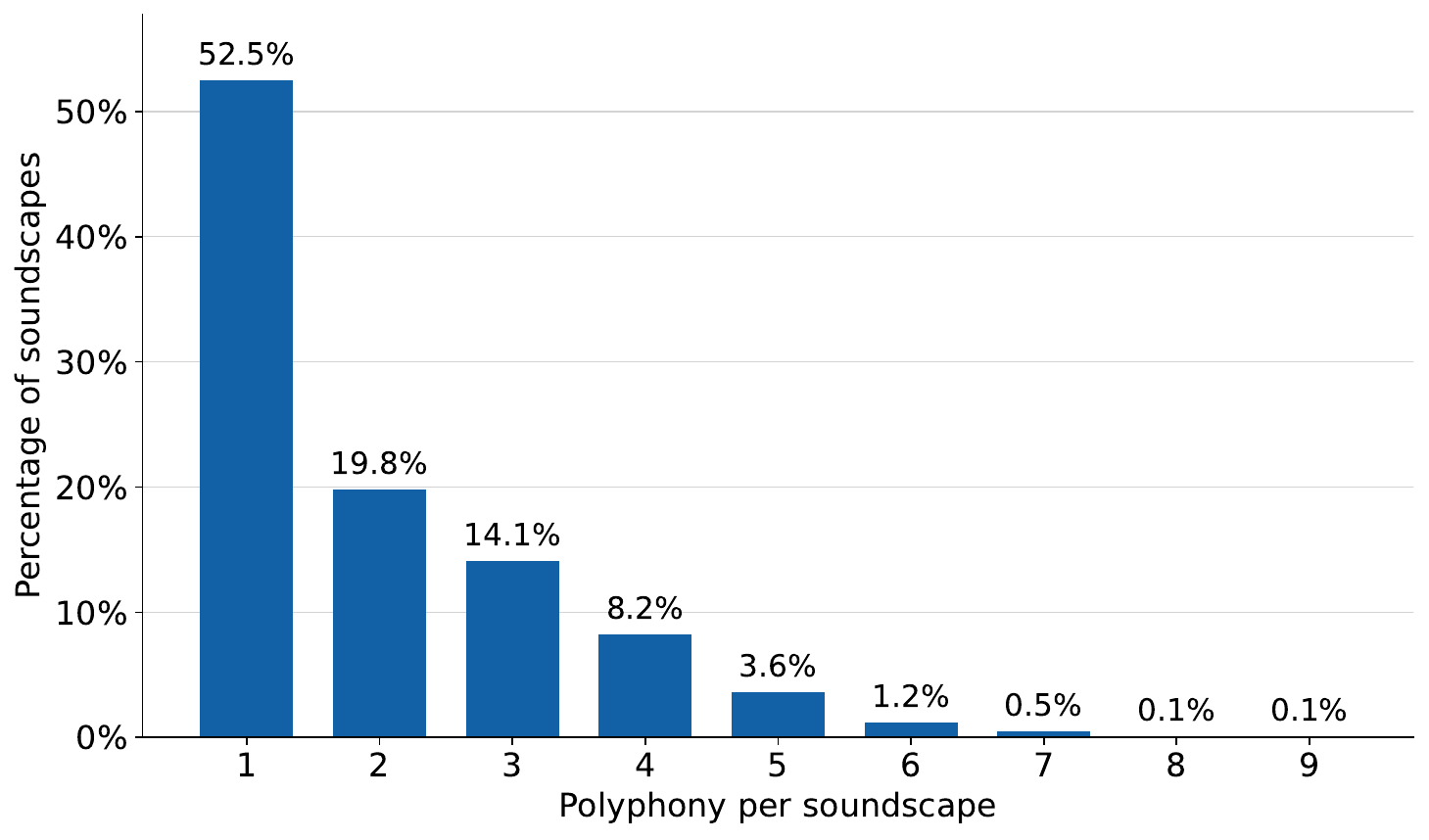}
        \caption{Distribution of polyphony per soundscape.}
        \label{fig:polyphony_distribution}
    \end{subfigure}
    \caption{Source and soundscape statistics in MixCardBench.}
    \label{fig:mixcard_stats}
\end{figure}

\begin{figure}[ht]
    \centering
    \includegraphics[width=\linewidth]{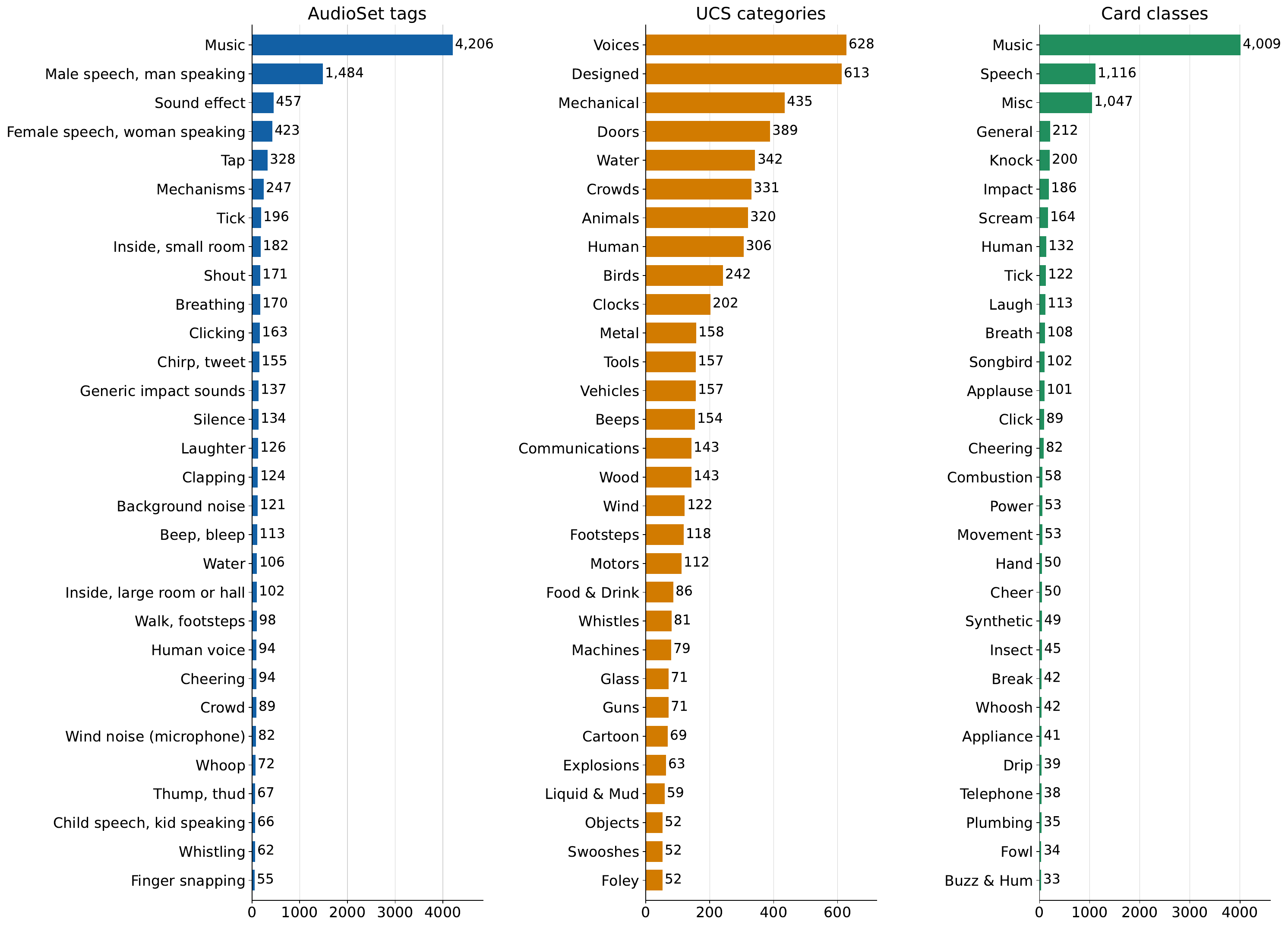}
    \caption{Distribution of AudioSet tags and UCS categories used in our data generation pipeline. Card classes shows the top 30 final classes in our MixCards. The plots are showing top 30.}
    \label{fig:class_distribution}
\end{figure}

\footnotesize
\begin{table}[H]
\centering
\caption{Number of Polyphony (sources) by domain.}
\label{tab:Polyphony_per_domain_counts}
\begin{tabular}{lcccccccccc}
\toprule
Event type 
& \makebox[0.6cm][c]{1}
& \makebox[0.6cm][c]{2}
& \makebox[0.6cm][c]{3}
& \makebox[0.6cm][c]{4}
& \makebox[0.6cm][c]{5}
& \makebox[0.6cm][c]{6}
& \makebox[0.6cm][c]{7}
& \makebox[0.6cm][c]{8}
& \makebox[0.6cm][c]{9}
& Total \\
\midrule
Music & 2520 & 720 & 395 & 226 & 95 & 37 & 9 & 4 & 3 & 4009 \\
Speech & 34 & 302 & 380 & 230 & 115 & 35 & 16 & 2 & 2 & 1116 \\ 
Sound effects & 9 & 906 & 688 & 401 & 178 & 57 & 22 & 5 & 3 & 2269 \\
\midrule
All soundscapes & 2563 & 967 & 688 & 401 & 178 & 57 & 22 & 5 & 3 & 4884 \\
\bottomrule
\end{tabular}
\end{table}

\subsection{Statistics by Audio Domain}
\label{app:field-stats}
\begin{figure}[H]
    \centering
    \includegraphics[width=\linewidth]{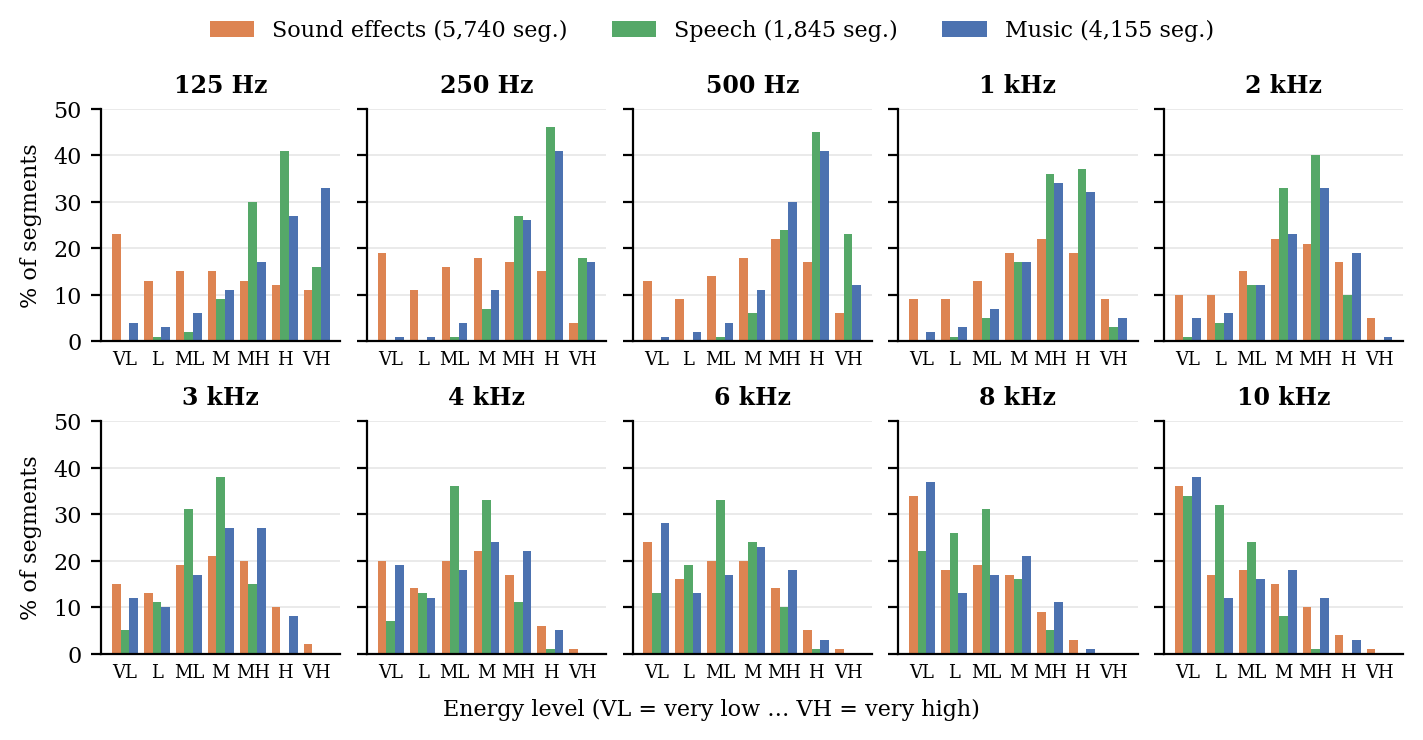}
    \caption{Distribution of frequency profile bands in the final benchmark. This is computed at the segment level and not at the soundscape level.}
    \label{fig:frequency_profile_bands_dist}
\end{figure}

Tables~\ref{tab:dist-music}--\ref{tab:dist-sfx} provide the tag distribution of every field of the released MixCards per card type. Figure~\ref{fig:frequency_profile_bands_dist} gives the frequency-profile level per band. Percentages are shares of cards (loudness: shares of segments).

\begingroup
\footnotesize
\begin{longtable}{@{}p{2.5cm}p{11cm}@{}}
\caption{Music cards (4009 cards in the final set): tag distribution per field. Percentages are the share of cards carrying the tag (loudness: share of segments); closed fields list every tag of the vocabulary, open fields the most frequent ones.}
\label{tab:dist-music} \\
\toprule
field & tags (share of cards) \\
\midrule
\endfirsthead

\multicolumn{2}{@{}l}{\small\tablename~\thetable{} (continued)} \\
\toprule
field & tags (share of cards) \\
\midrule
\endhead

\midrule
\multicolumn{2}{r@{}}{\emph{Continued on next page}} \\
\endfoot

\bottomrule
\endlastfoot

\textbf{genre (FMA metadata)} \newline {\scriptsize 135 distinct, top 15; 2.45 tags per card} & Experimental (23\%), Electronic (21\%), Avant-Garde (10\%), Ambient (10\%), Folk (9\%), Experimental Pop (7\%), Rock (7\%), Ambient Electronic (7\%), Pop (6\%), Lo-Fi (6\%), Improv (5\%), Jazz (5\%), Electroacoustic (5\%), Indie-Rock (4\%), Hip-Hop (4\%) \\ \addlinespace
\textbf{mood (GEMS-9)} \newline {\scriptsize 9 distinct; 1.42 tags per card; 0.6\% empty} & tension (39\%), calmness (27\%), joyful activation (21\%), power (15\%), nostalgia (15\%), tenderness (13\%), sadness (7\%), solemnity (5\%), amazement (0\%) \\ \addlinespace
\textbf{perceived audio quality} \newline {\scriptsize 13 distinct; 3.12 tags per card} & low-fidelity (48\%), dry (44\%), thin (43\%), clear (36\%), bass-heavy (27\%), room reverb (22\%), low background noise (21\%), muffled (18\%), distorted (15\%), high-fidelity (13\%), high background noise (11\%), reverb-heavy (7\%), harsh (7\%) \\ \addlinespace
\textbf{instruments} \newline {\scriptsize 53 distinct, top 20; 2.34 tags per card; 0.6\% empty} & drums (40\%), synthesizer (35\%), guitar (28\%), bass guitar (26\%), vocals (22\%), electric guitar (14\%), drum machine (13\%), piano (9\%), percussion (7\%), strings (5\%), violin (5\%), flute (3\%), accordion (3\%), glockenspiel (3\%), saxophone (2\%), choir (2\%), acoustic guitar (2\%), trumpet (2\%), cymbals (2\%), electric piano (1\%) \\ \addlinespace
\textbf{music descriptors} \newline {\scriptsize 36 distinct, top 20; 4.11 tags per card; 0.0\% empty} & repetitive (34\%), melodic (33\%), rhythmic (27\%), sparse (26\%), layered (22\%), raw (19\%), synthetic (18\%), percussive (18\%), looping (17\%), arpeggiated (16\%), evolving (15\%), sustained (15\%), detuned (15\%), delicate (14\%), glitchy (13\%), lively (10\%), intricate (10\%), groovy (10\%), plucked (9\%), frenetic (8\%) \\ \addlinespace
\textbf{tempo label} \newline {\scriptsize 6 distinct; 1.00 tags per card} & very slow (7\%), slow (28\%), medium (15\%), fast (16\%), very fast (19\%), no tempo (15\%) \\ \addlinespace
\textbf{loudness label} \newline {\scriptsize 1.04 segments per card} & extremely quiet (4\%), very quiet (6\%), quiet (15\%), medium (28\%), loud (33\%), very loud (14\%) \\
\end{longtable}
\endgroup

\begingroup
\footnotesize
\begin{longtable}{@{}p{2.5cm}p{11cm}@{}}
\caption{Speech cards (1116 cards in the final set): tag distribution per field. Percentages are the share of cards carrying the tag (loudness: share of segments); closed fields list every tag of the vocabulary, open fields the most frequent ones.}
\label{tab:dist-speech} \\
\toprule
field & tags (share of cards) \\
\midrule
\endfirsthead

\multicolumn{2}{@{}l}{\small\tablename~\thetable{} (continued)} \\
\toprule
field & tags (share of cards) \\
\midrule
\endhead

\midrule
\multicolumn{2}{r@{}}{\emph{Continued on next page}} \\
\endfoot

\bottomrule
\endlastfoot

\textbf{perceived audio quality} \newline {\scriptsize 11 distinct; 3.29 tags per card} & dry (79\%), muffled (77\%), low-fidelity (66\%), low background noise (34\%), clear (33\%), thin (24\%), room reverb (11\%), bass-heavy (4\%), high background noise (1\%), harsh (0\%), distorted (0\%) \\ \addlinespace
\textbf{manner (open)} \newline {\scriptsize 115 distinct, top 15; 2.28 tags per card} & conversational (37\%), monotonous (36\%), slow (36\%), steady (19\%), breathy (11\%), relaxed (11\%), resonant (8\%), raspy (6\%), deliberate (5\%), nasal (4\%), hesitant (4\%), flat (3\%), casual (3\%), soft (2\%), calm (2\%) \\ \addlinespace
\textbf{intelligibility words} \newline {\scriptsize 5 distinct; 1.46 tags per card} & highly intelligible (95\%), clean channel (43\%), mostly intelligible (5\%), poor channel quality (3\%), degraded intelligibility (0\%) \\ \addlinespace
\textbf{emotion words} \newline {\scriptsize 5 distinct; 1.01 tags per card} & neutral (67\%), excited/happy (16\%), angry/stressed (12\%), sad/subdued (3\%), calm (2\%) \\ \addlinespace
\textbf{language} \newline {\scriptsize 1 distinct; 1.00 tags per card} & en (100\%) \\ \addlinespace
\textbf{loudness label} \newline {\scriptsize 1.65 segments per card} & extremely quiet (1\%), very quiet (4\%), quiet (14\%), medium (33\%), loud (36\%), very loud (13\%) \\
\end{longtable}
\endgroup
\begingroup
\footnotesize
\begin{longtable}{@{}p{2.5cm}p{11cm}@{}}
\caption{Sound-effect cards (4493 cards in the final set): tag distribution per field. Percentages are the share of cards carrying the tag (loudness: share of segments); closed fields list every tag of the vocabulary, open fields the most frequent ones.}
\label{tab:dist-sfx} \\
\toprule
field & tags (share of cards) \\
\midrule
\endfirsthead

\multicolumn{2}{@{}l}{\small\tablename~\thetable{} (continued)} \\
\toprule
field & tags (share of cards) \\
\midrule
\endhead

\midrule
\multicolumn{2}{r@{}}{\emph{Continued on next page}} \\
\endfoot

\bottomrule
\endlastfoot

\textbf{class (source library)} \newline {\scriptsize 147 distinct, top 20; 1.00 tags per card} & Misc (23\%), General (5\%), Knock (4\%), Impact (4\%), Scream (4\%), Human (3\%), Tick (3\%), Laugh (3\%), Breath (2\%), Songbird (2\%), Applause (2\%), Click (2\%), Cheering (2\%), Combustion (1\%), Power (1\%), Movement (1\%), Hand (1\%), Cheer (1\%), Synthetic (1\%), Insect (1\%) \\ \addlinespace
\textbf{temporal descriptors} \newline {\scriptsize 13 distinct; 2.02 tags per card} & impact/abrupt (55\%), single event (37\%), sustained (28\%), continuous (27\%), intermittent (24\%), rising (9\%), periodic (7\%), rhythmic (5\%), fade out (5\%), oscillating (2\%), falling (2\%), fade in (1\%), sustained with occasional bursts (0\%) \\ \addlinespace
\textbf{plurality} \newline {\scriptsize 3 distinct; 1.00 tags per card} & one (79\%), multiple (18\%), none (3\%) \\ \addlinespace
\textbf{perceived audio quality} \newline {\scriptsize 13 distinct; 3.29 tags per card} & dry (73\%), low background noise (53\%), low-fidelity (47\%), thin (43\%), clear (43\%), bass-heavy (23\%), muffled (11\%), harsh (9\%), high-fidelity (9\%), high background noise (8\%), distorted (6\%), room reverb (4\%), reverb-heavy (1\%) \\ \addlinespace
\textbf{sound descriptors (open)} \newline {\scriptsize 451 distinct, top 15; 3.08 tags per card} & sharp (43\%), metallic (27\%), high-pitched (22\%), hollow (19\%), rumbling (10\%), low-pitched (10\%), rapid (7\%), deep (7\%), piercing (6\%), resonant (6\%), percussive (5\%), harsh (5\%), mechanical (5\%), steady (4\%), scraping (4\%) \\ \addlinespace
\textbf{sounding object (open)} \newline {\scriptsize 1123 distinct, top 15; 1.09 tags per card; 3.5\% empty} & crowd (5\%), human (3\%), wind (2\%), water (2\%), hands (2\%), door (2\%), metal (2\%), machine (1\%), alarm clock (1\%), door knocker (1\%), female (1\%), wood (1\%), relay (1\%), glass (1\%), synth (1\%) \\ \addlinespace
\textbf{sounding action (open)} \newline {\scriptsize 485 distinct, top 15; 1.27 tags per card; 1.4\% empty} & clicking (4\%), knocking (4\%), screaming (3\%), clapping (3\%), ticking (3\%), cheering (3\%), beeping (3\%), whistling (3\%), breathing (2\%), running (2\%), laughing (2\%), humming (2\%), rattling (2\%), singing (2\%), calling (2\%) \\ \addlinespace
\textbf{loudness label} \newline {\scriptsize 1.28 segments per card} & extremely quiet (6\%), very quiet (11\%), quiet (18\%), medium (26\%), loud (27\%), very loud (11\%) \\
\end{longtable}
\endgroup

\section{Example cards}\label{app:F_card-examples}
Figures~\ref{fig:card-music}--\ref{fig:card-sfx} present one MixCard of each type for the same scene. The music, the speech and one sound effect source is reproduced as released.

\par\medskip
\noindent\textbf{Music card, scene \texttt{7FTYb5K3uxk\_22000}}\par\nobreak\noindent\rule{\linewidth}{0.3pt}\vspace{-2pt}
{\scriptsize\begin{verbatim}
{
 "Event_ID": "7FTYb5K3uxk_22000",
 "segments": [["00:00.000", "00:08.034"]],
 "loudness": [[-33.2, "Very quiet"]],
 "frequency_profile": [
  {
   "start_time": 0.0,
   "end_time": 8.034,
   "bands_db": {
    "125": 26.5185,
    "250": 24.1402,
    "500": 25.5061,
    "1000": 24.0791,
    "2000": 21.2974,
    "3000": 18.0465,
    "4000": 14.8284,
    "6000": 12.2992,
    "8000": 8.9824,
    "10000": 6.8215
   },
   "bands_label": {
    "125": "medium",
    "250": "medium",
    "500": "medium",
    "1000": "medium",
    "2000": "medium",
    "3000": "medium-low",
    "4000": "medium-low",
    "6000": "medium-low",
    "8000": "low",
    "10000": "low"
   }
  }
 ],
 "genre": ["Indie-Rock", "Rock"],
 "mood": ["joyful activation", "power"],
 "perceived_audio_quality": ["low-fidelity", "dry", "thin", "muffled", "low background
   noise", "distorted"],
 "instruments": ["synthesizer", "bass guitar", "electric guitar", "drums", "vocals"],
 "music_descriptors": ["repetitive", "looping", "layered", "melodic"],
 "tempo": [119, "medium"],
 "short_caption": "Energetic indie-rock plays",
 "descriptive_caption": "A layered, melodic indie-rock track featuring synthesizer, bass,
   electric guitar, drums, and vocals loops repetitively at a medium tempo, delivering a
   powerful and joyful feel through a dry, low-fidelity, and distorted mix."
}
\end{verbatim}}\vspace{-6pt}\noindent\rule{\linewidth}{0.3pt}
\captionof{figure}{Music card of scene \texttt{7FTYb5K3uxk\_22000}.}\label{fig:card-music}
\medskip

\par\medskip
\noindent\textbf{Speech card, scene \texttt{7FTYb5K3uxk\_22000}}\par\nobreak\noindent\rule{\linewidth}{0.3pt}\vspace{-2pt}
{\scriptsize\begin{verbatim}
{
 "Event_ID": "7FTYb5K3uxk_22000",
 "Speech_stem_files": ["foreground0_Speech.wav"],
 "segments": [["00:00.000", "00:08.046"]],
 "loudness": [[-19.7, "Loud"]],
 "frequency_profile": [
  {
   "start_time": 0.0,
   "end_time": 8.046,
   "bands_db": {
    "125": 41.3912,
    "250": 40.0127,
    "500": 40.0989,
    "1000": 33.7436,
    "2000": 25.2864,
    "3000": 17.4382,
    "4000": 23.1175,
    "6000": 27.1334,
    "8000": 26.5359,
    "10000": 22.2535
   },
   "bands_label": {
    "125": "high",
    "250": "high",
    "500": "high",
    "1000": "medium-high",
    "2000": "medium",
    "3000": "medium-low",
    "4000": "medium",
    "6000": "medium-high",
    "8000": "medium",
    "10000": "medium"
   }
  }
 ],
 "median_f0_hz": [[0.0, 8.046, 127.1]],
 "perceived_audio_quality": ["low-fidelity", "dry", "muffled", "low background noise"],
 "manner": ["monotonous", "loud"],
 "intelligibility": ["highly intelligible"],
 "emotion": ["neutral"],
 "transcript": {
  "language": "en",
  "speaker_count": 1,
  "segments": [
   {
    "start": 0.292,
    "end": 0.873,
    "text": "and many more."
   },
   {
    "start": 1.013,
    "end": 5.462,
    "text": "Call Flatel today to get telephone, internet, voice over IP and long distance
      1-888-777-6561."
   }
  ]
 },
 "short_caption": "Monotonous deep voice speaking",
 "descriptive_caption": "A deep-voiced speaker delivers a monotonous, loud advertisement for
   Flatel's telecommunications services throughout the clip."
}
\end{verbatim}}\vspace{-6pt}\noindent\rule{\linewidth}{0.3pt}
\captionof{figure}{Speech card of scene \texttt{7FTYb5K3uxk\_22000}.}\label{fig:card-speech}
\medskip

\par\medskip
\noindent\textbf{Sound-effect card (Telephone), scene \texttt{7FTYb5K3uxk\_22000}}\par\nobreak\noindent\rule{\linewidth}{0.3pt}\vspace{-2pt}
{\scriptsize\begin{verbatim}
{
 "Event_ID": "7FTYb5K3uxk_22000",
 "Class": "Telephone",
 "SFX_stem_files": ["foreground2_Telephone.wav"],
 "segments": [["00:00.975", "00:02.206"]],
 "loudness": [[-37.5, "Very quiet"]],
 "frequency_profile": [
  {
   "start_time": 0.975,
   "end_time": 2.206,
   "bands_db": {
    "125": -22.6624,
    "250": -19.1229,
    "500": -0.7228,
    "1000": 24.9273,
    "2000": 21.6539,
    "3000": 14.1515,
    "4000": 5.3187,
    "6000": 2.7309,
    "8000": 0.1679,
    "10000": -7.8755
   },
   "bands_label": {
    "125": "very-low",
    "250": "very-low",
    "500": "very-low",
    "1000": "medium",
    "2000": "medium",
    "3000": "medium-low",
    "4000": "low",
    "6000": "low",
    "8000": "very-low",
    "10000": "very-low"
   }
  }
 ],
 "sources": [
  {
   "filename": "COMTelph_Phone Ring 06_PSE_SCSE_gc8Ea-s001.wav",
   "keywords": "Telephone Ring, Electronic, Pings, Chirps, Melodies",
   "description": "Electronic telephone ringing with pings and chirps in melodic sequence",
   "library": "PSE Audio Dataset",
   "category": "Communications",
   "subcategory": "Telephone",
   "segments": [[0.975, 2.206]]
  }
 ],
 "sounding_object": ["telephone"],
 "sounding_action": ["ringing"],
 "temporal_descriptors": ["periodic", "oscillating"],
 "plurality": "one",
 "perceived_audio_quality": ["clear", "dry", "thin"],
 "sound_descriptors": ["electronic", "high-pitched"],
 "short_caption": "Thin electronic phone ring",
 "descriptive_caption": "A thin, high-pitched electronic telephone rings in a periodic,
   oscillating sequence."
}
\end{verbatim}}\vspace{-6pt}\noindent\rule{\linewidth}{0.3pt}
\captionof{figure}{Sound-effect card (Telephone) of scene \texttt{7FTYb5K3uxk\_22000}.}\label{fig:card-sfx}
\medskip

\section{Evaluation protocol}
\label{app:G_eval_protocol}

To comprehensively evaluate the quality and accuracy of generated information cards across speech, music, and sound effects, we categorize the evaluated aspects into five distinct fields. Each field represents a critical dimension of audio understanding, ranging from high-level semantic descriptions to objective, low-level acoustic properties. We define five corresponding evaluation protocols, one per column type. Each protocol produces a normalized score in $[0,1]$, ensuring that performance can be easily aggregated, compared, and interpreted across different data formats and audio types.

\textbf{Sound Events.} To evaluate the temporal localization of active sound events, we follow TAC~\citep{kumarTACTimestampedAudio2026} and adapt two standard Sound Event Detection (SED) metrics. First, we adopt an event-based (collar-based) F1 score to evaluate the precise capture of event boundaries, utilizing a 1-second tolerance window for both onset and offset times. Second, we compute a segment-based F1 score using a 100-millisecond window to accurately measure the predicted duration and continuous temporal coverage of the events.

\textbf{Ordinal Values.} For ordinal labels, we calculate a normalized ordinal accuracy score. Rather than treating categories as independent classes, this metric degrades smoothly based on the distance between the predicted level and the ground truth. For an array of $N$ items, each having $L$ possible ordered levels, the score is calculated as:
\begin{equation}
    \label{eq:ordinal}
    s = 1 - \frac{1}{N} \sum_{i=1}^{N} \frac{\vert{}\hat{o}_i - o_i\vert{}}{L - 1}
\end{equation}

\textbf{Numerical Values.} For continuous scalar fields, we first compute the raw Mean Absolute Error (MAE) between the prediction $\hat{y}$ and the ground truth $y$. To project these unbounded errors onto our unified $[0,1]$ scale, we apply a $\sigma$-normalized transformation:

\begin{equation}
    \label{eq:norm_mae}
    s = \mathrm{clip}\!\left(1 - \frac{|\hat{y} - y|}{4\sigma},\ 0,\ 1\right)
\end{equation}

Here, $\sigma$ represents the robust dispersion of that specific variable across the dataset, calculated as the median absolute deviation. By scaling the error by $4\sigma$, perfect predictions score $1$, while outputs that deviate significantly are smoothly penalized toward $0$.

\textbf{Tag-like Descriptors.} For fields containing unstructured lists of words or categorical descriptors, we evaluate the predicted tag-sets using a single-pass LLM judge. Rather than relying on rigid exact lexical matching, the LLM is prompted with the predicted list and the reference list simultaneously to compute a single, holistic agreement score in $[0,1]$. By reasoning over both sets as a whole, the judge intelligently credits valid synonyms and near-synonyms (e.g., sound $\leftrightarrow$ noise, metallic $\leftrightarrow$ industrial) while jointly penalizing misses (reference tags with no match) and spurious extras (predicted tags matching nothing). To ground the evaluation, the judge is provided with a field-specific hint (e.g., ``sound-source nouns'' or ``acoustic-quality adjectives'').

\textbf{Captions.} We evaluate free-text description fields for overall caption adequacy using the LLM judge. The judge rates the prediction against the reference across four core criteria: correctness, relevance, completeness, and clarity~\cite{kumar2026mmaupro}. Each criterion is scored in $[0,1]$, and the per-sample score is their mean.

\textbf{LLM as a judge.} We evaluated three different LLM judges and found that they produced largely consistent model rankings. We ultimately selected Llama-3-8B-Instruct \citep{llama3modelcard} to use a judge from a different model family than the evaluated LALMs, reducing potential evaluation bias. Table~\ref{tab:judges} reports each model's scores and ranks under the three LLM judges, together with the Spearman correlation between each judge's ranking and that of Llama-3-8B-Instruct.

\begin{table*}[ht]
\centering
\small
\caption{Model scores and ranks (in parentheses) under the three LLM judges. $^\dagger$ marks a rank that differs from Llama-3-8B's.}
\label{tab:judges}
\begin{tabular}{lrrrrrr}
\toprule
 & \multicolumn{3}{c}{Stage 1: sound-effect segment F1} & \multicolumn{3}{c}{Stage 2: judged-field score} \\
\cmidrule(lr){2-4} \cmidrule(lr){5-7}
Model & Llama-3-8B & Qwen2.5-7B & Qwen3-4B & Llama-3-8B & Qwen2.5-7B & Qwen3-4B \\
\midrule
Qwen3-Omni & 0.374 (1) & 0.337 (1) & 0.224 (1) & 0.577 (1) & 0.569 (1) & 0.643 (1) \\
Qwen2.5-Omni & 0.276 (2) & 0.252 (2) & 0.164 (2) & 0.454 (2) & 0.444 (2) & 0.512 (2) \\
AF 3 & 0.259 (3) & 0.239 (3) & 0.164 (3) & 0.060 (6) & 0.052 (6) & 0.059 (6) \\
Music Flamingo & 0.084 (6) & 0.069 (6) & 0.054 (6) & 0.327 (3) & 0.310 (3) & 0.349 (4)$^\dagger$ \\
AF Next & 0.208 (4) & 0.184 (4) & 0.145 (4) & 0.291 (5) & 0.262 (5) & 0.310 (5) \\
Kimi-Audio & 0.200 (5) & 0.150 (5) & 0.088 (5) & 0.327 (4) & 0.305 (4) & 0.384 (3)$^\dagger$ \\
\midrule
Spearman $\rho$ & -- & 1.00 & 1.00 & -- & 1.00 & 0.94 \\
\bottomrule
\end{tabular}
\end{table*}

\section{Extended Results}
\label{app:extended_results}

\subsection{Human Validation}
\begin{figure}[H]
    \centering
    \includegraphics[width=\linewidth]{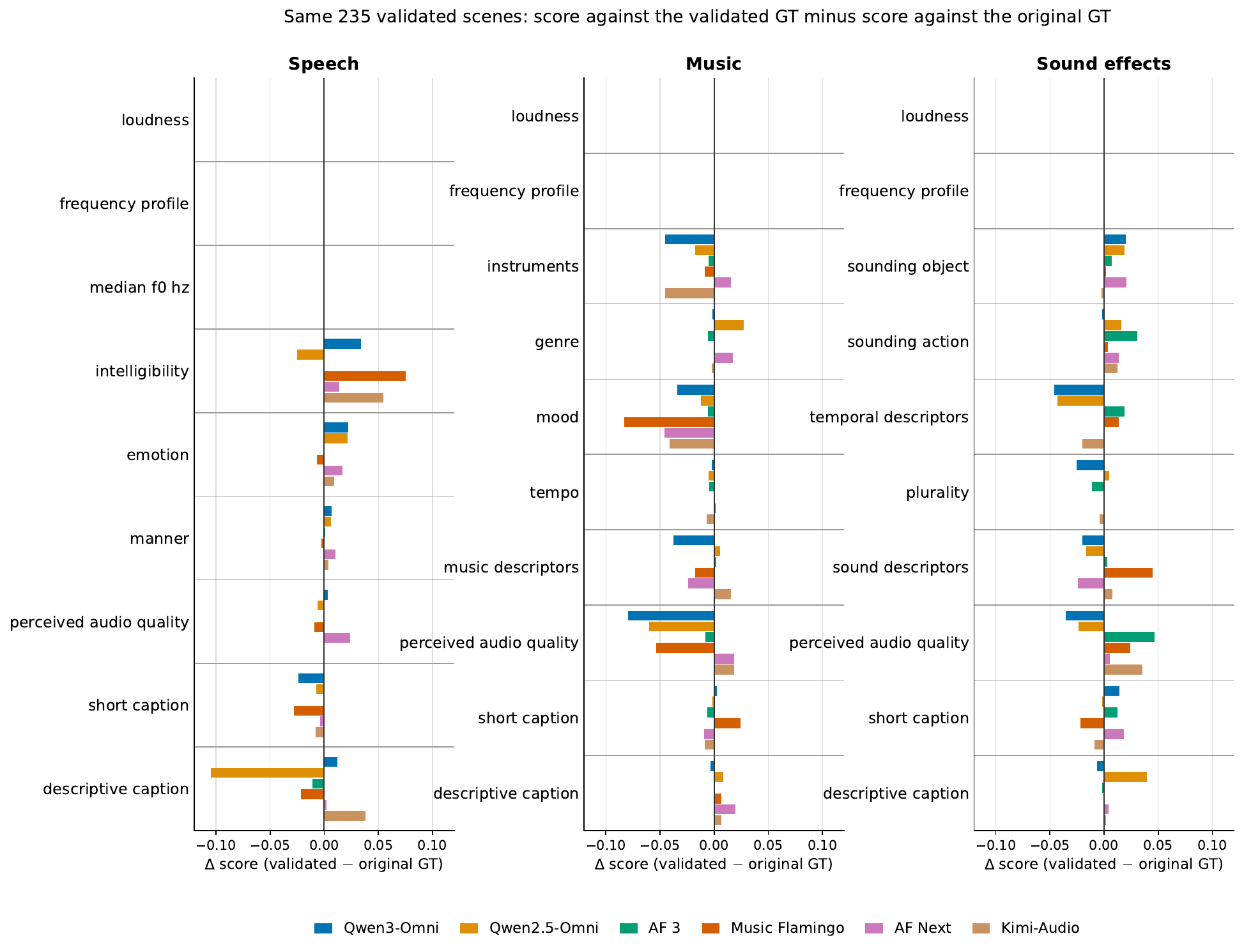}
    \caption{Change in model performance when scored against the human-validated ground truth instead of the generated ground truth.}
    \label{fig:validated_vs_original}
\end{figure}

\subsection{Model Performance}

\begin{figure}[H]
    \centering
    \includegraphics[width=\linewidth]{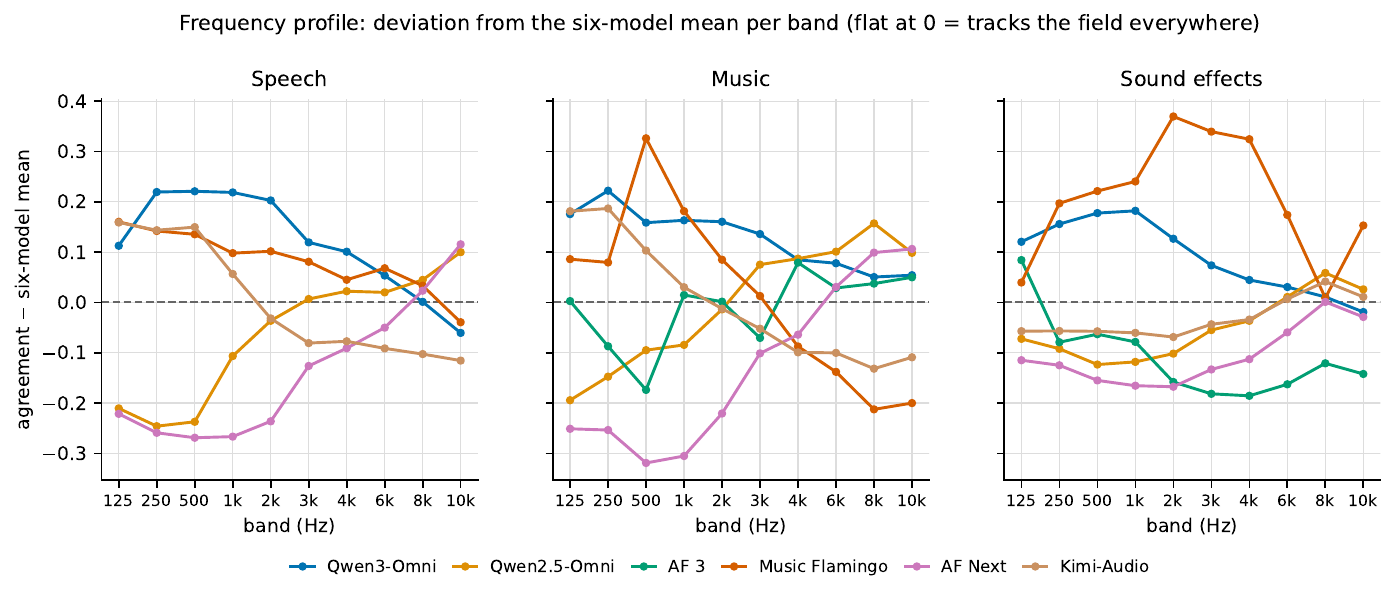}
    \caption{Frequency profile agreement per band, shown as each model's deviation from the six-model mean.}
    \label{fig:frequency_bands}
\end{figure}

\begin{table*}[H]
\centering
\small
\caption{Source inventory F1 (segment-based, 100\,ms; event-based, 1\,s collar) and mean card field score, per card type, on 4884 scenes. Best per column in bold.}
\label{tab:overview}
\begin{tabular}{lrrrrrrrrr}
\toprule
 & \multicolumn{3}{c}{Segment F1} & \multicolumn{3}{c}{Event F1} & \multicolumn{3}{c}{Card field score} \\
\cmidrule(lr){2-4} \cmidrule(lr){5-7} \cmidrule(lr){8-10}
Model & Sp. & Mu. & SFX & Sp. & Mu. & SFX & Sp. & Mu. & SFX \\
\midrule
Qwen3-Omni & \textbf{0.639} & 0.932 & \textbf{0.374} & \textbf{0.510} & \textbf{0.858} & \textbf{0.239} & \textbf{0.648} & \textbf{0.631} & \textbf{0.612} \\
Qwen2.5-Omni & 0.599 & 0.886 & 0.276 & 0.185 & 0.785 & 0.103 & 0.466 & 0.456 & 0.517 \\
AF 3 & 0.577 & 0.858 & 0.259 & 0.129 & 0.676 & 0.075 & 0.017 & 0.039 & 0.127 \\
Music Flamingo & 0.334 & 0.894 & 0.084 & 0.065 & 0.365 & 0.030 & 0.309 & 0.291 & 0.319 \\
AF Next & 0.487 & \textbf{0.937} & 0.208 & 0.072 & 0.674 & 0.066 & 0.298 & 0.324 & 0.443 \\
Kimi-Audio & 0.218 & 0.870 & 0.200 & 0.053 & 0.758 & 0.060 & 0.421 & 0.364 & 0.327 \\
\bottomrule
\end{tabular}
\end{table*}

\begin{table*}[H]
\centering
\small
\caption{Score on the fields every card type carries, per card type, on 4884 scenes. Best model per row in bold.}
\label{tab:common_fields}
\begin{tabular}{llrrrrrr}
\toprule
 & & Qwen3-Omni & Qwen2.5-Omni & AF 3 & Music Flamingo & AF Next & Kimi-Audio \\
\midrule
\multirow{3}{*}{loudness} & Speech & \textbf{0.822} & 0.455 & 0.001 & 0.652 & 0.619 & 0.620 \\
 & Music & \textbf{0.803} & 0.559 & 0.064 & 0.688 & 0.738 & 0.440 \\
 & Sound effects & \textbf{0.803} & 0.530 & 0.041 & 0.497 & 0.654 & 0.355 \\
\midrule
\multirow{3}{*}{frequency profile} & Speech & \textbf{0.732} & 0.503 & 0.000 & 0.006 & 0.364 & 0.533 \\
 & Music & \textbf{0.740} & 0.610 & 0.000 & 0.000 & 0.462 & 0.578 \\
 & Sound effects & \textbf{0.734} & 0.593 & 0.011 & 0.002 & 0.501 & 0.595 \\
\midrule
\multirow{3}{*}{perceived audio quality} & Speech & \textbf{0.598} & 0.453 & 0.024 & 0.362 & 0.229 & 0.481 \\
 & Music & \textbf{0.697} & 0.571 & 0.038 & 0.499 & 0.412 & 0.460 \\
 & Sound effects & \textbf{0.659} & 0.573 & 0.131 & 0.419 & 0.460 & 0.393 \\
\midrule
\multirow{3}{*}{short caption} & Speech & \textbf{0.410} & 0.304 & 0.003 & 0.177 & 0.153 & 0.216 \\
 & Music & \textbf{0.388} & 0.236 & 0.012 & 0.124 & 0.122 & 0.053 \\
 & Sound effects & \textbf{0.238} & 0.192 & 0.020 & 0.130 & 0.188 & 0.047 \\
\midrule
\multirow{3}{*}{descriptive caption} & Speech & \textbf{0.676} & 0.450 & 0.020 & 0.272 & 0.258 & 0.307 \\
 & Music & \textbf{0.702} & 0.392 & 0.000 & 0.291 & 0.298 & 0.339 \\
 & Sound effects & \textbf{0.392} & 0.287 & 0.036 & 0.210 & 0.222 & 0.096 \\
\bottomrule
\end{tabular}
\end{table*}

\begin{table*}[H]
\centering
\small
\caption{Score on the fields unique to each card type, on 4884 scenes. Best model per row in bold.}
\label{tab:unique_fields}
\begin{tabular}{llrrrrrr}
\toprule
 & & Qwen3-Omni & Qwen2.5-Omni & AF 3 & Music Flamingo & AF Next & Kimi-Audio \\
\midrule
\multirow{4}{*}{Speech} & median f0 hz & \textbf{0.849} & 0.451 & 0.000 & 0.006 & 0.441 & 0.512 \\
 & intelligibility & \textbf{0.537} & 0.458 & 0.040 & 0.399 & 0.066 & 0.233 \\
 & emotion & \textbf{0.709} & 0.670 & 0.036 & 0.534 & 0.237 & 0.458 \\
 & manner & \textbf{0.498} & 0.447 & 0.025 & 0.376 & 0.315 & 0.433 \\
\midrule
\multirow{5}{*}{Music} & instruments & \textbf{0.688} & 0.538 & 0.040 & 0.285 & 0.335 & 0.451 \\
 & genre & 0.467 & \textbf{0.478} & 0.100 & 0.292 & 0.202 & 0.272 \\
 & mood & \textbf{0.710} & 0.563 & 0.109 & 0.409 & 0.341 & 0.477 \\
 & tempo & \textbf{0.510} & 0.223 & 0.007 & 0.016 & 0.063 & 0.205 \\
 & music descriptors & \textbf{0.606} & 0.393 & 0.025 & 0.308 & 0.273 & 0.367 \\
\midrule
\multirow{5}{*}{Sound effects} & sounding object & \textbf{0.531} & 0.497 & 0.167 & 0.424 & 0.409 & 0.309 \\
 & sounding action & \textbf{0.627} & 0.613 & 0.206 & 0.514 & 0.514 & 0.475 \\
 & temporal descriptors & \textbf{0.719} & 0.563 & 0.153 & 0.395 & 0.346 & 0.429 \\
 & plurality & 0.886 & \textbf{0.892} & 0.336 & 0.186 & 0.751 & 0.248 \\
 & sound descriptors & \textbf{0.533} & 0.435 & 0.168 & 0.417 & 0.386 & 0.327 \\
\bottomrule
\end{tabular}
\end{table*}

\begin{figure}[H]
    \centering
    \begin{subfigure}[b]{0.48\textwidth}
        \centering
        \includegraphics[width=\textwidth]{figures/fig6_precision_recall_segment.pdf} 
        \label{fig:sub1}
    \end{subfigure}
    \hfill
    \begin{subfigure}[b]{0.48\textwidth}
        \centering
        \includegraphics[width=\textwidth]{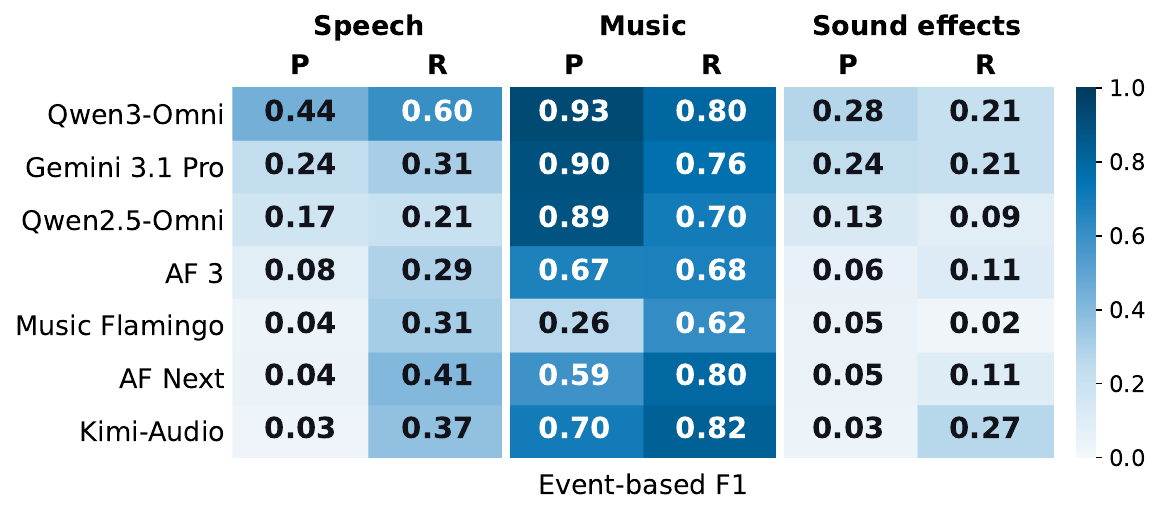}
        \label{fig:sub2}
    \end{subfigure}
    \caption{Stage-1 precision (P) and recall (R) per card type.}
    \label{fig:main_figure}
\end{figure}

\begin{figure}[ht]
    \centering
    \includegraphics[width=\linewidth]{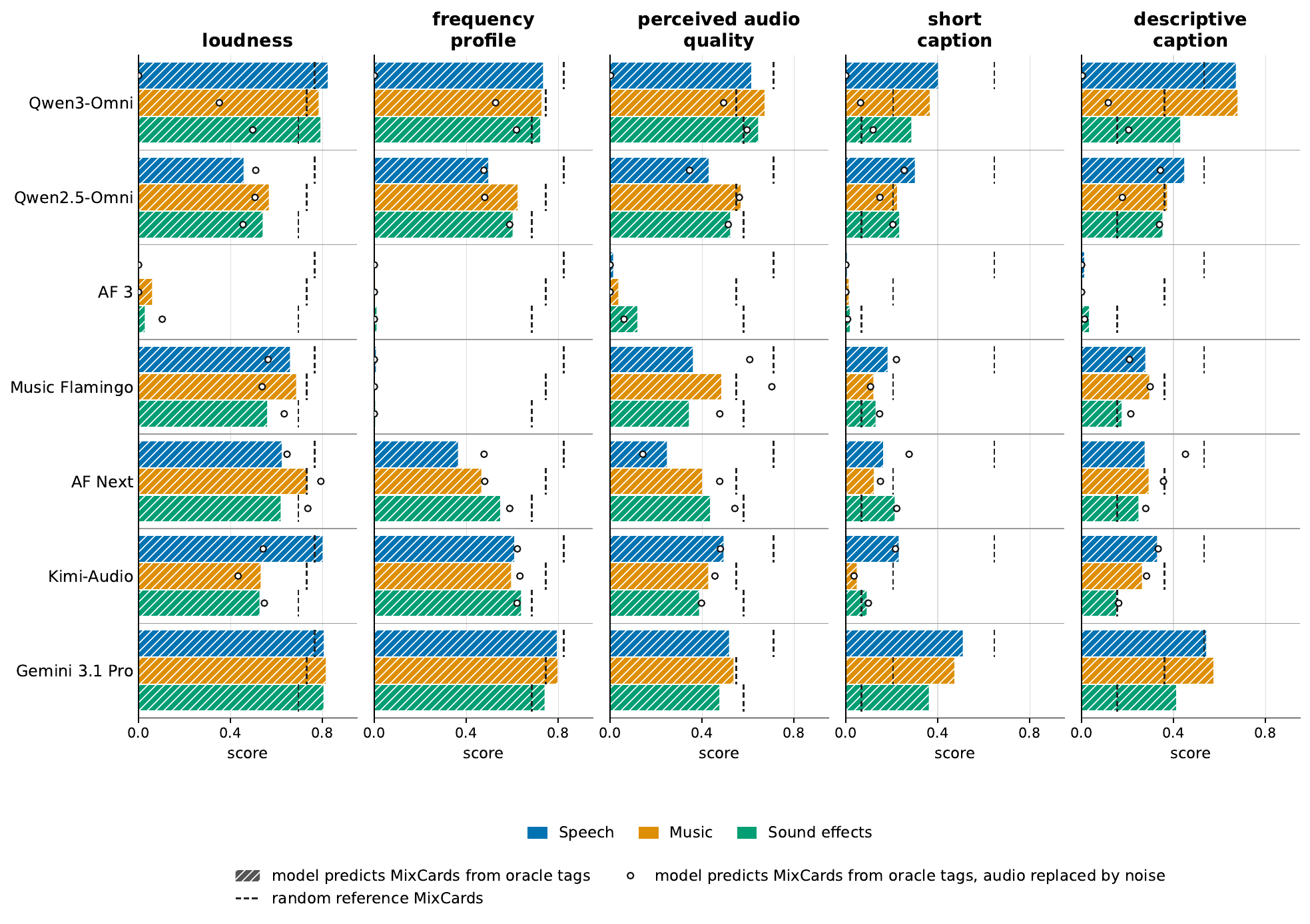}
    \caption{Model performance on the five fields common to all card types, compared across speech, music and sound effects.}
    \label{fig:common_fields}
\end{figure}

\begin{figure}[ht]
    \centering
    \includegraphics[width=\linewidth]{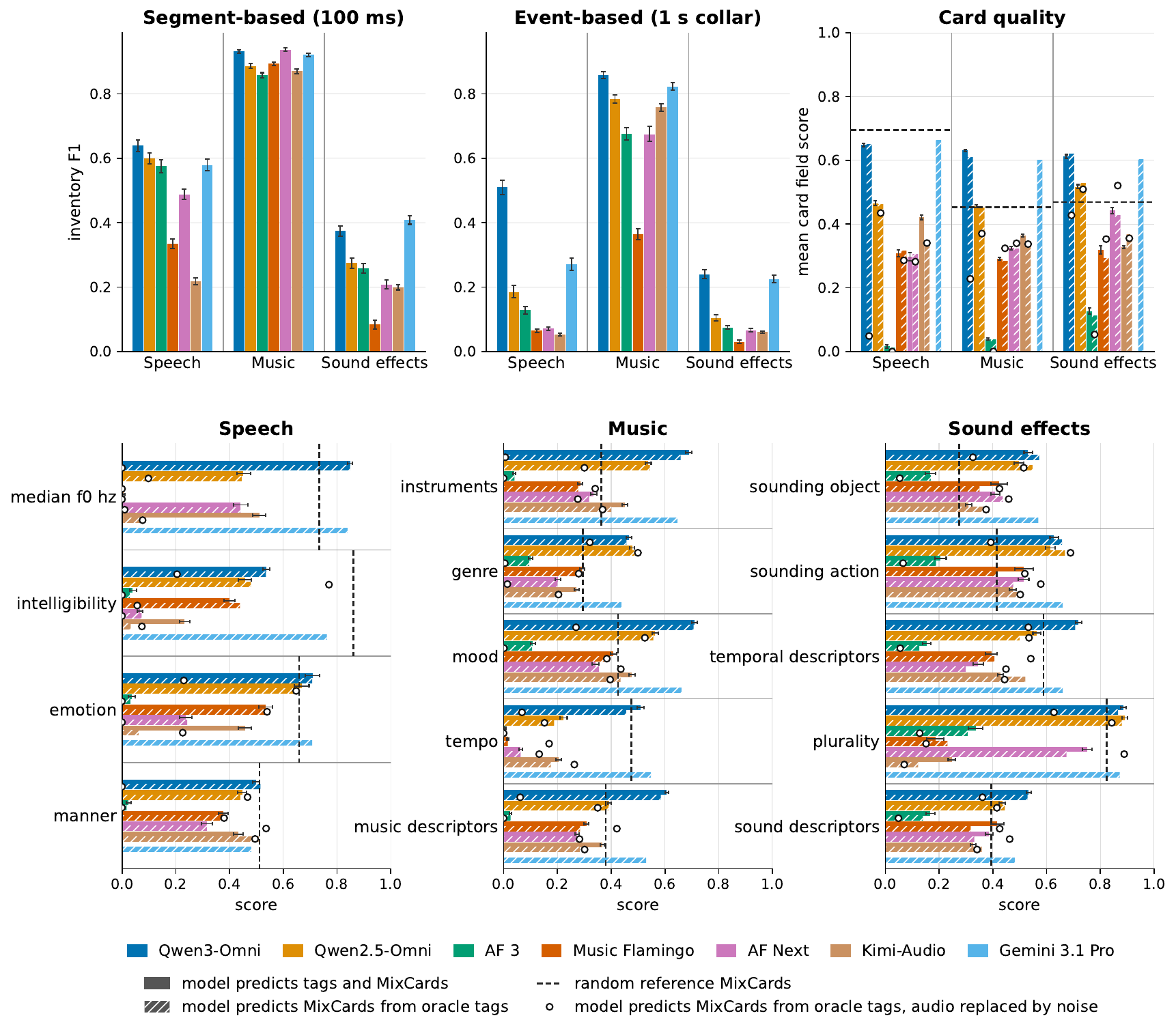}
    \caption{MixCardBench performance with oracle tags, i.e. conditioned on ground truth tags, and predicted tags.}
    \label{fig:oracle_predicted_unique_fields}
\end{figure}

\begin{figure}[ht]
    \centering
    \includegraphics[width=\linewidth]{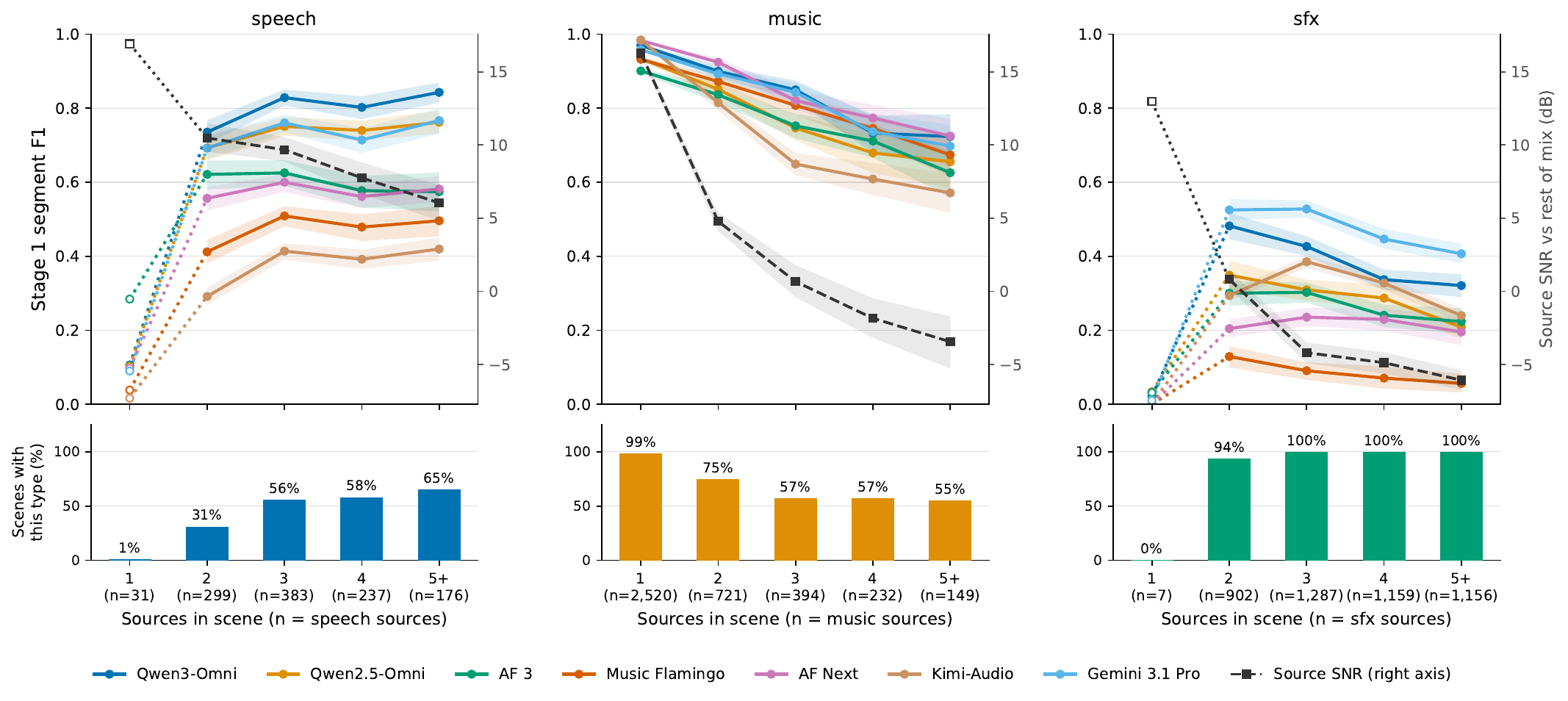}
    \caption{Model performance on detection, compared across speech, music and sound effects at different polyphony levels.}
    \label{fig:polyphony_modality_snr_prevalence}
\end{figure}

\begin{table}[t]
\centering
\caption{Results on MixCardBench-val (565 cards). Stage 1 is the source detection F1 (event-based and segment-based). Stage 2 is the mean card field score, overall and per card type, for MixCards conditioned on oracle tags (i.e. ground truth stage 1 predictions) or on the model's predicted tags.}
\label{tab:val565_closed_models}
\small
\setlength{\tabcolsep}{4pt}
\begin{tabular}{l cc cccc cccc}
\toprule
 & \multicolumn{2}{c}{Stage 1} & \multicolumn{4}{c}{Stage 2, oracle tags} & \multicolumn{4}{c}{Stage 2, predicted tags} \\
\cmidrule(lr){2-3} \cmidrule(lr){4-7} \cmidrule(lr){8-11}
Model & Event & Segment & All & Speech & Music & SFX & All & Speech & Music & SFX \\
\midrule
Qwen3-Omni & \textbf{0.416} & \textbf{0.680} & \textbf{0.607} & \textbf{0.655} & 0.579 & \textbf{0.598} & \textbf{0.616} & \underline{0.647} & \underline{0.599} & \textbf{0.608} \\
Gemini 3.1 Pro & \underline{0.305} & \underline{0.656} & \underline{0.601} & 0.646 & \underline{0.616} & 0.571 & \underline{0.603} & \textbf{0.658} & \textbf{0.625} & 0.562 \\
GPT-Audio-Mini & 0.159 & 0.544 & 0.550 & 0.658 & 0.485 & 0.541 & 0.555 & 0.673 & 0.481 & 0.545 \\
Gemini 3.8 Live & 0.214 & 0.592 & 0.573 & 0.658 & 0.565 & 0.502 & 0.581 & 0.678 & 0.573 & 0.499 \\
\bottomrule
\end{tabular}
\end{table}

\section{Prompts}\label{app:J_prompts}
This appendix lists every prompt used to generate MixCards (Appendix~\ref{app:prompts:listen} and~\ref{app:prompts:caption}) and to run the benchmark inference (Appendix~\ref{app:prompts:bench}). The text is copied verbatim from the code. Items in angle brackets, such as \texttt{<start>}, are values filled in per card at run time. A listening-stage call sends the system prompt, the user prompt and the isolated audio of the source. A caption-stage call sends the system prompt, three worked examples (omitted here) and the card rendered as text. A benchmark call sends the system prompt, the user prompt and the full 10-second scene.

\subsection{Ground-truth generation: listening stage (Qwen3-Omni-30B-A3B-Instruct)}\label{app:prompts:listen}
\subsubsection{Music: mood}
Qwen3-Omni listens to the music stem and picks one or two mood categories.

\noindent\begin{minipage}{\linewidth}
\noindent\textbf{System prompt}\par\nobreak\noindent\rule{\linewidth}{0.3pt}\vspace{-2pt}
{\scriptsize\begin{verbatim}
You are an expert music analyst. You identify the MOOD - the emotional character the music
  expresses or evokes, NOT its genre, instruments, or production quality.
\end{verbatim}}\vspace{-6pt}\noindent\rule{\linewidth}{0.3pt}\par\medskip
\end{minipage}
\par\medskip
\noindent\textbf{User prompt}\par\nobreak\noindent\rule{\linewidth}{0.3pt}\vspace{-2pt}
{\scriptsize\begin{verbatim}
Listen to this ~10-second music excerpt and choose the MOOD categories that best describe
  the emotional character the music expresses or evokes, from the fixed vocabulary below. Do
  NOT use any word that is not in the vocabulary.

Vocabulary (music-emotion categories, with definitions):
- amazement: feeling of wonder and happiness
- solemnity: feeling of transcendence, inspiration, thrills
- tenderness: sensuality, affect, feeling of love
- nostalgia: dreamy, melancholic, sentimental feelings
- calmness: relaxation, serenity, meditativeness
- power: feeling strong, heroic, triumphant, energetic
- joyful activation: feels like dancing, bouncy feeling, animated, amused
- tension: nervous, impatient, irritated, eerie, ominous, sense of dread or menace
- sadness: depressed, sorrowful

Context already extracted (judge with your ears; do NOT just repeat this):
- Genre: <FMA genre>, <FMA genre>
- Loudness: ['<LUFS>', '<label>']
- Measured frequency-band energy (band centre Hz: dB, 125 Hz to 10 kHz), per segment:
  seg0[125:<dB>, 250:<dB>, 500:<dB>, 1000:<dB>, 2000:<dB>, 3000:<dB>, 4000:<dB>, 6000:<dB>,
  8000:<dB>, 10000:<dB>]

Reply with ONLY a comma-separated list of 1-2 category names from the vocabulary, the
  dominant one first. No sentences, no explanation.
\end{verbatim}}\vspace{-6pt}\noindent\rule{\linewidth}{0.3pt}\par\medskip

\subsubsection{Music: perceived audio quality}
Qwen3-Omni rates the recording quality of the music stem.

\noindent\begin{minipage}{\linewidth}
\noindent\textbf{System prompt}\par\nobreak\noindent\rule{\linewidth}{0.3pt}\vspace{-2pt}
{\scriptsize\begin{verbatim}
You are an expert audio engineer and critical listener. You judge the PERCEIVED AUDIO
  QUALITY of music: the production, technical, and channel/source characteristics you can
  actually hear, NOT the genre, mood, or instruments.
\end{verbatim}}\vspace{-6pt}\noindent\rule{\linewidth}{0.3pt}\par\medskip
\end{minipage}
\par\medskip
\noindent\begin{minipage}{\linewidth}
\noindent\textbf{User prompt}\par\nobreak\noindent\rule{\linewidth}{0.3pt}\vspace{-2pt}
{\scriptsize\begin{verbatim}
Listen to this ~10-second music excerpt and describe ONLY its PERCEIVED AUDIO QUALITY by
  choosing tags from the fixed vocabulary below. Do NOT tag genre, mood, or instruments. Do
  NOT use any word that is not in the vocabulary.

Vocabulary, grouped by axis. Within an axis, pick AT MOST ONE tag (Frequency balance: up to
  two if not contradictory). Only tag what is CLEARLY audible; skip an axis entirely if
  nothing on it clearly applies.
- Fidelity: low-fidelity | clear | high-fidelity
- Space: dry | room reverb | reverb-heavy
- Frequency balance: bass-heavy | muffled | thin | harsh
- Drive: distorted
- Noise: low background noise | high background noise

Context already extracted (judge with your ears; do NOT just repeat this):
- Genre: <FMA genre>, <FMA genre>
- Loudness: ['<LUFS>', '<label>']
- Measured frequency-band energy (band centre Hz: dB, 125 Hz to 10 kHz), per segment:
  seg0[125:<dB>, 250:<dB>, 500:<dB>, 1000:<dB>, 2000:<dB>, 3000:<dB>, 4000:<dB>, 6000:<dB>,
  8000:<dB>, 10000:<dB>]

Reply with ONLY a comma-separated list of tags from the vocabulary above, most prominent
  first. No sentences, no explanation.
\end{verbatim}}\vspace{-6pt}\noindent\rule{\linewidth}{0.3pt}\par\medskip
\end{minipage}
\par\medskip

\subsubsection{Music: instruments}
Qwen3-Omni names the instruments it hears in the music stem.

\noindent\begin{minipage}{\linewidth}
\noindent\textbf{System prompt}\par\nobreak\noindent\rule{\linewidth}{0.3pt}\vspace{-2pt}
{\scriptsize\begin{verbatim}
You are an expert musician with a precise ear for instrumentation. You identify the
  INSTRUMENTS you can actually hear playing, NOT genre, mood, or production quality.
\end{verbatim}}\vspace{-6pt}\noindent\rule{\linewidth}{0.3pt}\par\medskip
\end{minipage}
\par\medskip
\noindent\textbf{User prompt}\par\nobreak\noindent\rule{\linewidth}{0.3pt}\vspace{-2pt}
{\scriptsize\begin{verbatim}
Listen to this ~10-second music excerpt and list ONLY the INSTRUMENTS you can actually hear
  playing, choosing from the fixed vocabulary below. Do NOT use any word that is not in the
  vocabulary.

Vocabulary (grouped for readability; pick from anywhere in the list):
- Keys: piano | electric piano | organ | harpsichord | synthesizer | accordion
- Guitars & plucked: guitar | electric guitar | acoustic guitar | bass guitar | ukulele |
  banjo | mandolin | lute | sitar | harp
- Bowed strings: violin | viola | cello | double bass | strings
- Brass: trumpet | trombone | tuba | french horn | brass
- Woodwinds: flute | piccolo | clarinet | oboe | bassoon | saxophone | bagpipes | harmonica
  | woodwinds
- Drums & percussion: drums | drum machine | percussion | cymbals | timpani | congas | tabla
  | darbuka | timbales | shaker | triangle | gong | handpan
- Pitched percussion: glockenspiel | xylophone | marimba | vibraphone | bell | chimes
- Voice: vocals | choir | whistling | beatboxing

Rule: name the MOST SPECIFIC instrument you can confidently identify. Use a family term
  (guitar, strings, brass, woodwinds, percussion, choir) ONLY when a whole section is
  playing or the specific instrument cannot be determined. Never use both a family term and
  one of its members for the same sound.
For guitars, commit to "acoustic guitar" or "electric guitar" whenever you can tell; use
  plain "guitar" only when genuinely unsure.
"drum machine" = programmed / electronic beats; "drums" = an acoustic drum kit.

Context already extracted (judge with your ears; do NOT just repeat this):
- Genre: <FMA genre>, <FMA genre>
- Loudness: ['<LUFS>', '<label>']
- Measured frequency-band energy (low / low-mid / high-mid / high), per segment:
  seg0[125:<dB>, 250:<dB>, 500:<dB>, 1000:<dB>, 2000:<dB>, 3000:<dB>, 4000:<dB>, 6000:<dB>,
  8000:<dB>, 10000:<dB>]

Reply with ONLY a comma-separated list of 1-6 instruments from the vocabulary, most
  prominent first. No sentences, no explanation.
\end{verbatim}}\vspace{-6pt}\noindent\rule{\linewidth}{0.3pt}\par\medskip

\subsubsection{Music: descriptors}
Qwen3-Omni describes the character of the music stem with tags from a fixed list.

\noindent\begin{minipage}{\linewidth}
\noindent\textbf{System prompt}\par\nobreak\noindent\rule{\linewidth}{0.3pt}\vspace{-2pt}
{\scriptsize\begin{verbatim}
You are an expert music critic. You give MUSIC DESCRIPTORS - words for the character,
  texture, or feel of the music itself that are NOT genre labels, NOT moods/emotions, NOT
  instrument names, and NOT production quality.
\end{verbatim}}\vspace{-6pt}\noindent\rule{\linewidth}{0.3pt}\par\medskip
\end{minipage}
\par\medskip
\noindent\begin{minipage}{\linewidth}
\noindent\textbf{User prompt}\par\nobreak\noindent\rule{\linewidth}{0.3pt}\vspace{-2pt}
{\scriptsize\begin{verbatim}
Listen to this ~10-second music excerpt and give MUSIC DESCRIPTOR tags for the
  character/texture/feel of the music (no genre, no mood, no instruments, no production
  quality), choosing from the fixed vocabulary below. Do NOT use any word that is not in the
  vocabulary.

Vocabulary:
rhythmic | layered | repetitive | sparse | sustained | dissonant | syncopated | percussive |
  minimalist | complex | arpeggiated | synthetic | glitchy | gritty | delicate | plucked |
  raw | melodic | warm | groovy | bright | abrupt | harmonic | high-pitched | rapid |
  intricate | punchy | orchestral | unpolished | detuned | lively | looping | evolving |
  static | ornamental | frenetic

Context already extracted (judge with your ears; do NOT just repeat this):
- Genre: <FMA genre>, <FMA genre>
- Loudness: ['<LUFS>', '<label>']
- Measured frequency-band energy (low / low-mid / high-mid / high), per segment:
  seg0[125:<dB>, 250:<dB>, 500:<dB>, 1000:<dB>, 2000:<dB>, 3000:<dB>, 4000:<dB>, 6000:<dB>,
  8000:<dB>, 10000:<dB>]

Reply with ONLY a comma-separated list of 2-5 tags from the vocabulary, most prominent
  first. No sentences, no explanation.
\end{verbatim}}\vspace{-6pt}\noindent\rule{\linewidth}{0.3pt}\par\medskip
\end{minipage}
\par\medskip

\noindent\begin{minipage}{\linewidth}
\subsubsection{Speech: perceived audio quality}
Qwen3-Omni rates the recording quality of the speech stem; the user prompt repeats the music one with speech-specific wording.

\noindent\textbf{System prompt}\par\nobreak\noindent\rule{\linewidth}{0.3pt}\vspace{-2pt}
{\scriptsize\begin{verbatim}
You are an expert audio engineer and critical listener. You judge the PERCEIVED AUDIO
  QUALITY of speech recordings: the production, technical, and channel/source
  characteristics you can actually hear, NOT what is said, NOT the speaker's emotion, and
  NOT how intelligible the words are.
\end{verbatim}}\vspace{-6pt}\noindent\rule{\linewidth}{0.3pt}\par\medskip
\noindent\textbf{User prompt (identical to Music: perceived audio quality except these lines)}\par\nobreak\noindent\rule{\linewidth}{0.3pt}\vspace{-2pt}
{\scriptsize\begin{verbatim}
Listen to the SPEECH in this clip (one or more speakers) and describe ONLY its PERCEIVED
  AUDIO QUALITY by choosing tags from the fixed vocabulary below. Do NOT tag the spoken
  content, the emotion, or intelligibility. Do NOT use any word that is not in the
  vocabulary.
- Speech loudness per segment: [<start>-<end>s: <LUFS> LUFS]
- Measured speech frequency-band energy (band centre Hz: dB, 125 Hz to 10 kHz), per segment:
  seg0[125:<dB>, 250:<dB>, 500:<dB>, 1000:<dB>, 2000:<dB>, 3000:<dB>, 4000:<dB>, 6000:<dB>,
  8000:<dB>, 10000:<dB>]
- Median F0 (fundamental frequency) per speech segment: [<start>-<end>s: <Hz> Hz]
\end{verbatim}}\vspace{-6pt}\noindent\rule{\linewidth}{0.3pt}\par\medskip
\end{minipage}
\par\medskip

\subsubsection{Speech: manner}
Qwen3-Omni describes how the speech is delivered.

\noindent\begin{minipage}{\linewidth}
\noindent\textbf{System prompt}\par\nobreak\noindent\rule{\linewidth}{0.3pt}\vspace{-2pt}
{\scriptsize\begin{verbatim}
You are an expert speech analyst. You describe the MANNER of speech - delivery, timbre and
  prosody characteristics, NOT emotion labels, NOT what is said.
\end{verbatim}}\vspace{-6pt}\noindent\rule{\linewidth}{0.3pt}\par\medskip
\end{minipage}
\par\medskip
\noindent\begin{minipage}{\linewidth}
\noindent\textbf{User prompt}\par\nobreak\noindent\rule{\linewidth}{0.3pt}\vspace{-2pt}
{\scriptsize\begin{verbatim}
Listen to the SPEECH in this ~10-second clip (one or more speakers) and give tags for the
  MANNER of speech - the delivery / prosody characteristics only: pace, rhythm,
  articulation, vocal effort, tone of delivery (e.g. gruff, whispered, conversational,
  animated, hesitant, monotonous, breathy).
Do NOT give pitch words (low-pitched, high-pitched, deep - pitch is measured),
  recording/channel words (muffled, distorted, thin, reverberant - that is audio quality),
  or emotion words (calm, angry, happy, sad, stressed, neutral - emotion is measured
  separately).

Context already extracted (judge with your ears; do NOT just repeat this):
- Speech loudness per segment: [<start>-<end>s: <LUFS> LUFS]
- Measured speech frequency-band energy (band centre Hz: dB, 125 Hz to 10 kHz), per segment:
  seg0[125:<dB>, 250:<dB>, 500:<dB>, 1000:<dB>, 2000:<dB>, 3000:<dB>, 4000:<dB>, 6000:<dB>,
  8000:<dB>, 10000:<dB>]
- Median F0 (fundamental frequency) per speech segment: [<start>-<end>s: <Hz> Hz]

Reply with ONLY a comma-separated list of 2-5 tags, most prominent first. No sentences, no
  explanation.
\end{verbatim}}\vspace{-6pt}\noindent\rule{\linewidth}{0.3pt}\par\medskip
\end{minipage}
\par\medskip

\subsubsection{Sound effects: temporal descriptors}
Qwen3-Omni describes how the sound behaves over time.

\noindent\begin{minipage}{\linewidth}
\noindent\textbf{System prompt}\par\nobreak\noindent\rule{\linewidth}{0.3pt}\vspace{-2pt}
{\scriptsize\begin{verbatim}
You are an expert sound designer. You identify the TEMPORAL BEHAVIOR of a sound - how it
  occurs and evolves over the clip - NOT what the source is, NOT how many sources there are,
  and NOT its production quality.
\end{verbatim}}\vspace{-6pt}\noindent\rule{\linewidth}{0.3pt}\par\medskip
\end{minipage}
\par\medskip
\noindent\begin{minipage}{\linewidth}
\noindent\textbf{User prompt}\par\nobreak\noindent\rule{\linewidth}{0.3pt}\vspace{-2pt}
{\scriptsize\begin{verbatim}
Listen to this sound-effects clip and describe ONLY the TEMPORAL BEHAVIOR of the sound over
  the clip by choosing tags from the fixed vocabulary below. Do NOT use any word that is not
  in the vocabulary.

Vocabulary, grouped by axis. Pick EXACTLY ONE Occurrence tag; from the other axes pick AT
  MOST ONE tag each, only if clearly audible - skip an axis if nothing on it clearly
  applies.
- Occurrence: single event | intermittent | periodic | rhythmic | continuous
- Loudness dynamics: fade in | fade out | sustained | sustained with occasional bursts |
  impact/abrupt
- Timbre dynamics: rising | falling | oscillating

A sound recurring over time is a temporal pattern, not multiple sources - describe the
  recurrence here. Never count repetitions.

Context already extracted (judge with your ears; do NOT just repeat this):
- Category: <UCS category> / <UCS subcategory>
- Keywords: <library keywords>
- Description: <library description>
- Duration: <clip duration>s
- Events (start-end, loudness; more negative = quieter): ev0[<start>-<end>s, <LUFS>],
  ev1[...]
- Measured frequency-band energy (125 Hz -> 10 kHz), per event: ev0[125:<level>,
  250:<level>, ..., 10000:<level>], ev1[...]

Reply with ONLY a comma-separated list of tags from the vocabulary above, most prominent
  first; or the single word "none" if the audio gives no temporal evidence. No sentences, no
  explanation.
\end{verbatim}}\vspace{-6pt}\noindent\rule{\linewidth}{0.3pt}\par\medskip
\end{minipage}
\par\medskip

\subsubsection{Sound effects: plurality}
Qwen3-Omni says whether one, several or no countable sources sound at once.

\noindent\begin{minipage}{\linewidth}
\noindent\textbf{System prompt}\par\nobreak\noindent\rule{\linewidth}{0.3pt}\vspace{-2pt}
{\scriptsize\begin{verbatim}
You are an expert sound designer. You identify the PLURALITY of sound sources - how many
  distinct sources sound SIMULTANEOUSLY - NOT what the sources are, and NOT how a single
  source repeats over time.
\end{verbatim}}\vspace{-6pt}\noindent\rule{\linewidth}{0.3pt}\par\medskip
\end{minipage}
\par\medskip
\noindent\begin{minipage}{\linewidth}
\noindent\textbf{User prompt}\par\nobreak\noindent\rule{\linewidth}{0.3pt}\vspace{-2pt}
{\scriptsize\begin{verbatim}
Listen to this sound-effects clip and state how many distinct sources sound SIMULTANEOUSLY,
  as exactly ONE of these three words: none | one | multiple.

- Use "one" for a single source. One source that starts and stops repeatedly is still a
  single source - that belongs to temporal behavior, not plurality.
- Use "multiple" when several sources sound at once (two dogs, a crowd, overlapping
  footsteps).
- Use "none" when a source count is not applicable: a mass or diffuse phenomenon with no
  countable source (e.g. water, rain, wind, room ambience, synthetic tones).

The event count in the context is occurrences over time, never a source count.

Context already extracted (judge with your ears; do NOT just repeat this):
- Category: <UCS category> / <UCS subcategory>
- Keywords: <library keywords>
- Description: <library description>
- Duration: <clip duration>s
- Events (start-end, loudness; more negative = quieter): ev0[<start>-<end>s, <LUFS>],
  ev1[...]
- Measured frequency-band energy (125 Hz -> 10 kHz), per event: ev0[125:<level>,
  250:<level>, ..., 10000:<level>], ev1[...]

Reply with ONLY that single word. No sentences, no explanation.
\end{verbatim}}\vspace{-6pt}\noindent\rule{\linewidth}{0.3pt}\par\medskip
\end{minipage}
\par\medskip

\noindent\begin{minipage}{\linewidth}
\subsubsection{Sound effects: perceived audio quality}
Qwen3-Omni rates the recording quality of the sound-effect stem; the user prompt repeats the music one with sfx-specific wording.

\noindent\textbf{System prompt}\par\nobreak\noindent\rule{\linewidth}{0.3pt}\vspace{-2pt}
{\scriptsize\begin{verbatim}
You are an expert audio engineer and critical listener. You judge the PERCEIVED AUDIO
  QUALITY of a sound-effects recording: the production, technical, and channel/source
  characteristics you can actually hear, NOT what the sound is, NOT its sonic character, and
  NOT its mood.
\end{verbatim}}\vspace{-6pt}\noindent\rule{\linewidth}{0.3pt}\par\medskip
\noindent\textbf{User prompt (identical to Music: perceived audio quality except these lines)}\par\nobreak\noindent\rule{\linewidth}{0.3pt}\vspace{-2pt}
{\scriptsize\begin{verbatim}
Listen to this sound-effects clip and describe ONLY its PERCEIVED AUDIO QUALITY by choosing
  tags from the fixed vocabulary below. Do NOT tag what the source is or how the sound
  itself strikes the ear. Do NOT use any word that is not in the vocabulary.

Context already extracted (judge with your ears; do NOT just repeat this):
- Category: <UCS category> / <UCS subcategory>
- Keywords: <library keywords>
- Description: <library description>
- Duration: <clip duration>s
- Events (start-end, loudness; more negative = quieter): ev0[<start>-<end>s, <LUFS>],
  ev1[...]
- Measured frequency-band energy (125 Hz -> 10 kHz), per event: ev0[125:<level>,
  250:<level>, ..., 10000:<level>], ev1[...]


Reply with ONLY a comma-separated list of tags from the vocabulary above, most prominent
  first; or the single word "none" if the audio gives no evidence. No sentences, no
  explanation.
\end{verbatim}}\vspace{-6pt}\noindent\rule{\linewidth}{0.3pt}\par\medskip
\end{minipage}
\par\medskip

\noindent\begin{minipage}{\linewidth}
\subsubsection{Sound effects: sound descriptors}
Qwen3-Omni describes how the sound is perceived.

\noindent\textbf{System prompt}\par\nobreak\noindent\rule{\linewidth}{0.3pt}\vspace{-2pt}
{\scriptsize\begin{verbatim}
You are an expert sound designer. You characterize the SONIC CHARACTER of a sound with plain
  adjectives and adverbs - how it strikes the ear - NOT what the source is, NOT its mood or
  emotional effect, and NOT its recording quality.
\end{verbatim}}\vspace{-6pt}\noindent\rule{\linewidth}{0.3pt}\par\medskip
\noindent\textbf{User prompt}\par\nobreak\noindent\rule{\linewidth}{0.3pt}\vspace{-2pt}
{\scriptsize\begin{verbatim}
Listen to this sound-effects clip and give tags describing ONLY the SONIC CHARACTER of the
  sound itself - plain adjectives or adverbs, e.g. harsh, hollow, sharp, rumbling, hurried,
  gentle, metallic, scraping.

Context already extracted (judge with your ears; do NOT just repeat this):
- Category: <UCS category> / <UCS subcategory>
- Keywords: <library keywords>
- Description: <library description>
- Duration: <clip duration>s
- Events (start-end, loudness; more negative = quieter): ev0[<start>-<end>s, <LUFS>],
  ev1[...]
- Measured frequency-band energy (125 Hz -> 10 kHz), per event: ev0[125:<level>,
  250:<level>, ..., 10000:<level>], ev1[...]


Do not use mood words (e.g. eerie, cheerful) or channel/production words (e.g. muffled,
  reverberant) - only how the sound itself strikes the ear.

Reply with ONLY a comma-separated list of 2-5 tags, most prominent first, or the single word
  "none" if the audio gives no evidence. No sentences, no explanation.
\end{verbatim}}\vspace{-6pt}\noindent\rule{\linewidth}{0.3pt}\par\medskip
\end{minipage}
\par\medskip

\subsubsection{Sound effects: sounding object and action}
Qwen3-Omni keeps only the objects and actions from the source file's metadata that it can actually hear.

\noindent\begin{minipage}{\linewidth}
\noindent\textbf{System prompt}\par\nobreak\noindent\rule{\linewidth}{0.3pt}\vspace{-2pt}
{\scriptsize\begin{verbatim}
You are an expert sound designer. You identify WHICH SOURCES are audible in a sound-effects
  clip - the object making the sound and what it does - NOT how the sound behaves over time,
  NOT how many sources there are, and NOT its production quality.
\end{verbatim}}\vspace{-6pt}\noindent\rule{\linewidth}{0.3pt}\par\medskip
\end{minipage}
\par\medskip
\noindent\textbf{User prompt}\par\nobreak\noindent\rule{\linewidth}{0.3pt}\vspace{-2pt}
{\scriptsize\begin{verbatim}
Listen to this sound-effects clip and name ONLY the sounding objects and actions you can
  actually hear, choosing them from the source metadata below. Do NOT use any word that is
  not in the metadata.

The metadata covers the whole source recording, so it names more than this clip contains.
  Cutting it down is the task:
- Keep a term only if you clearly hear it here; drop every other one, object or action
  alike.
- Never add a source the metadata does not name, even if you think you hear it, and never
  swap a term for a broader, narrower, or synonymous one.
- objects: the noun(s) producing the sound. E.g. dog, microwave, wooden door.
- actions: what they do, in -ing form. E.g. barking, beeping, breaking. Inflecting a term
  you keep ("Snarls" -> "snarling") is the only rewording allowed.
- At most 3 of each, most prominent first.

Context already extracted (judge with your ears; do NOT just repeat this):
- Category: <UCS category> / <UCS subcategory>
- Keywords: <library keywords>
- Description: <library description>
- Duration: <clip duration>s
- Events (start-end, loudness; more negative = quieter): ev0[<start>-<end>s, <LUFS>],
  ev1[...]
- Measured frequency-band energy (125 Hz -> 10 kHz), per event: ev0[125:<level>,
  250:<level>, ..., 10000:<level>], ev1[...]


Reply with ONLY these two lines:
objects: <comma-separated list>
actions: <comma-separated list>
Write "none" on a line you can confirm nothing for. No sentences, no explanation.
\end{verbatim}}\vspace{-6pt}\noindent\rule{\linewidth}{0.3pt}\par\medskip

\subsection{Ground-truth generation: caption stage (Qwen3.6-35B-A3B-FP8)}\label{app:prompts:caption}
\subsubsection{Music card}
Qwen3.6 checks the music tags and writes the two captions.

\noindent\textbf{System prompt: instructions}\par\nobreak\noindent\rule{\linewidth}{0.3pt}\vspace{-2pt}
{\scriptsize\begin{verbatim}
You are an expert audio engineer validating and captioning the music in a 10-second
  soundscape. Each request gives you one music card, built from measured signal features and
  from tags predicted by models that listened to the isolated music.

A soundscape has at most one music card: it covers ALL the music in the clip, and each
  continuous stretch of music appears as its own event line. The card has:
- music events: one line per stretch of music, with its start-end time as MM:SS.mmm (from
  segments), its loudness as "LUFS, discrete label" (more negative LUFS = quieter; discrete
  label is one of Very loud / Loud / Medium / Quiet / Very quiet / Extremely quiet), and a
  frequency profile of ten band-energy labels. The ten labels run low to high frequency over
  the 125, 250, 500, 1000, 2000, 3000, 4000, 6000, 8000 and 10000 Hz bands, each a
  seven-level energy: very-low, low, medium-low, medium, medium-high, high, very-high.
- tempo: a bare string - either "{bpm}, {tempo word}" or "no tempo" / none detected.
- perceived_audio_quality, music_descriptors, instruments, mood: tag fields predicted by
  models that listened to the isolated music (see field descriptions below). genre is
  read-only caption context - never revise it or tempo.

Do two things with this card:
1. Check every tag in perceived_audio_quality, music_descriptors, instruments, and mood
  against the field descriptions below. MOVE a tag to whichever of these four fields it fits
  better - drop it rather than duplicate it if the target field already says the same thing.
  REMOVE a tag only where it contradicts another of the four fields, the genre, or the
  signal features (e.g. "high-fidelity" and "low-fidelity" both in perceived_audio_quality;
  a "calmness" mood on a Very loud, distorted event). Different wording for the same
  impression is NOT a contradiction. Do not invent tags. Then prune perceived_audio_quality,
  music_descriptors, and mood to the 2-5 most salient tags each - drop near-duplicates and
  weaker extras; instruments stays as the instruments heard (usually already short). genre
  and tempo are read-only context for the captions - never revise them.
2. Write short_caption and descriptive_caption from the CORRECTED fields.

Field descriptions. The four tag fields each have a FIXED vocabulary: a tag may only be
  kept, moved, or dropped - never reworded, normalized, or replaced with a synonym (keep
  "calmness", not "calm"; keep "low-fidelity", not "lo-fi"). A tag that belongs to no
  field's vocabulary is dropped.
- perceived_audio_quality: production/technical/channel character of the recording, without
  stating a cause. Vocabulary (at most one per axis): fidelity low-fidelity | clear |
  high-fidelity; space dry | room reverb | reverb-heavy; frequency balance bass-heavy |
  muffled | thin | harsh (up to two); drive distorted; noise low background noise | high
  background noise. Not mood, not genre, not arrangement.
- music_descriptors: how the music is put together - texture, arrangement, performance
  character. Vocabulary: rhythmic, layered, repetitive, sparse, sustained, dissonant,
  syncopated, percussive, minimalist, complex, arpeggiated, synthetic, glitchy, gritty,
  delicate, plucked, raw, melodic, warm, groovy, bright, abrupt, harmonic, high-pitched,
  rapid, intricate, punchy, orchestral, unpolished, detuned, lively, looping, evolving,
  static, ornamental, frenetic. Not mood, not genre, not an instrument name.
- instruments: the instruments heard, from the fixed 58-item instrument vocabulary (piano,
  electric piano, organ, harpsichord, synthesizer, accordion, guitar, electric guitar,
  acoustic guitar, bass guitar, ukulele, banjo, mandolin, lute, sitar, harp, violin, viola,
  cello, double bass, strings, trumpet, trombone, tuba, french horn, brass, flute, piccolo,
  clarinet, oboe, bassoon, saxophone, bagpipes, harmonica, woodwinds, drums, drum machine,
  percussion, cymbals, timpani, congas, tabla, darbuka, timbales, shaker, triangle, gong,
  handpan, glockenspiel, xylophone, marimba, vibraphone, bell, chimes, vocals, choir,
  whistling, beatboxing). Not genres or ensemble adjectives.
- mood: the emotional feel of the music, 1-2 categories from the fixed music-emotion
  vocabulary: amazement, solemnity, tenderness, nostalgia, calmness, power, joyful
  activation, tension, sadness. Not tempo or loudness words.
- short_caption: 2-4 words in the form [Adjective] [Noun] [Verb] - the adjective drawn from
  the genre, mood, or tempo; the noun the music, melody, or its leading instrument; the verb
  what it does over the clip. E.g. "Rock music fades"; "Happy melody plays"; "Driving
  indie-rock blares".
- descriptive_caption: one natural sentence describing the music - mood, genre, instruments,
  tempo feel, loudness character, recording character, and how it unfolds over the clip.

Caption rules:
- Ground every word in the card. Do not invent content the card does not state: no song
  title, artist, year, lyric, or language.
- Never echo a number or tier label (timestamp, LUFS number, BPM, band level) - translate
  them into listener language: the tempo word into pace wording, the discrete loudness label
  into how loud it sounds, band shape into spectral wording (e.g. bass-heavy, bright), and
  the number, length, and spacing of events into how the music unfolds (e.g. "plays
  throughout", "pausing briefly before resuming"). When tempo is "no tempo" / none detected,
  do not characterize the pace.
- Events are continuous stretches of music over time, not separate songs or tracks.
- Name the one or two most specific genres that fit, not the whole list. Do not force every
  tag into the captions: pick what a listener would notice first and let the rest go
  unmentioned.
\end{verbatim}}\vspace{-6pt}\noindent\rule{\linewidth}{0.3pt}\par\medskip

\noindent\begin{minipage}{\linewidth}
\noindent\textbf{System prompt: output rules}\par\nobreak\noindent\rule{\linewidth}{0.3pt}\vspace{-2pt}
{\scriptsize\begin{verbatim}
Formatting:
- Output only a JSON object with exactly these 6 keys, in order: perceived_audio_quality,
  music_descriptors, instruments, mood, short_caption, descriptive_caption.
- perceived_audio_quality, music_descriptors, instruments, and mood are each a single plain
  string - the corrected tags for that field as a comma-separated list, or an empty string
  if nothing supports the field. For perceived_audio_quality, music_descriptors, and mood,
  prefer 2-5 tags each.
- short_caption and descriptive_caption are plain strings.
- No prose, no markdown, no code fences, and do not repeat the card's other fields.
\end{verbatim}}\vspace{-6pt}\noindent\rule{\linewidth}{0.3pt}\par\medskip
\end{minipage}
\par\medskip

\subsubsection{Speech card (with transcript)}
Qwen3.6 checks the speech tags and writes the two captions, using the transcript.

\noindent\textbf{System prompt: instructions}\par\nobreak\noindent\rule{\linewidth}{0.3pt}\vspace{-2pt}
{\scriptsize\begin{verbatim}
You are an expert audio engineer validating and captioning the speech in a 10-second
  soundscape. Each request gives you one speech card, built from measured signal features, a
  transcript from an ASR pass, and tags predicted by models that listened to the isolated
  speech.

A soundscape has at most one speech card: it covers ALL the speech in the clip, and every
  separate utterance appears as its own event line. The card has:
- speech events: one line per utterance, with its start-end time as MM:SS.mmm (from
  segments), its loudness as "LUFS, discrete label" (more negative LUFS = quieter; discrete
  label is one of Very loud / Loud / Medium / Quiet / Very quiet / Extremely quiet), the
  median voice pitch in Hz, and a frequency profile of ten band-energy labels. The ten
  labels run low to high frequency over the 125, 250, 500, 1000, 2000, 3000, 4000, 6000,
  8000 and 10000 Hz bands, each a seven-level energy: very-low, low, medium-low, medium,
  medium-high, high, very-high.
- transcript: an ASR transcript of what is said, one segment per utterance, each with its
  own time span and a detected language for the whole card. It may be missing, empty, or too
  garbled to carry meaning.
- intelligibility and emotion: MEASURED fields (SQUIM intelligibility estimates, audeering
  arousal/dominance/valence) whose word tags are fixed by thresholds. Copy them EXACTLY as
  given - never add, drop, or change a word in these two fields.
- perceived_audio_quality, manner: tag fields predicted by a model that listened to the
  isolated speech (see field descriptions below).

Do two things with this card:
1. Copy intelligibility and emotion unchanged. Check every tag in perceived_audio_quality
  and manner against the field descriptions below. MOVE a tag to whichever of these two
  fields it fits better - drop it rather than duplicate it if the target field already says
  the same thing. REMOVE a tag only where it contradicts another of the four fields or the
  signal features (e.g. an "animated, expressive" manner on a Very quiet event with degraded
  intelligibility). Different wording for the same impression is NOT a contradiction. Keep
  every other tag unchanged: the model heard the clip, and a plausible tag is a valid
  observation, not an error.
2. Write short_caption and descriptive_caption from the CORRECTED fields.

Field descriptions:
- intelligibility: MEASURED (SQUIM). Words come from fixed thresholds on STOI/PESQ:
  highly/mostly intelligible or degraded intelligibility, optionally clean channel or poor
  channel quality. Copy exactly as given; a card with no channel word simply has no channel
  word.
- perceived_audio_quality: production/technical/channel character of the recording, without
  stating a cause. FIXED vocabulary (at most one per axis): fidelity low-fidelity | clear |
  high-fidelity; space dry | room reverb | reverb-heavy; frequency balance bass-heavy |
  muffled | thin | harsh (up to two); drive distorted; noise low background noise | high
  background noise. Keep, move, or drop tags - never reword or replace a vocabulary tag with
  a synonym, and never write a word outside this list. Not delivery, not emotion.
- emotion: MEASURED (audeering arousal/dominance/valence mapped to one word: neutral,
  excited/happy, angry/stressed, sad/subdued, calm, ...). Copy exactly as given.
- manner: delivery characteristics ONLY - pace, rhythm, articulation, vocal effort, tone of
  delivery (e.g. gruff, whispered, conversational, animated, hesitant, monotonous, breathy,
  articulate). manner must NOT contain: pitch words (low-pitched, high-pitched, deep,
  mid-range - pitch is the measured median F0), recording/channel words (clear, muffled,
  distorted, thin, reverberant, low-fidelity - that is perceived_audio_quality; say
  "articulate" for clear diction), or emotion words (calm, angry, happy, sad, stressed,
  excited, neutral, relaxed as a mood - that is the measured emotion field). Drop such tags
  from manner; never reword them; move a channel word to perceived_audio_quality only if it
  is on that field's list.
- short_caption: 2-4 words naming the most salient character of the speech, without gender -
  use pitch when useful. E.g. "Angry deep voice shouting".
- descriptive_caption: one natural sentence describing the speech, what about (quoted when
  brief), how it sounds (including pitch from median F0), and how it unfolds over the clip -
  without labeling speaker gender.

Caption rules:
- Ground every word in the card. Do not invent content the card does not state: no spoken
  words, topics, names, or languages beyond what the transcript gives you.
- manner must not contradict the measured emotion: drop a manner tag that asserts the
  opposite mood of the emotion field (e.g. "calm, relaxed" next to emotion "angry/stressed",
  or "animated, energetic" next to "sad/subdued"). The emotion field itself is never
  changed.
- Do not label the speaker's gender (no man/woman/male/female/boy/girl). Prefer
  gender-neutral wording (speaker, voice, someone). As a compromise, convey speaker
  character via median F0 as pitch wording (deep / low / mid / high voice) instead of
  gender.
- Never echo a number or tier label (timestamp, LUFS number, Hz, band level) - translate
  them into listener language: median F0 into deep-voice/high-voice wording, band shape into
  spectral wording, and the number, length, and spacing of events into how the speech
  unfolds (e.g. "speaks throughout", "in two short bursts", "a brief remark near the end").
- Events are utterances over time, not separate speakers. 
- Do not force every tag into the captions: pick what a listener would notice first and let
  the rest go unmentioned.
- Transcript handling: incorporate what is said into descriptive_caption. Quote the
  transcript verbatim only when an utterance is just a few words; otherwise paraphrase the
  topic in a short phrase (e.g. "talking about a recipe"). If the language is not English,
  name the language and do not quote or translate. If the transcript is missing, empty, or
  too garbled to carry meaning, do not mention spoken content at all. Spoken content may
  come ONLY from the transcript - never invent words, topics, names, or languages beyond it.
  The transcript is caption material only: never use it to remove or change a tag - spoken
  words do not determine delivery or emotion.
\end{verbatim}}\vspace{-6pt}\noindent\rule{\linewidth}{0.3pt}\par\medskip

\noindent\begin{minipage}{\linewidth}
\noindent\textbf{System prompt: output rules}\par\nobreak\noindent\rule{\linewidth}{0.3pt}\vspace{-2pt}
{\scriptsize\begin{verbatim}
Formatting:
- Output only a JSON object with exactly these 6 keys, in order: intelligibility,
  perceived_audio_quality, emotion, manner, short_caption, descriptive_caption.
- intelligibility, perceived_audio_quality, emotion, and manner are each a single plain
  string - the corrected tags for that field as a comma-separated list, or an empty string
  if nothing supports the field.
- short_caption and descriptive_caption are plain strings.
- No prose, no markdown, no code fences, and do not repeat the card's other fields.
\end{verbatim}}\vspace{-6pt}\noindent\rule{\linewidth}{0.3pt}\par\medskip
\end{minipage}
\par\medskip

\noindent\begin{minipage}{\linewidth}
\subsubsection{Speech card (without transcript)}
Same as above without the transcript: the captions describe only how the speech sounds.

\noindent\textbf{System prompt: instructions (identical to Speech card (with transcript) except these lines)}\par\nobreak\noindent\rule{\linewidth}{0.3pt}\vspace{-2pt}
{\scriptsize\begin{verbatim}
You are an expert audio engineer validating and captioning the speech in a 10-second
  soundscape. Each request gives you one speech card, built from measured signal features
  and tags predicted by models that listened to the isolated speech. NO transcript is
  provided: you do not know what is said.
- descriptive_caption: one natural sentence describing the speech - how it sounds (including
  pitch from median F0), its delivery and emotion, and how it unfolds over the clip -
  without labeling speaker gender and without any spoken content.
- Ground every word in the card. Do not invent content the card does not state: no spoken
  words, no topics, no names, no language - the card carries no transcript, so the captions
  describe only how the speech sounds, never what it says.
- There is no transcript (the card shows "transcript: none"). Never mention, guess, or hint
  at spoken content, topic, or language.
\end{verbatim}}\vspace{-6pt}\noindent\rule{\linewidth}{0.3pt}\par\medskip
\end{minipage}
\par\medskip

\subsubsection{Sound-effect card}
Qwen3.6 checks the sound-effect tags and writes the two captions.

\noindent\textbf{System prompt: instructions}\par\nobreak\noindent\rule{\linewidth}{0.3pt}\vspace{-2pt}
{\scriptsize\begin{verbatim}
You are an expert sound designer writing structured metadata for a sound-effects library.
  Each request gives you one sfx card for one sound event:
- metadata: the source file's text (Keywords, Description, Library, Filename) and its UCS
  category and subcategory. Read-only context for the captions - never revise it.
- sfx events: one line per sound event, with its start-end time as MM:SS.mmm (from
  segments), its loudness as "LUFS, discrete label" (more negative LUFS = quieter; discrete
  label is one of Very loud / Loud / Medium / Quiet / Very quiet / Extremely quiet), and its
  frequency profile as ten band-energy labels (125 Hz to 10 kHz, low to high; each a
  seven-level energy: very-low, low, medium-low, medium, medium-high, high, very-high).
- sounding_object, sounding_action, temporal_descriptors, plurality,
  perceived_audio_quality, sound_descriptors: six tag fields predicted by a model that
  listened to the isolated sound event (see field descriptions below). sounding_object and
  sounding_action were chosen by that listener from the metadata's own wording, keeping only
  what it could hear in this clip - the metadata covers the whole source recording and names
  more than the clip contains.

From these you write the full audiocard body:
1. sounding_object and sounding_action come from the listener, which picked them out of the
  metadata and kept only what it could hear. Two edits are open to you, and both only ever
  take terms OUT.
   (a) MOVE a term to a field where it fits better. sound_descriptors, the open field, is
     the usual destination - a "metallic" sitting in sounding_object says how the sound
     strikes the ear, not what produced it, so it belongs there. temporal_descriptors,
     plurality and perceived_audio_quality are CLOSED vocabularies: they can never receive a
     word that is not already on their list. A term that IS on the list may move there (a
     "rhythmic" in sounding_action is a temporal_descriptors word); one that is not goes to
     sound_descriptors instead, or is dropped if it describes no sonic character.
   (b) Where two kept terms name the same source ("lion, wild cat"), keep the clearer one.
   Nothing may come the other way: NEVER add an object or action the metadata names but the
     listener left out - those are the sources it could not hear in this clip, and putting
     them back is the one error these fields must not make - and never move a tag IN from
     another field. Never swap a term for a broader, narrower, or synonymous one. An empty
     field stays empty - write "".
2. Check every tag in the other four fields - temporal_descriptors, plurality,
  perceived_audio_quality, sound_descriptors - against the field descriptions below. MOVE a
  tag to another field where it is a better fit (e.g. "rushed" in perceived_audio_quality
  fits sound_descriptors better) - but never into sounding_object or sounding_action, which
  take nothing in from anywhere. Remove a tag only where it contradicts the metadata (e.g. a
  "muffled" perceived_audio_quality tag when the metadata describes a clean, close-mic'd
  recording; a "playful" sound_descriptors tag on a metadata-described distress call);
  different wording for the same source is not a contradiction. Do not invent tags. Keep all
  other tags unchanged, as the audio model heard the clip, and plausible tags the metadata
  is silent on are valid observations, not errors. Then prune sound_descriptors - counting
  anything moved in here, whether from these fields or from step 1 - to the 2-5 most salient
  tags, dropping near-duplicates and weaker extras. 2-5 is a ceiling and a preference, not a
  quota: if removals leave fewer than two, leave it short rather than inventing a tag to
  reach two.
3. Predict the short_caption and descriptive_caption from the corrected fields. The source
  identity comes from sounding_object and sounding_action: a caption may not name a source
  those two leave out, even when the metadata describes it. Use the metadata only for
  setting and manner that the kept fields already support (a caged lion may still be "in a
  cage"), and take the rest of the detail from the other fields.

Read signal features perceptually, and never echo a timestamp, LUFS value, band center, tier
  label, or discrete loudness label in any body field - translate them into listener
  language:
- number, length, and spacing of events should inform temporal_descriptors.
- the discrete loudness label should translate into how loud it sounds.
- band shape should inform spectral wording.
The events list counts occurrences over time, not sources at once: never infer plurality
  from the number of events.

Ground every field in the card. Do not add specifics none of it states, e.g. no brand or
  model names, no place names, and no source counts that are not given. If nothing supports
  a field, output an empty string "".

The body has these fields, in this order:
- sounding_object: noun(s) producing the sound, at most 3. E.g. dog, dogs, microwave, plate,
  wooden door. Comes from the listening pass: the metadata cut down to what is audible here,
  never the full metadata. A term that is really a quality or character word moves out to
  the field it belongs in; nothing ever moves in (see step 1).
- sounding_action: verb(s) for what the source does, in -ing form, at most 3. E.g. barking,
  beeping, breaking. Same source and same rule: terms may leave, none may enter.
- temporal_descriptors: how the sound occurs and evolves over the clip. FIXED vocabulary,
  grouped by axis: occurrence single event | intermittent | periodic | rhythmic | continuous
  (exactly one); loudness dynamics fade in | fade out | sustained | sustained with
  occasional bursts | impact/abrupt (at most one); timbre dynamics rising | falling |
  oscillating (at most one). Never count repetitions and never write a word outside this
  list.
- plurality: how many sources of sound at once, not multiple occurrences over time. Exactly
  one of: none | one | multiple. "one" = a single source (a single source that repeats over
  time is still "one"; the recurrence belongs in temporal_descriptors); "multiple" = several
  sources at once; "none" = a mass or diffuse phenomenon with no countable source (water,
  rain, wind, room ambience).
- perceived_audio_quality: perceived quality of the recording (production, technical,
  channel/source), without stating a cause. FIXED vocabulary (at most one per axis):
  fidelity low-fidelity | clear | high-fidelity; space dry | room reverb | reverb-heavy;
  frequency balance bass-heavy | muffled | thin | harsh (up to two); drive distorted; noise
  low background noise | high background noise. Not mood or genre.
For temporal_descriptors and perceived_audio_quality, tags may only be kept, moved, or
  dropped - never reworded, normalized, or replaced with a synonym; a tag outside the target
  field's vocabulary is dropped (or moved to sound_descriptors if it describes sonic
  character). plurality is a single required value, not a tag list: carry it over unchanged
  unless the metadata clearly contradicts it, in which case replace it with the correct one
  of none | one | multiple. Never drop it, never move it to another field, and never leave
  it empty.
- sound_descriptors: non-mood, non-genre adjectives or adverbs describing the sound. E.g.
  harsh, hurried, furiously.
- short_caption: 2-4 words. E.g. "Dog barks angrily"; "Hurried, loud footsteps".
- descriptive_caption: one natural sentence. E.g. "A group of angry dogs bark repeatedly in
  a cathedral."
\end{verbatim}}\vspace{-6pt}\noindent\rule{\linewidth}{0.3pt}\par\medskip

\noindent\begin{minipage}{\linewidth}
\noindent\textbf{System prompt: output rules}\par\nobreak\noindent\rule{\linewidth}{0.3pt}\vspace{-2pt}
{\scriptsize\begin{verbatim}
Formatting:
- Output only a JSON object with exactly these 8 keys, in the field order listed above:
  sounding_object, sounding_action, temporal_descriptors, plurality,
  perceived_audio_quality, sound_descriptors, short_caption, descriptive_caption.
- Every value is a plain string. For multiple values, comma-separate them inside the one
  string. Never output a JSON list/array. For sound_descriptors, prefer 2-5 tags; for
  sounding_object and sounding_action, at most 3 each.
- No prose, no markdown, no code fences, and do not repeat the card's other fields.
\end{verbatim}}\vspace{-6pt}\noindent\rule{\linewidth}{0.3pt}\par\medskip
\end{minipage}
\par\medskip

\noindent\begin{minipage}{\linewidth}
\subsubsection{Re-captioning human-validated cards}
For human-validated cards, the tag-checking step of each prompt above is replaced by this instruction to copy the tags unchanged (music and sound-effect versions shown; speech is analogous).

\noindent\textbf{Music: replaced step}\par\nobreak\noindent\rule{\linewidth}{0.3pt}\vspace{-2pt}
{\scriptsize\begin{verbatim}
The tag fields of this card were checked and corrected by human reviewers: they are FINAL.
  Do two things with this card:
1. Copy perceived_audio_quality, music_descriptors, instruments and mood EXACTLY as given -
  the same words in the same order. Never add, drop, move, reword, reorder or prune a tag,
  even if a tag seems to contradict another field, the genre or the signal features; an
  empty field stays empty. genre and tempo are read-only context for the captions.
2. Write short_caption and descriptive_caption from these fields.
\end{verbatim}}\vspace{-6pt}\noindent\rule{\linewidth}{0.3pt}\par\medskip
\noindent\textbf{Sound effects: replaced step}\par\nobreak\noindent\rule{\linewidth}{0.3pt}\vspace{-2pt}
{\scriptsize\begin{verbatim}
From these you write the full audiocard body:
1. All six tag fields (sounding_object, sounding_action, temporal_descriptors, plurality,
  perceived_audio_quality, sound_descriptors) were checked and corrected by human reviewers
  and are FINAL: copy each one EXACTLY as given - the same words in the same order. Never
  add, drop, move, reword, reorder or prune a term, and never restore anything from the
  metadata; an empty field stays empty - write "".
2. Predict the short_caption and descriptive_caption from these fields.
\end{verbatim}}\vspace{-6pt}\noindent\rule{\linewidth}{0.3pt}\par\medskip
\end{minipage}
\par\medskip

\subsubsection{Sound-effect revision pass}
A second pass over each sound-effect card that removes proper nouns and repetition counts.

\noindent\begin{minipage}{\linewidth}
\noindent\textbf{System prompt: instructions}\par\nobreak\noindent\rule{\linewidth}{0.3pt}\vspace{-2pt}
{\scriptsize\begin{verbatim}
You are an expert sound designer revising one finished sfx card. Each request gives you the
  card's published fields. This is a minimal-edit revision pass: apply ONLY the edits below,
  and copy every word the edits do not touch verbatim. Do not re-tag, re-order, prune, or
  re-caption beyond what the edits require.

Edits:
1. sounding_object: remove proper nouns - brand names, product model names and years,
  franchise names, place names, personal names, and library names. Replace each with its
  generic head noun ("2001 Jeep Cherokee" -> "car"; "iPhone" -> "phone"; "Boeing 747" ->
  "jet airliner"); if no generic noun fits, keep as is. Common nouns that are NOT proper
  nouns, e.g. species, materials, and object types (lion, oak door, chickens) stay.
2. sounding_action: put every verb in lowercase -ing form ("Snarls" -> "snarling"; "knock"
  -> "knocking"). This is a wording fix only - never add a verb, never remove one, and never
  replace one with a different verb. The field comes from a listening pass that kept only
  the actions it could hear, so a verb that is missing is missing on purpose.
3. Captions: rewrite only the phrases affected by edits 1-2 - generalize the same proper
  nouns. Independently, quantize any exact repetition count that appears in a caption
  ("knocks four times" -> "knocks several times"): the tag fields carry no counts, so a
  caption must not assert one.

If an edit empties a field, output an empty string "". Never invent new tags or new caption
  content.
\end{verbatim}}\vspace{-6pt}\noindent\rule{\linewidth}{0.3pt}\par\medskip
\end{minipage}
\par\medskip
\noindent\begin{minipage}{\linewidth}
\noindent\textbf{System prompt: output rules}\par\nobreak\noindent\rule{\linewidth}{0.3pt}\vspace{-2pt}
{\scriptsize\begin{verbatim}
Formatting:
- Output only a JSON object with exactly these 8 keys, in order: sounding_object,
  sounding_action, temporal_descriptors, plurality, perceived_audio_quality,
  sound_descriptors, short_caption, descriptive_caption.
- Every value is a plain string; comma-separate multiple values inside the one string. Never
  output a JSON list/array. plurality is exactly one of "none", "one", "multiple".
- Fields untouched by the edits must be byte-identical to the input card.
- No prose, no markdown, no code fences.
\end{verbatim}}\vspace{-6pt}\noindent\rule{\linewidth}{0.3pt}\par\medskip
\end{minipage}
\par\medskip

\subsection{Benchmark inference prompts}\label{app:prompts:bench}
Every evaluated model receives the same four prompts.
\subsubsection{Stage 1: source inventory}
The evaluated model lists the sources in the scene and when each is audible.

\noindent\begin{minipage}{\linewidth}
\noindent\textbf{System prompt}\par\nobreak\noindent\rule{\linewidth}{0.3pt}\vspace{-2pt}
{\scriptsize\begin{verbatim}
You are an expert audio scene analyst. You identify the distinct sound sources present in an
  audio recording and when each one is audible.
\end{verbatim}}\vspace{-6pt}\noindent\rule{\linewidth}{0.3pt}\par\medskip
\end{minipage}
\par\medskip
\noindent\begin{minipage}{\linewidth}
\noindent\textbf{User prompt}\par\nobreak\noindent\rule{\linewidth}{0.3pt}\vspace{-2pt}
{\scriptsize\begin{verbatim}
Listen to this 10-second audio scene. It is a mixture that may contain speech, music, and
  sound effects, possibly overlapping.

List every DISTINCT sound source you can hear, and for each one the time span(s) during
  which it is audible. Rules:
- Each source has a type: speech, music, or sound effect.
- Sounds that belong to one semantic source count as ONE source (e.g. an engine and rolling
  wheels together are one source: a car). A constant featureless background noise bed is NOT
  a source; ignore it. But sustained environmental sound that characterizes a place (e.g.
  room tone, ventilation hum) IS a sound-effect source.
- All speech in the scene is one source; all music in the scene is one source; each distinct
  sound-effect source is its own entry.
- Time spans are in seconds from the start of the clip (0.0 to 10.0). A source that occurs
  several separate times gets several spans, separated by semicolons.

Reply with ONLY a numbered list, one source per line, in this exact format:
1. <type> | <brief name of the source> | <start>-<end>; <start>-<end>
No other text.
\end{verbatim}}\vspace{-6pt}\noindent\rule{\linewidth}{0.3pt}\par\medskip
\end{minipage}
\par\medskip

\subsubsection{Stage 2: music card}
The evaluated model fills a music card for one source it listed in Stage 1; \texttt{\{SOURCE\}} is its own name for that source.

\noindent\begin{minipage}{\linewidth}
\noindent\textbf{System prompt}\par\nobreak\noindent\rule{\linewidth}{0.3pt}\vspace{-2pt}
{\scriptsize\begin{verbatim}
You are an expert music analyst, musician and audio engineer with a precise ear. You
  describe ONLY the requested music source inside a mixed audio scene, judging with your
  ears: its mood, instruments, musical descriptors and perceived audio quality are DISTINCT
  fields - never repeat one in another.
\end{verbatim}}\vspace{-6pt}\noindent\rule{\linewidth}{0.3pt}\par\medskip
\end{minipage}
\par\medskip
\noindent\textbf{User prompt}\par\nobreak\noindent\rule{\linewidth}{0.3pt}\vspace{-2pt}
{\scriptsize\begin{verbatim}
Listen to this 10-second audio scene (a mixture that may contain speech, music and sound
  effects). Focus ONLY on this source: MUSIC - {SOURCE}. Ignore all other sounds in the
  mixture.

Fill a metadata card for this music with exactly the following fields. Judge with your ears.
  Where a field has a fixed vocabulary, do NOT use any word that is not in it. Keep every
  field distinct - do not repeat the same information across fields.

- "segments": the time span(s) where this source is audible, as [["MM:SS.mmm", "MM:SS.mmm"],
  ...]
- "loudness": loudness per segment as [[<LUFS value>, "<label>"], ...], label from:
  Extremely quiet | Very quiet | Quiet | Medium | Loud | Very loud
- "frequency_profile": energy distribution of this source per segment across ten bands (125,
  250, 500, 1000, 2000, 3000, 4000, 6000, 8000, 10000 Hz), as [{"bands_label": {"125":
  "<level>", "250": "<level>", "500": "<level>", "1000": "<level>", "2000": "<level>",
  "3000": "<level>", "4000": "<level>", "6000": "<level>", "8000": "<level>", "10000":
  "<level>"}}, ...] with levels from: very-low | low | medium-low | medium | medium-high |
  high | very-high
- "perceived_audio_quality": tags from this fixed vocabulary, grouped by axis - within an
  axis pick AT MOST ONE tag (Frequency balance: up to two if not contradictory); skip an
  axis if nothing clearly applies. Fidelity: low-fidelity | clear | high-fidelity. Space:
  dry | room reverb | reverb-heavy. Frequency balance: bass-heavy | muffled | thin | harsh.
  Drive: distorted. Noise: low background noise | high background noise. Most prominent
  first, as a list
- "music_descriptors": 2-5 tags for the character/texture/feel of the music (no genre, no
  mood, no instruments, no production quality) from this fixed vocabulary: rhythmic |
  layered | repetitive | sparse | sustained | dissonant | syncopated | percussive |
  minimalist | complex | arpeggiated | synthetic | glitchy | gritty | delicate | plucked |
  raw | melodic | warm | groovy | bright | abrupt | harmonic | high-pitched | rapid |
  intricate | punchy | orchestral | unpolished | detuned | lively | looping | evolving |
  static | ornamental | frenetic. Most prominent first, as a list
- "instruments": 1-6 instruments you can actually hear playing, most prominent first, from
  this fixed vocabulary: piano | electric piano | organ | harpsichord | synthesizer |
  accordion | guitar | electric guitar | acoustic guitar | bass guitar | ukulele | banjo |
  mandolin | lute | sitar | harp | violin | viola | cello | double bass | strings | trumpet
  | trombone | tuba | french horn | brass | flute | piccolo | clarinet | oboe | bassoon |
  saxophone | bagpipes | harmonica | woodwinds | drums | drum machine | percussion | cymbals
  | timpani | congas | tabla | darbuka | timbales | shaker | triangle | gong | handpan |
  glockenspiel | xylophone | marimba | vibraphone | bell | chimes | vocals | choir |
  whistling | beatboxing. Name the MOST SPECIFIC instrument you can confidently identify;
  family terms (guitar, strings, brass, woodwinds, percussion, choir) ONLY for whole
  sections or when the specific instrument cannot be determined; never both a family term
  and one of its members for the same sound. For guitars, commit to "acoustic guitar" or
  "electric guitar" whenever you can tell. "drum machine" = programmed/electronic beats;
  "drums" = an acoustic drum kit. As a list
- "genre": the musical genre label(s), using Free Music Archive genre names (e.g. Rock,
  Punk, Folk, Ambient Electronic, Hip-Hop, Experimental Pop), most specific first, as a list
- "mood": 1-2 categories, dominant first, from this fixed vocabulary (music-emotion
  categories): amazement (wonder and happiness) | solemnity (transcendence, inspiration,
  thrills) | tenderness (sensuality, affect, love) | nostalgia (dreamy, melancholic,
  sentimental) | calmness (relaxation, serenity, meditativeness) | power (strong, heroic,
  triumphant, energetic) | joyful activation (like dancing, bouncy, animated, amused) |
  tension (nervous, impatient, irritated, eerie, ominous, dread or menace) | sadness
  (depressed, sorrowful). As a list
- "tempo": the tempo as [<BPM>, "<label>"] with label from: very slow | slow | medium | fast
  | very fast; or the string "no tempo" if there is no discernible beat
- "short_caption": 2-4 words in the form [Adjective] [Noun] [Verb] - the adjective from the
  genre, mood, or tempo; the noun the music, melody, or its leading instrument; the verb
  what it does. E.g. "Rock music fades"; "Happy melody plays"; "Driving indie-rock blares"
- "descriptive_caption": one natural sentence describing the music - mood, genre,
  instruments, tempo feel, loudness character, recording character, and how it unfolds over
  the clip. Never echo numbers or tier labels; translate them into listener language

Reply with ONLY a single JSON object containing exactly these keys, in this order. No
  sentences outside the JSON, no explanation.
\end{verbatim}}\vspace{-6pt}\noindent\rule{\linewidth}{0.3pt}\par\medskip

\subsubsection{Stage 2: speech card}
The evaluated model fills a speech card for one source it listed in Stage 1.

\noindent\begin{minipage}{\linewidth}
\noindent\textbf{System prompt}\par\nobreak\noindent\rule{\linewidth}{0.3pt}\vspace{-2pt}
{\scriptsize\begin{verbatim}
You are an expert speech analyst and audio engineer with a precise ear. You describe ONLY
  the requested speech inside a mixed audio scene, judging with your ears: intelligibility,
  perceived audio quality, emotion and manner of speech are DISTINCT fields - never repeat
  one in another. What is said belongs ONLY in the transcript field (and, briefly, in the
  descriptive caption) - never in the tag fields.
\end{verbatim}}\vspace{-6pt}\noindent\rule{\linewidth}{0.3pt}\par\medskip
\end{minipage}
\par\medskip
\noindent\textbf{User prompt}\par\nobreak\noindent\rule{\linewidth}{0.3pt}\vspace{-2pt}
{\scriptsize\begin{verbatim}
Listen to this 10-second audio scene (a mixture that may contain speech, music and sound
  effects). Focus ONLY on this source: SPEECH - {SOURCE}. Ignore all other sounds in the
  mixture.

Fill a metadata card for the SPEECH with exactly the following fields. Judge with your ears.
  Where a field has a fixed vocabulary, do NOT use any word that is not in it. Keep every
  field distinct - do not repeat the same information across fields.

- "segments": the time span(s) where this source is audible, as [["MM:SS.mmm", "MM:SS.mmm"],
  ...]
- "loudness": loudness per segment as [[<LUFS value>, "<label>"], ...], label from:
  Extremely quiet | Very quiet | Quiet | Medium | Loud | Very loud
- "frequency_profile": energy distribution of this source per segment across ten bands (125,
  250, 500, 1000, 2000, 3000, 4000, 6000, 8000, 10000 Hz), as [{"bands_label": {"125":
  "<level>", "250": "<level>", "500": "<level>", "1000": "<level>", "2000": "<level>",
  "3000": "<level>", "4000": "<level>", "6000": "<level>", "8000": "<level>", "10000":
  "<level>"}}, ...] with levels from: very-low | low | medium-low | medium | medium-high |
  high | very-high
- "median_f0_hz": the median fundamental frequency (voice pitch) per speech segment, as
  [[start, end, Hz], ...] with start/end in seconds
- "perceived_audio_quality": tags from this fixed vocabulary, grouped by axis - within an
  axis pick AT MOST ONE tag (Frequency balance: up to two if not contradictory); skip an
  axis if nothing clearly applies. Fidelity: low-fidelity | clear | high-fidelity. Space:
  dry | room reverb | reverb-heavy. Frequency balance: bass-heavy | muffled | thin | harsh.
  Drive: distorted. Noise: low background noise | high background noise. Do NOT tag the
  spoken content, the emotion, or intelligibility. Most prominent first, as a list
- "manner": tags for the MANNER of speech - delivery only: pace, rhythm, articulation, vocal
  effort, tone of delivery (e.g. gruff, whispered, conversational, animated, hesitant,
  monotonous, breathy, articulate). Do NOT use pitch words (low-pitched, high-pitched, deep
  - pitch belongs to median_f0_hz), recording/channel words (clear, muffled, distorted,
  thin, reverberant - that is perceived_audio_quality), or emotion words (calm, angry,
  happy, sad, stressed, neutral - that is emotion). 2-5 tags, most prominent first, as a
  list
- "intelligibility": how understandable the speech is and how clean its channel is, as
  {"words": [..]}. Exactly one of: highly intelligible | mostly intelligible | degraded
  intelligibility; optionally followed by one of: clean channel | poor channel quality
  (leave the channel word out when neither clearly applies)
- "emotion": the speaker's emotional state, as {"words": [..]} with exactly one of: neutral
  | excited/happy | angry/stressed | sad/subdued | calm
- "transcript": what is said, as {"language": "<ISO 639-1 code, e.g. en>", "segments":
  [{"start": <seconds>, "end": <seconds>, "text": "<verbatim words>"}, ...]}, one segment
  per utterance. If you cannot make out the words, use an empty "segments" list - never
  invent words
- "short_caption": 2-4 words naming the most salient character of the speech, without gender
  - use pitch when useful. E.g. "Angry deep voice shouting"
- "descriptive_caption": one natural sentence describing the speech: what it is about
  (quoted only when brief), how it sounds (including pitch as deep/low/mid/high voice), and
  how it unfolds over the clip - without labeling speaker gender (no
  man/woman/male/female/boy/girl)

Reply with ONLY a single JSON object containing exactly these keys, in this order. No
  sentences outside the JSON, no explanation.
\end{verbatim}}\vspace{-6pt}\noindent\rule{\linewidth}{0.3pt}\par\medskip

\subsubsection{Stage 2: sound-effect card}
The evaluated model fills a sound-effect card for one source it listed in Stage 1.

\noindent\begin{minipage}{\linewidth}
\noindent\textbf{System prompt}\par\nobreak\noindent\rule{\linewidth}{0.3pt}\vspace{-2pt}
{\scriptsize\begin{verbatim}
You are an expert sound designer describing one sound-effect source inside a mixed audio
  scene by listening. Base every field on what you hear. The sounding object, its action,
  temporal behavior (how it occurs and evolves over the clip), plurality (how many distinct
  sources sound simultaneously), sonic character (how it strikes the ear) and perceived
  audio quality (channel/production, without naming a cause) are DISTINCT fields - never
  repeat one in another.
\end{verbatim}}\vspace{-6pt}\noindent\rule{\linewidth}{0.3pt}\par\medskip
\end{minipage}
\par\medskip
\noindent\textbf{User prompt}\par\nobreak\noindent\rule{\linewidth}{0.3pt}\vspace{-2pt}
{\scriptsize\begin{verbatim}
Listen to this 10-second audio scene (a mixture that may contain speech, music and sound
  effects). Focus ONLY on this source: SOUND EFFECT - {SOURCE}. Ignore all other sounds in
  the mixture.

Fill a metadata card for this sound with exactly the following fields. Judge with your ears.
  Where a field has a fixed vocabulary, do NOT use any word that is not in it. Keep every
  field distinct - do not repeat the same information across fields.

- "segments": the time span(s) where this source is audible, as [["MM:SS.mmm", "MM:SS.mmm"],
  ...]
- "loudness": loudness per segment as [[<LUFS value>, "<label>"], ...], label from:
  Extremely quiet | Very quiet | Quiet | Medium | Loud | Very loud
- "frequency_profile": energy distribution of this source per segment across ten bands (125,
  250, 500, 1000, 2000, 3000, 4000, 6000, 8000, 10000 Hz), as [{"bands_label": {"125":
  "<level>", "250": "<level>", "500": "<level>", "1000": "<level>", "2000": "<level>",
  "3000": "<level>", "4000": "<level>", "6000": "<level>", "8000": "<level>", "10000":
  "<level>"}}, ...] with levels from: very-low | low | medium-low | medium | medium-high |
  high | very-high
- "sounding_object": noun(s) producing the sound - generic names, no
  brand/model/place/personal names. E.g. dog, dogs, microwave, plate, wooden door. As a list
- "sounding_action": verb(s) for what the source does. E.g. barking, beeping, breaking. As a
  list
- "temporal_descriptors": the TEMPORAL BEHAVIOR of the sound over the clip, from this fixed
  vocabulary grouped by axis - pick EXACTLY ONE Occurrence tag; from the other axes pick AT
  MOST ONE tag each, only if clearly audible. Occurrence: single event | intermittent |
  periodic | rhythmic | continuous. Loudness dynamics: fade in | fade out | sustained |
  sustained with occasional bursts | impact/abrupt. Timbre dynamics: rising | falling |
  oscillating. A sound recurring over time is a temporal pattern, not multiple sources -
  describe the recurrence here. Never count repetitions. As a list
- "plurality": how many distinct sources sound SIMULTANEOUSLY - exactly one of none | one |
  multiple. One source that starts and stops repeatedly is still one source (that belongs to
  temporal behavior). Use none for a mass or diffuse phenomenon with no countable source
  (water, rain, wind, room ambience)
- "perceived_audio_quality": tags from this fixed vocabulary, grouped by axis - within an
  axis pick AT MOST ONE tag (Frequency balance: up to two if not contradictory); skip an
  axis if nothing clearly applies. Fidelity: low-fidelity | clear | high-fidelity. Space:
  dry | room reverb | reverb-heavy. Frequency balance: bass-heavy | muffled | thin | harsh.
  Drive: distorted. Noise: low background noise | high background noise. NOT what the sound
  is, NOT its mood. Most prominent first, as a list
- "sound_descriptors": tags describing ONLY the SONIC CHARACTER of the sound itself - plain
  adjectives or adverbs, e.g. harsh, hollow, sharp, rumbling, hurried, gentle, metallic,
  scraping. Do not use mood words (e.g. eerie, cheerful) or channel/production words (e.g.
  muffled, reverberant) - only how the sound itself strikes the ear. 2-5 tags, most
  prominent first, as a list
- "short_caption": 2-4 words. E.g. "Dog barks angrily"; "Hurried, loud footsteps"
- "descriptive_caption": one natural sentence. E.g. "A group of angry dogs bark repeatedly
  in a cathedral." Ground every word in what you hear; no brand or model names, no place
  names

Reply with ONLY a single JSON object containing exactly these keys, in this order. No
  sentences outside the JSON, no explanation.
\end{verbatim}}\vspace{-6pt}\noindent\rule{\linewidth}{0.3pt}\par\medskip
\color{black}

\end{document}